\documentstyle[12pt,epic,misc]{article}
\begin{document}
\title{\bf POLYNOMIAL IDENTITIES\\ 
FOR HYPERMATRICES}

\author{{\bf Victor Tapia}\\
\\
{\it Departamento de Matem\'aticas}\\
{\it Universidad Nacional de Colombia}\\
{\it Bogot\'a, Colombia}\\
\\
{\tt tapiens@ciencias.unal.edu.co}}

\maketitle

\begin{abstract}
We develop an algorithm to construct algebraic invariants for hypermatrices. We then construct hyperdeterminants and exhibit a generalization of the Cayley--Hamilton theorem for hypermatrices.
\end{abstract}

\section{Introduction}

Hypermatrices appear in several contexts in mathematics \cite{WE,WZ1,WZ2} and in applications such as in the quantum mechanics of entangled states \cite{ALP,CKW}, and image processing \cite{AS1,AS2}. Important mathematical problems in the above applications are the construction of algebraic invariants associated to hypermatrices and the determination of the minimal number of algebraically independent invariants. In this work we address these last two problems. 

For ordinary matrices the algebraic invariants associated to a matrix ${\bf a}$ can be obtained as traces of powers of the given matrix. According to the Cayley--Hamilton theorem only a finite number of these powers is linearly independent and therefore only a finite number of algebraic invariants is linearly independent. A more convenient set of invariants is given by the discriminants which are suitable combinations of traces and are constructed in terms of alternating products with the unit matrix ${\bf I}$. 

Therefore, algebraic invariants (discriminants) can be constructed by considering all possible products among ${\bf a}$'s and ${\bf I}$'s and these products are in a one--to--one correspondence with semi--magic squares of rank 2 (a semi--magic square is a square array of numbers such that the sum of element in each row and column gives the same result). The discriminants can be obtained in terms of semi--magic squares by counting all possible permutations of indices and for practical purposes the discriminants can be constructed using a graphical algorithm in terms of grids which we develop here.

We then apply the above algorithm to hyper--matrices. We restrict our considerations to the fourth--rank case. We obtain the corresponding discriminants, the determinant and the Cayley--Hamilton theorem. 

\section{Matrix Calculus}

\subsection{Basic Definitions}

A matrix ${\bf a}$ is a $d\times d$ square array of numbers $a_{ij}$, with $i,j=1,\cdots,d$. Let ${\bf a}$ and ${\bf b}$ be two matrices with components $a_{ij}$ and $b_{ij}$, respectively. The matrix addition ``$+$'' is defined by the resulting matrix ${\bf c}={\bf a}+{\bf b}$ with components

\begin{equation}
c_{ij}=\,a_{ij}+b_{ij}\,.
\label{001}
\end{equation}

\noindent The matrix multiplication ``$\cdot$'' is defined by the resulting matrix ${\bf c}={\bf a}\cdot{\bf b}$ with components

\begin{equation}
c_{ij}=\sum_{k=1}^d\,a_{ik}\,b_{kj}\,.
\label{002}
\end{equation}

\noindent This operation is the usual Cartesian product among rows and columns. Therefore, we have a ring.

The unit element ${\bf I}$ for the matrix multiplication, ${\bf a}\cdot{\bf I}={\bf I}\cdot{\bf a}={\bf a}$, has components $I_{ij}$ given by

\begin{equation}
I_{ij}=\cases{1\,,&for $i=j$\cr0\,,&otherwise\cr}\,.
\label{003}
\end{equation}

The inverse matrix ${\bf a}^{-1}$ is a matrix satisfying

\begin{equation}
{\bf a}^{-1}\,\cdot\,{\bf a}={\bf a}\,\cdot\,{\bf a}^{-1}={\bf I}\,.
\label{004}
\end{equation}

\noindent In terms of componentes we have

\begin{equation}
\sum_{k=1}^d\,(a^{-1})_{ik}\,a_{kj}=\sum_{k=1}^d\,a_{ik}\,(a^{-1})_{kj}=I_{ij}\,.
\label{005}
\end{equation}

\noindent The conditions for the existence of the inverse ${\bf a}^{-1}$ will be given later on. If the inverse ${\bf a}^{-1}$ exists then ${\bf a}$ is said to be a regular matrix.

The product of a matrix ${\bf a}$ with itself, that is ${\bf a}^2$, is the matrix with components

\begin{equation}
({\bf a}^2)_{ij}=\sum_{k=1}^d\,a_{ik}\,a_{kj}\,.
\label{006}
\end{equation} 

\noindent Products of ${\bf a}$ of an order $s$, ${\bf a}^s$, have the components

\begin{equation}
({\bf a}^s)_{ij}=\underbrace{\sum_{k_1=1}^d\,\cdots\,\sum_{k_{s-1}=1}^d}_{s-1\,{\rm times}}\,
\underbrace{a_{ik_1}\,\cdots\,a_{k_{s-1}j}}_{s\,{\rm times}}\,.
\label{007}
\end{equation}

\noindent By definition ${\bf a}^0={\bf I}$ and ${\bf a}^1={\bf a}$. Furthermore

\begin{equation}
{\bf a}^p\,\cdot\,{\bf a}^q={\bf a}^{p+q}\,.
\label{008}
\end{equation}

For a matrix ${\bf a}$ the trace is given by

\begin{equation}
\left<{\bf a}\right>={\rm trace}({\bf a})=\sum_{i=1}^d\,a_{ii}\,.
\label{009}
\end{equation}

\noindent The trace of ${\bf a}^s$ is given by

\begin{equation}
\left<{\bf a}^s\right>={\rm trace}({\bf a}^s)=\sum_{i=1}^d\,({\bf a}^s)_{ii}\,.
\label{010}
\end{equation}

\noindent Furthermore, ${\rm trace}({\bf a}^0)=d$. 

Let ${\bf u}$ be a regular matrix. Then we can define the similarity, or affine, transformation

\begin{equation}
{\bf a}'={\bf u}^{-1}\,\cdot\,{\bf a}\,\cdot\,{\bf u}\,.
\label{011}
\end{equation}

\noindent We can then verify that

\begin{equation}
\left<{\bf a}'^s\right>=\left<{\bf a}^s\right>\,.
\label{012}
\end{equation}

\noindent Therefore, the traces are invariant under similarity transformations and this set of invariants can be used to characterise the matrix ${\bf a}$.

Discriminants are a set of invariants which are given in term of traces by the relation among the coefficients in a Fourier expansion and the coefficients in a Taylor expansion. The first discriminants are given by

\begin{eqnarray}
c_0({\bf a})&=&1\,,\nonumber\\
c_1({\bf a})&=&\left<{\bf a}\right>\,,\nonumber\\
c_2({\bf a})&=&{1\over2}\,\left[\left<{\bf a}\right>^2-\left<{\bf a}^2\right>\right]\,,
\nonumber\\
c_3({\bf a})&=&{1\over{3!}}\,\left[\left<{\bf a}\right>^3-3\,\left<{\bf a}\right>\,\left<{\bf a
}^2\right>+2\,\left<{\bf a}^3\right>\right]\,,\nonumber\\
c_4({\bf a})&=&{1\over{4!}}\,\left[\left<{\bf a}\right>^4-6\,\left<{\bf a}\right>^2\,\left<{\bf a}^2\right>+8\,\left<{\bf a}\right>\,\left<{\bf a}^3\right>+3\,\left<{\bf a}^2\right>^2-6\,
\left<{\bf a}^4\right>\right]\,,\nonumber\\
c_5({\bf a})&=&{1\over{5!}}\,\left[\left<{\bf a}\right>^5-10\,\left<{\bf a}\right>^3\,\left<{\bf a}^2\right>+15\,\left<{\bf a}\right>\,\left<{\bf a}^2\right>^2+20\,\left<{\bf a}\right>^2\,
\left<{\bf a}^3\right>\right.\nonumber\\
&&\quad\quad\left.-20\,\left<{\bf a}^2\right>\,\left<{\bf a}^3\right>-30\,\left<{\bf a}\right>\,
\left<{\bf a}^4\right>+24\,\left<{\bf a}^5\right>\right]\,,\nonumber\\
c_6({\bf a})&=&{1\over{6!}}\,\left[\left<{\bf a}\right>^6-15\,\left<{\bf a}\right>^4\,\left<{\bf a}^2\right>+45\,\left<{\bf a}\right>^2\,\left<{\bf a}^2\right>^2-15\,\left<{\bf a}^2\right>^3
\right.\nonumber\\
&&\quad\quad+40\,\left<{\bf a}\right>^3\,\left<{\bf a}^3\right>-120\,\left<{\bf a}\right>\,
\left<{\bf a}^2\right>\,\left<{\bf a}^3\right>+40\,\left<{\bf a}^3\right>^2\nonumber\\
&&\quad\quad\left.-90\,\left<{\bf a}\right>^2\,\left<{\bf a}^4\right>+90\,\left<{\bf a}^2\right>
\,\left<{\bf a}^4\right>+144\,\left<{\bf a}\right>\,\left<{\bf a}^5\right>-120\,\left<{\bf a}^6
\right>\right]\,,
\label{013}
\end{eqnarray}

\noindent etc. Only the first $d$ discriminants are non--trivial, while the discriminants of an order higher than $d$ are identically zero, $c_s({\bf a})\equiv0$, for $s>d$. Therefore, the number of invariants characterising a $d\times d$ matrix ${\bf a}$ is finite and given by $d$. This means, by the way, that only the first $d$ traces of a matrix ${\bf a}$ are linearly independent. However, discriminants are a more convenient set of invariants than the traces since only the first $d$ discriminants are non trivial.

Let us consider the derivatives $\partial c_s({\bf a})/\partial{\bf a}$. For the first values of $s$ we obtain

\begin{eqnarray}
{{\partial c_1({\bf a})}\over{\partial{\bf a}}}&=&{\bf I}\,,\nonumber\\
{{\partial c_2({\bf a})}\over{\partial{\bf a}}}&=&\left<{\bf a}\right>\,{\bf I}-{\bf a}\,,
\nonumber\\
{{\partial c_3({\bf a})}\over{\partial{\bf a}}}&=&{1\over2}\,\left[\left<{\bf a}\right>^2-\left<{\bf a}^2\right>\right]\,{\bf I}-\left<{\bf a}\right>\,{\bf a}+{\bf a}^2\nonumber\\
{{\partial c_4({\bf a})}\over{\partial{\bf a}}}&=&{1\over{3!}}\,\left[\left<{\bf a}\right>^3-3\,
\left<{\bf a}\right>\,\left<{\bf a}^2\right>+2\,\left<{\bf a}^3\right>\right]\,{\bf I}-{1\over2}
\,\left[\left<{\bf a}\right>^2-\left<{\bf a}^2\right>\right]\,{\bf a}+\left<{\bf a}\right>\,{\bf a}^2-{\bf a}^3\,,
\label{014}
\end{eqnarray}

\noindent etc. We can then verify that

\begin{equation}
{{\partial c_s({\bf a})}\over{\partial{\bf a}}}=(-1)^{s-1}\,\sum_{k=0}^{s-1}\,(-1)^k\,c_k({\bf a})\,{\bf a}^{s-k-1}\,.
\label{015}
\end{equation}

\noindent Let us define the polynomials

\begin{equation}
P^d_{{\bf a}}(\lambda)=c_d({\bf a}-\lambda\,{\bf I})\,,
\label{016}
\end{equation}

\noindent where $\lambda$ is a scalar. By means of a Taylor expansion around ${\bf a}$, and using (\ref{011}), we obtain

\begin{equation}
P^d_{{\bf a}}(\lambda)=\sum_{k=0}^d\,(-1)^k\,c_k({\bf a})\,\lambda^{d-k}=P^{d-1}_{{\bf a}}
(\lambda)\,\lambda+(-1)^d\,c_d({\bf a})\,.
\label{017}
\end{equation}

\noindent For the first values of $d$ we obtain

\begin{eqnarray}
P^1_{{\bf a}}(\lambda)&=&\lambda-c_1({\bf a})\,,\nonumber\\
P^2_{{\bf a}}(\lambda)&=&\lambda^2-c_1({\bf a})\,\lambda+c_2({\bf a})\,,\nonumber\\
P^3_{{\bf a}}(\lambda)&=&\lambda^3-c_1({\bf a})\,\lambda^2+c_2({\bf a})\,\lambda-c_3({\bf a})
\,,\nonumber\\
P^4_{{\bf a}}(\lambda)&=&\lambda^4-c_1({\bf a})\,\lambda^3+c_2({\bf a})\,\lambda^2-c_3({\bf a})
\,\lambda+c_4({\bf a})\,.
\label{018}
\end{eqnarray}

\noindent $P^d_{{\bf a}}(\lambda)$ is known as the characteristic polynomial associated to the matrix ${\bf a}$. We can write the polynomials (\ref{014}) with ${\bf a}$ as its argument, namely ${\bf P}^d_{{\bf a}}({\bf a})=P^d_{{\bf a}}({\bf a})$. Then, we obtain

\begin{equation}
{\bf P}^d_{{\bf a}}({\bf a})=\sum_{k=0}^d\,(-1)^k\,c_k({\bf a})\,{\bf a}^{d-k}={\bf P}^{d-1}_{{\bf a}}({\bf a})\,\cdot\,{\bf a}+(-1)^d\,c_d({\bf a})\,{\bf I}\,.
\label{019}
\end{equation}

\noindent Comparison with (\ref{012}) shows that

\begin{equation}
{\bf P}^d_{{\bf a}}({\bf a})=(-1)^d\,{{\partial c_{d+1}({\bf a})}\over{\partial{\bf a}}}\,.
\label{020}
\end{equation}

\noindent Therefore, eq. (\ref{016}) can be rewritten as

\begin{equation}
{\bf P}^d_{{\bf a}}({\bf a})=(-1)^{d-1}\,\left[{{\partial c_d({\bf a})}\over{\partial{\bf a}}}
\,\cdot\,{\bf a}-c_d({\bf a})\,{\bf I}\right]\,.
\label{021}
\end{equation}

The powers of a matrix ${\bf a}$ are not all linearly independent. They are related by the Cayley--Hamilton theorem:

\bigskip

{\bf Theorem.} (Cayley--Hamilton) {\it A $d\times d$ matrix ${\bf a}$ is a root of its characteristic polynomial, that is ${\bf P}^d_{{\bf a}}({\bf a})\equiv0$.}

\bigskip

\noindent Therefore, the Cayley--Hamilton theorem states that there exists a relation of the form

\begin{equation}
{{\partial c_d({\bf a})}\over{\partial{\bf a}}}\,\cdot\,{\bf a}-c_d({\bf a})\,{\bf I}\equiv0\,.
\label{022}
\end{equation}

\noindent If $c_d({\bf a})\not=0$, then there exist an inverse matrix ${\bf a}^{-1}$ which is given by

\begin{equation}
{\bf a}^{-1}={1\over{c_d({\bf a})}}\,{{\partial c_d({\bf a})}\over{\partial{\bf a}}}\,.
\label{023}
\end{equation}

\noindent In terms of components this inverse matrix is given by

\begin{equation}
({\bf a}^{-1})_{ij}={1\over{c_d({\bf a})}}\,{{\partial c_d({\bf a})}\over{\partial a_{ij}}}\,.
\label{024}
\end{equation}

\noindent We have now two algorithms to compute the inverse of a matrix ${\bf a}$. The first one is based on eq. (\ref{006}). We have $n^2$ unknows $(a^{-1})_{ij}$ and $n^2$ equations. The condition for this equation to have a solution is $a=\det({\bf a})\not=0$. Then the solution is (\ref{024}). The second algorithm is based on the direct construction of $c_d({\bf a})$. If $c_d({\bf a})\not=0$ then the inverse matrix is given by (\ref{023}). For second--rank matrices the two algorithms are equivalent and give the same result. However, for higher--rank matrices only the second algorithm admits a generalization.

An additional set of relevant matrices is given by

\begin{equation}
{\bf c}_s({\bf a})={{\partial c_s({\bf a})}\over{\partial{\bf I}}}\,.
\label{025}
\end{equation}

\noindent This expression is an abbreviation for the following operation. Let us consider a second matrix ${\bf b}$ with components $b_{ij}$ and define ${\bf c}={\bf b}\cdot{\bf a}$. Then, eq. (\ref{024}) is an abbreviation for

\begin{equation}
{\bf c}_s({\bf a})=\left({{\partial c_s({\bf c})}\over{\partial{\bf b}}}\right)_{{\bf b}={\bf I}}\,.
\label{026}
\end{equation}

\noindent In terms of components we have

\begin{equation}
[{\bf c}_s({\bf a})]_{ij}=\left({{\partial c_s({\bf c})}\over{\partial b_{ij}}}\right)_{{\bf b}
={\bf I}}\,.
\label{027}
\end{equation}

\noindent For the first values of $s$ the result is

\begin{eqnarray}
{\bf c}_1({\bf a})&=&{\bf a}\nonumber\\
{\bf c}_2({\bf a})&=&\left<{\bf a}\right>\,{\bf a}-{\bf a}^2\nonumber\\
{\bf c}_3({\bf a})&=&{1\over2}\,\left[\left<{\bf a}\right>^2-\left<{\bf a}^2\right>\right]\,{\bf a}-\left<{\bf a}\right>\,{\bf a}^2+{\bf a}^3\nonumber\\
{\bf c}_4({\bf a})&=&{1\over{3!}}\,\left[\left<{\bf a}\right>^3-3\,\left<{\bf a}\right>\,\left<{
\bf a}^2\right>+2\,\left<{\bf a}^3\right>\right]\,{\bf a}-{1\over2}\,\left[\left<{\bf a}\right>
^2-\left<{\bf a}^2\right>\right]\,{\bf a}^2+\left<{\bf a}\right>\,{\bf a}^3-{\bf a}^4\,.
\label{028}
\end{eqnarray}

\noindent We can verify that

\begin{equation}
{\bf c}_s({\bf a})=-(-1)^s\,{\bf P}^{s-1}_{{\bf a}}({\bf a})\,\cdot\,{\bf a}\,.
\label{029}
\end{equation}

\noindent For $s=d$ we obtain

\begin{equation}
{\bf c}_d({\bf a})=-(-1)^d\,{\bf P}^{d-1}_{{\bf a}}({\bf a})\,\cdot\,{\bf a}=c_d({\bf a})\,{\bf I}\,,
\label{030}
\end{equation}

\noindent where we have used (\ref{016}) and (\ref{018}). Therefore

\begin{equation}
{{\partial c_d({\bf a})}\over{\partial{\bf I}}}-c_d({\bf a})\,{\bf I}=0\,,
\label{031}
\end{equation}

\noindent is a statement equivalent to the Cayley--Hamilton theorem (\ref{020}). Let us observe that $c_d({\bf a})$ is symmetric under the interchange of ${\bf a}$ and ${\bf I}$. Therefore, in the same way in which the inverse of the matrix ${\bf a}$ is defined by (\ref{021}), the inverse of the matrix ${\bf I}$ is defined by 

\begin{equation}
{\bf I}^{-1}={\bf I}={1\over{c_d({\bf a})}}\,{{\partial c_d({\bf a})}\over{\partial{\bf I}}}\,,
\label{032}
\end{equation}

\noindent which is equivalent to (\ref{029}). For the first values of $d$ the result is

\begin{eqnarray}
{\bf a}-c_1({\bf a})\,{\bf I}&\equiv&0\,,\nonumber\\
{\bf a}^2-c_1({\bf a})\,{\bf a}+c_2({\bf a})\,{\bf I}&\equiv&0\,,\nonumber\\
{\bf a}^3-c_1({\bf a})\,{\bf a}^2+c_2({\bf a})\,{\bf a}-c_3({\bf a})\,{\bf I}&\equiv&0\,,
\nonumber\\
{\bf a}^4-c_1({\bf a})\,{\bf a}^3+c_2({\bf a})\,{\bf a}^2-c_3({\bf a})\,{\bf a}+c_4({\bf a})
\,{\bf I}&\equiv&0\,.
\label{033}
\end{eqnarray}

A further set of relevant quantities is obtained as follows. Let us define

\begin{equation}
t_s({\bf a})={1\over s}\,\left<{\bf a}^s\right>\,.
\label{034}
\end{equation}

\noindent Using (\ref{024}) and (\ref{027}) we now obtain

\begin{equation}
{{\partial t_s({\bf a})}\over{\partial{\bf I}}}={\bf a}^s\,.
\label{035}
\end{equation}

\noindent This is the definition of the matrix multiplication in terms of discriminants which will use for the generalization to fourth--rank matrices. For this purpose, however, it is necessary to write (\ref{034}) in terms of discriminants. For the first values of $s$ we obtain

\begin{eqnarray}
t_1({\bf a})&=&c_1({\bf a})\,,\nonumber\\
t_2({\bf a})&=&{1\over2}\,c_1^2({\bf a})-c_2({\bf a})\,,\nonumber\\
t_3({\bf a})&=&{1\over3}\,c_1^3({\bf a})-c_1({\bf a})\,c_2({\bf a})+c_3({\bf a})\,,\nonumber\\
t_4({\bf a})&=&{1\over4}\,c_1^4({\bf a})-c_1^2({\bf a})\,c_2({\bf a})+c_1({\bf a})\,c_3({\bf a})+{1\over2}\,c_2^2({\bf a})-c_4({\bf a})\,.
\label{036}
\end{eqnarray}

\subsection{Index Notation}

For the purposes of generalisations to higher--rank matrices the notation above is not adequate. We have found that an index notation similar to that of tensor analysis is more convenient. In this case is necessary to distinguish between covariant and contravariant indices. According to this scheme, the matrix ${\bf a}$ of the section above will be now a second--rank covariant matrix $a_{ij}$. Analogously, a second--rank contravariant matrix ${\bf b}$ has the components $b^{ij}$.

In order to obtain the index version of the matrix multiplication for covariant matrices, let us consider a second--rank contravariant matrix ${\bf I}$ with components

\begin{equation}
I^{ij}=\cases{1\,,&if $i=j$,\cr0\,,&otherwise.\cr}
\label{037}
\end{equation}

\noindent Then the matrix multiplication is given by

\begin{equation}
c_{ij}=a_{ik}\,I^{kl}\,b_{lj}\,.
\label{038}
\end{equation}

\noindent (The summation convention over repeated indices is assumed.) In this sense, ${\bf I}$ behaves as a metric. With these conventions we can reproduce all the definitions and results of the previous section, but now product and traces are defined and constructed with $I^{ij}$.

The similarity transformations are now constructed in terms of a matrix ${\bf u}$ of mixed covariance with components ${u^i}_j$ and the inverse ${\bf u}^{-1}$ with components ${(u^{-1})^i}_j$, such that

\begin{equation}
{u^i}_k\,{(u^{-1})^k}_j={(u^{-1})^i}_k\,{u^k}_j=\delta^i_j\,.
\label{039}
\end{equation}

\noindent Then

\begin{eqnarray}
{\bf a}'&=&{\bf u}\,\cdot\,{\bf a}\,\cdot\,{\bf u}\,,\nonumber\\
{\bf b}'&=&{\bf u}^{-1}\,\cdot\,{\bf b}\,\cdot\,{\bf u}^{-1}\,,
\label{040}
\end{eqnarray}

\noindent or, in terms of components,

\begin{eqnarray}
(a')_{ij}&=&a_{kl}\,{u^k}_i\,{u^l}_j\,,\nonumber\\
(b')^{ij}&=&b^{kl}\,{(u^{-1})^i}_k\,{(u^{-1})^j}_l\,.
\label{041}
\end{eqnarray}

\noindent Then, the discriminants (\ref{012}) are invariant under the similarity transformations (\ref{039}).

\subsection{Alternating Products and Discriminants}

In terms of the components of the unit matrix ${\bf I}$ we can define the following symbols

\begin{equation}
q_s^{i_1j_1\cdots i_s j_s}=P_j\,I^{i_1j_1}\,\cdots\,I^{i_s j_s}={1\over{s!}}\,I^{|[i_1j_1}\,
\cdots\,I^{i_s j_s]|}\,,
\label{042}
\end{equation}

\noindent where $P_j$ denotes the sum over all permutations with respect to the second index $j$; $|[\cdots]|$ denotes complete antisymmetry with respect to the indices $j$'s or, equivalently, with respect to the indices $i$'s. Due to the antisymmetry the symbols ${\bf q}$ are non trivial only for $s\leq d$. For the first values of $s$ ${\bf q}$ is given by

\begin{eqnarray}
q_1^{ij}&=&I^{ij}\,,\nonumber\\
q_2^{i_1j_1i_2j_2}&=&{1\over2}\,(I^{i_1j_1}\,I^{i_2j_2}-I^{i_1j_2}\,I^{i_2j_1})\,,\nonumber\\
q_3^{i_1j_1i_2j_2i_3j_3}&=&{1\over{3!}}\,[I^{i_1j_1}\,I^{i_2j_2}\,I^{i_3j_3}-(I^{i_1j_1}\,I^{i_2
j_3}\,I^{i_3j_2}+I^{i_1j_3}\,I^{i_2j_2}\,I^{i_3j_1}\nonumber\\
&&+I^{i_1j_2}\,I^{i_2j_1}\,I^{i_3j_3})+(I^{i_1j_2}\,I^{i_2j_3}\,I^{i_3j_1}+I^{i_1j_3}\,I^{i_2j_1}\,I^{i_3j_2})]\,.
\label{043}
\end{eqnarray}

\noindent Then, the discriminants are given by

\begin{equation}
c_s({\bf a})=q_s^{i_1j_1\cdots i_s j_s}\,a_{i_1j_1}\,\cdots\,a_{i_s j_s}\,.
\label{044}
\end{equation}

Let us observe that

\begin{equation}
q_d^{i_1\cdots i_d j_1\cdots j_d}={1\over{d!}}\,I^{|[i_1j_1}\,\cdots\,I^{i_d j_d]|}={1\over{d!}}
\,\epsilon^{i_1\cdots i_d}\,\epsilon^{j_1\cdots j_d}\,.
\label{045}
\end{equation}

\noindent Therefore

\begin{equation}
c_d({\bf a})={1\over{d!}}\,\epsilon^{i_1\cdots i_d}\,\epsilon^{j_1\cdots j_d}\,a_{i_1j_1}\,
\cdots\,a_{i_d j_d}\,.
\label{046}
\end{equation}

\noindent In matrix calculus the usual definition of the determinant is given by

\begin{equation}
\det({\bf a})={1\over{d!}}\,\epsilon^{i_1\cdots i_d}\,\epsilon^{j_1\cdots j_d}\,a_{i_1j_1}\,
\cdots\,a_{i_d j_d}\,.
\label{047}
\end{equation}

\noindent Therefore

\begin{equation}
c_d({\bf a})=\det({\bf a})\,.
\label{048}
\end{equation}

\noindent Let $a=\det({\bf a})$. If $a\not=0$, then eq. (\ref{020}) can be rewritten as

\begin{equation}
{\bf a}^{-1}={1\over a}\,{{\partial a}\over{\partial{\bf a}}}\,.
\label{049}
\end{equation}

\noindent In terms of components

\begin{equation}
a^{ij}={1\over a}\,{{\partial a}\over{\partial a_{ij}}}\,.
\label{050}
\end{equation}

\noindent Therefore

\begin{equation}
a^{ij}={1\over{(d-1)!}}\,{1\over a}\,\epsilon^{i i_1\cdots i_{(d-1)}}\,\epsilon^{j j_1\cdots j
_{(d-1)}}\,a_{i_1j_1}\,\cdots\,a_{i_{(d-1)}j_{(d-1)}}\,.
\label{051}
\end{equation}

\noindent We can verify that

\begin{equation}
a^{ik}\,a_{jk}=\delta^i_j\,.
\label{052}
\end{equation}

The relations (\ref{050}--\ref{051}) are the usual definitions of the inverse of a matrix in matrix calculus.

\subsection{Semi--Magic Squares}

Monomial algebraic invariants are characterised by the way in which the indices of ${\bf a}$ are contracted with the indices of ${\bf I}$. This contraction scheme can be represented by a semi--magic square. In order to construct algebraic invariants we consider products of ${\bf a}$'s and ${\bf I}$'s. In order for the result to be an invariant all indices must be contracted. This means that we must consider an equal number $n$ of ${\bf a}$'s and ${\bf I}$'s. Since both ${\bf a}$ and ${\bf I}$ have two indices each, ${\bf a}$ can be contracted at most with 2 indices belonging to ${\bf I}$'s and the same is true for ${\bf I}$'s. The way in which the $n$ ${\bf a}$'s are contracted with the $n$ ${\bf I}$'s can be represented by an $n\times n$ square array of numbers ${\bf s}$, where the components $s_{IJ}$ denote the number of contractions between the $I$th ${\bf a}$ and the $J$th ${\bf I}$. Graphically

\begin{equation}
{\bf s}=\bordermatrix{&{\bf I}_1&\cdots&{\bf I}_n\cr
{\bf a}_1&s_{11}&\cdots&s_{1n}\cr
\vdots&\vdots&&\vdots\cr 
{\bf a}_n&s_{n1}&\cdots&s_{nn}\cr}\,.
\label{053}
\end{equation}

\noindent If the $I$th ${\bf a}$ is contracted once with the $J$th ${\bf I}$, then we write $s_{IJ}=1$; If the $I$th ${\bf a}$ is contracted twice with the $J$th ${\bf I}$, then we write $s_{IJ}=2$; If there is no contraction between the $I$th ${\bf a}$ and the $J$th ${\bf I}$, then we write $s_{IJ}=0$. Since all indices must be  contracted the sum of the elements of each row and each column must be equal to 2, that is,

\begin{eqnarray}
\sum_{I=1}^n\,s_{IJ}&=&2\,,\nonumber\\
\sum_{J=1}^n\,s_{IJ}&=&2\,.
\label{054}
\end{eqnarray}

\noindent Arrays with this property are known as semi--magic squares \cite{St}. Therefore, the number of possible algebraic invariants is determined by the number $H_n(2)$ of semi--magic squares ${\bf s}$. Semi--magic squares of rank $r$ are defined by the relations

\begin{eqnarray}
\sum_{I=1}^n\,s_{IJ}&=&r\,,\nonumber\\
\sum_{J=1}^n\,s_{IJ}&=&r\,.
\label{055}
\end{eqnarray}

\noindent The number of different possible semi--magic squares $H_n(r)$ is given by \cite{McM,SS,St}

\begin{eqnarray}
H_1(r)&=&1\,,\nonumber\\
H_2(r)&=&r+1\,,\nonumber\\
H_3(r)&=&6+15\,\left(\matrix{r-1\cr1\cr}\right)+19\,\left(\matrix{r-1\cr2\cr}\right)+12\,\left(
\matrix{r-1\cr3\cr}\right)+3\,\left(\matrix{r-1\cr4\cr}\right)\nonumber\\
&=&{1\over8}\,\left(r^4+6\,r^3+15\,r^2+18\,r+8\right)\,,\nonumber\\
&=&{1\over2}\,R\,(R+1)\,,\nonumber\\
H_4(r)&=&24+258\,\left(\matrix{r-1\cr1\cr}\right)+1468\,\left(\matrix{r-1\cr2\cr}\right)
+4945\,\left(\matrix{r-1\cr3\cr}\right)\nonumber\\
&&+10532\,\left(\matrix{r-1\cr4\cr}\right)+14620\,\left(\matrix{r-1\cr5\cr}\right)+13232\,
\left(\matrix{r-1\cr6\cr}\right)\nonumber\\
&&+7544\,\left(\matrix{r-1\cr7\cr}\right)+2464\,\left(\matrix{r-1\cr8\cr}\right)+352\,\left(
\matrix{r-1\cr9\cr}\right)\,.
\label{056}
\end{eqnarray}

\noindent where $R=(r+1)(r+2)/2$.

For $r=2$ the result is: $H_1(2)=1$, $H_2(2)=3$, $H_3(2)=21$, $H_4(2)=282$. For $n=1$ and $n=2$ the corresponding semi--magic squares are given by

\begin{eqnarray}
s_{2,1}&=&\{(2)\}\,,\nonumber\\
\nonumber\\
s_{2,2}&=&\left\{\left(\matrix{2&0\cr0&2\cr}\right)\,,\,\left(\matrix{0&2\cr2&0\cr}\right)\,,\,
\left(\matrix{1&1\cr1&1\cr}\right)\right\}\,,
\label{057}
\end{eqnarray}

\noindent etc. Since each column and each row represent the same matrix, semi--magic squares which are related by the permutation of rows and/or columns represent the same algebraic invariant. Therefore, semi--magic squares can be classifed into equivalence classes related by permutations of rows and/or columns. Therefore, we need to take care only of the representatives $s_{r,n}=\{s_i,i=1,\cdots,p_r(n)\}$ for each equivalence class, where $p_r(n)$ is the number of equivalence classes for rank $r$ and order $n$. For $r=2$ the number of equivalence classes $p_2(n)$ is given by the number of integer partitions of $n$, taht is, $p(n)$. For the first values of $n$ the representatives of each equivalence class are

\begin{eqnarray}
s_{2,1}&=&\{(2)\}\,,\nonumber\\
\nonumber\\
s_{2,2}&=&\left\{\left(\matrix{2&0\cr0&2\cr}\right)\,,\,\left(\matrix{1&1\cr1&1\cr}\right)
\right\}\,,\nonumber\\
\nonumber\\
s_{2,3}&=&\left\{\left(\matrix{2&0&0\cr0&2&0\cr0&0&2\cr}\right)\,,\,
\left(\matrix{2&0&0\cr0&1&1\cr0&1&1\cr}\right)\,,\,
\left(\matrix{0&1&1\cr1&0&1\cr1&1&0\cr}\right)\right\}\,,\nonumber\\
\nonumber\\
s_{2,4}&=&\left\{\left(\matrix{2&0&0&0\cr0&2&0&0\cr0&0&2&0\cr0&0&0&2\cr}\right)\,,\,
\left(\matrix{2&0&0&0\cr0&2&0&0\cr0&0&1&1\cr0&0&1&1\cr}\right)\,,\,
\left(\matrix{2&0&0&0\cr0&0&1&1\cr0&1&0&1\cr0&1&1&0\cr}\right)\right.\,,\nonumber\\
&&\quad\left.\left(\matrix{1&1&0&0\cr1&1&0&0\cr0&0&1&1\cr0&0&1&1\cr}\right)\,,\,
\left(\matrix{1&1&0&0\cr0&1&1&0\cr0&0&1&1\cr1&0&0&1\cr}\right)\right\}\,.
\label{060}
\end{eqnarray}

The algebraic invariants which each semi--magic square represents are given by

\begin{eqnarray}
(2)&=&\left<{\bf a}\right>\,,\nonumber\\
\nonumber\\
\left(\matrix{2&0\cr0&2\cr}\right)&=&\left<{\bf a}\right>^2\,,\quad\,
\left(\matrix{1&1\cr1&1\cr}\right)=\left<{\bf a}^2\right>\,,\nonumber\\
\nonumber\\
\left(\matrix{2&0&0\cr0&2&0\cr0&0&2\cr}\right)&=&\left<{\bf a}\right>^3\,,\quad\,
\left(\matrix{2&0&0\cr0&1&1\cr0&1&1\cr}\right)=\left<{\bf a}\right>\,\left<{\bf a}^2\right>\,,
\quad\,\left(\matrix{0&1&1\cr1&0&1\cr1&1&0\cr}\right)=\left<{\bf a}^3\right>\,,\nonumber\\
\nonumber\\
\left(\matrix{2&0&0&0\cr0&2&0&0\cr0&0&2&0\cr0&0&0&2\cr}\right)&=&\left<{\bf a}\right>^4\,,\quad
\,\left(\matrix{2&0&0&0\cr0&2&0&0\cr0&0&1&1\cr0&0&1&1\cr}\right)=\left<{\bf a}\right>^2\,\left<{
\bf a}^2\right>\,,\nonumber\\
\left(\matrix{2&0&0&0\cr0&0&1&1\cr0&1&0&1\cr0&1&1&0\cr}\right)&=&\left<{\bf a}\right>\,\left<{
\bf a}^3\right>\,,\quad\,\left(\matrix{1&1&0&0\cr1&1&0&0\cr0&0&1&1\cr0&0&1&1\cr}\right)=\left<{
\bf a}^2\right>^2\,,\nonumber\\
\left(\matrix{1&1&0&0\cr0&1&1&0\cr0&0&1&1\cr1&0&0&1\cr}\right)&=&\left<{\bf a}^4\right>\,.
\label{061}
\end{eqnarray}

Let us observe that block semi--magic squares can be decomposed in terms of lower order semi-magic squares as

\begin{eqnarray}
\left(\matrix{2&0\cr0&2\cr}\right)&=&(2)^2\,,\nonumber\\
\nonumber\\
\left(\matrix{2&0&0\cr0&2&0\cr0&0&2\cr}\right)&=&(2)^3\,,\quad\,
\left(\matrix{2&0&0\cr0&1&1\cr0&1&1\cr}\right)=(2)\,\left(\matrix{1&1\cr1&1\cr}\right)\,,
\nonumber\\
\nonumber\\
\left(\matrix{2&0&0&0\cr0&2&0&0\cr0&0&2&0\cr0&0&0&2\cr}\right)&=&(2)^4\,,\quad\,
\left(\matrix{2&0&0&0\cr0&2&0&0\cr0&0&1&1\cr0&0&1&1\cr}\right)=(2)^2\,\left(\matrix{1&1\cr1&1
\cr}\right)\,,\nonumber\\
\left(\matrix{2&0&0&0\cr0&0&1&1\cr0&1&0&1\cr0&1&1&0\cr}\right)&=&(2)\,\left(\matrix{0&1&1\cr
1&0&1\cr1&1&0\cr}\right)\,,\quad\,
\left(\matrix{1&1&0&0\cr1&1&0&0\cr0&0&1&1\cr0&0&1&1\cr}\right)=\left(\matrix{1&1\cr1&1\cr}
\right)^2\,.
\label{062}
\end{eqnarray}

\subsection{Permutations}

Each algebraic invariant is represented uniquely by a semi--magic square. Therefore, semi--magic squares are adequate to represent the different algebraic invariants. A discriminant of order $n$ is a linear combinations of the semi--magic squares of order $n$. In order to determine the coefficients of this linear combination we proceed as follows.

A matrix ${\bf a}$ can be contracted with a second matrix ${\bf a}$ in a number of ways which depend on the possible permutations, $\pi(n)$, of the indices in the second matrix. For each possible permutation we obtain a signed semi--meagic square $k_i$, $i=1,\cdots,\pi(n)$. The discriminant is then given by the general formula

\begin{equation}
c_n({\bf a})={1\over{n!}}\,\sum_{i=1}^{\pi(n)}\,k_i\,,
\label{063}
\end{equation}

\noindent where $\pi(n)$ is the number of possible outcomes $k_i$ after considering all possible permutations. The numerical factor $1/n!$ in front of the previous expression is conventional.

The signed semi--magic squares $k_i$ are obtained as follows. For $n=2$ the possible permutations are

\begin{equation}
\pi=\left\{\left(\matrix{1&0\cr0&1\cr}\right)_+,\,\left(\matrix{0&1\cr1&0\cr}\right)_-\right\}\,,
\label{064}
\end{equation}

\noindent where the subindex denotes the parity of the given permutation; Therefore, $\pi(2)=2$. We can always choose to keep fixed the indices of the first matrix and represent it by a square with 1's as entries in the diagonal. For $n=2$ we have

\begin{equation}
\pi_0=\left(\matrix{1&0\cr0&1\cr}\right)\,.
\label{065}
\end{equation}

\noindent The possible outcomes are

\begin{eqnarray}
k_+=\pi_0\oplus\pi_+=\left(\matrix{1&0\cr0&1\cr}\right)\oplus\left(\matrix{1&0\cr0&1\cr}\right)_+&=&\left(\matrix{2&0\cr0&2\cr}\right)\,,\nonumber\\
k_-=\pi_0\oplus\pi_-=\left(\matrix{1&0\cr0&1\cr}\right)\oplus\left(\matrix{0&1\cr1&0\cr}\right)_
-&=&-\left(\matrix{1&1\cr1&1\cr}\right)\,
\label{066}
\end{eqnarray}

\noindent Then, the discriminant is given by

\begin{equation}
c_2({\bf a})={1\over2}\,\left[\left(\matrix{2&0\cr0&2\cr}\right)-\left(\matrix{1&1\cr1&1\cr}
\right)\right]={1\over2}\,\left[\left<{\bf a}\right>^2-\left<{\bf a}^2\right>\right]\,.
\label{067}
\end{equation}

For $n=3$ there are 6 possible permutations, $\pi(3)=6$, given by

\begin{eqnarray}
\pi&=&\left\{\left(\matrix{1&0&0\cr0&1&0\cr0&0&1\cr}\right)_+\,,\,\left(\matrix{1&0&0\cr0&0&1\cr0&1&0\cr}\right)_-\,,\,\left(\matrix{0&0&1\cr0&1&0\cr1&0&0\cr}\right)_-\right.\,,\nonumber\\
&&\quad\left.\left(\matrix{0&1&0\cr1&0&0\cr0&0&1\cr}\right)_-\,,\,\left(\matrix{0&1&0
\cr0&0&1\cr1&0&0\cr}\right)_+\,,\,\left(\matrix{0&0&1\cr1&0&0\cr0&1&0\cr}\right)_+\right\}\,.
\label{068}
\end{eqnarray}

\noindent The possible outcomes are

\begin{eqnarray}
k_1&=&\pi_0\oplus\pi_1=\left(\matrix{1&0&0\cr0&1&0\cr0&0&1\cr}\right)\oplus\left(\matrix{1&0&0
\cr0&1&0\cr0&0&1\cr}\right)_+=\left(\matrix{2&0&0\cr0&2&0\cr0&0&2\cr}\right)\,,\nonumber\\
k_2&=&\pi_0\oplus\pi_2=\left(\matrix{1&0&0\cr0&1&0\cr0&0&1\cr}\right)\oplus\left(\matrix{1&0&0
\cr0&0&1\cr0&1&0\cr}\right)_-=-\left(\matrix{2&0&0\cr0&1&1\cr0&1&1\cr}\right)\,,\nonumber\\
k_3&=&\pi_0\oplus\pi_3=\left(\matrix{1&0&0\cr0&1&0\cr0&0&1\cr}\right)\oplus\left(\matrix{0&0&1
\cr0&1&0\cr1&0&0\cr}\right)_-=-\left(\matrix{1&0&1\cr0&2&0\cr1&0&1\cr}\right)\,,\nonumber\\
k_4&=&\pi_0\oplus\pi_4=\left(\matrix{1&0&0\cr0&1&0\cr0&0&1\cr}\right)\oplus\left(\matrix{0&1&0
\cr1&0&0\cr0&0&1\cr}\right)_-=-\left(\matrix{1&1&0\cr1&1&0\cr0&0&2\cr}\right)\,,\nonumber\\
k_5&=&\pi_0\oplus\pi_5=\left(\matrix{1&0&0\cr0&1&0\cr0&0&1\cr}\right)\oplus\left(\matrix{0&1&0
\cr0&0&1\cr1&0&0\cr}\right)_+=\left(\matrix{1&1&0\cr0&1&1\cr1&0&1\cr}\right)\,,\nonumber\\
k_6&=&\pi_0\oplus\pi_6=\left(\matrix{1&0&0\cr0&1&0\cr0&0&1\cr}\right)\oplus\left(\matrix{0&0&1
\cr1&0&0\cr0&1&0\cr}\right)_+=\left(\matrix{1&0&1\cr1&1&0\cr0&1&1\cr}\right)\,.
\label{069}
\end{eqnarray}

\noindent The discriminant is given by

\begin{eqnarray}
c_3({\bf a})&=&{1\over{3!}}\,\left[\left(\matrix{2&0&0\cr0&2&0\cr0&0&2\cr}\right)-3\,\left(
\matrix{2&0&0\cr0&1&1\cr0&1&1\cr}\right)+2\,\left(\matrix{1&1&0\cr0&1&1\cr1&0&1\cr}\right)
\right]\nonumber\\
&=&{1\over{3!}}\,\left[\left<{\bf a}\right>^3-3\,\left<{\bf a}\right>\,\left<{\bf a}^2\right>
+2\,\left<{\bf a}^3\right>\right]\,.
\label{070}
\end{eqnarray}

For $n=4$ there are 24 possible permutations, $\pi(4)=24$, namely,

\begin{eqnarray}
\pi&=&\left\{\left(\matrix{1&0&0&0\cr0&1&0&0\cr0&0&1&0\cr0&0&0&1\cr}\right)_+\,,\,
\left(\matrix{1&0&0&0\cr0&1&0&0\cr0&0&0&1\cr0&0&1&0\cr}\right)_-\,,\,
\left(\matrix{1&0&0&0\cr0&0&0&1\cr0&0&1&0\cr0&1&0&0\cr}\right)_-\,,\,
\left(\matrix{1&0&0&0\cr0&1&0&0\cr0&1&0&0\cr0&0&0&1\cr}\right)_-\right.\,,\nonumber\\
&&\quad\left(\matrix{0&0&0&1\cr0&1&0&0\cr0&0&1&0\cr1&0&0&0\cr}\right)_-\,,\,
\left(\matrix{0&0&1&0\cr0&1&0&0\cr1&0&0&0\cr0&0&0&1\cr}\right)_-\,,\,
\left(\matrix{0&1&0&0\cr1&0&0&0\cr0&0&1&0\cr0&0&0&1\cr}\right)_-\,,\,
\left(\matrix{0&1&0&0\cr1&0&0&0\cr0&0&0&1\cr0&0&1&0\cr}\right)_+\,,\nonumber\\
&&\quad\left(\matrix{0&0&1&0\cr0&0&0&1\cr1&0&0&0\cr0&1&0&0\cr}\right)_+\,,\,
\left(\matrix{0&0&0&1\cr0&0&1&0\cr0&1&0&0\cr1&0&0&0\cr}\right)_+\,,\,
\left(\matrix{1&0&0&0\cr0&0&1&0\cr0&0&0&1\cr0&1&0&0\cr}\right)_+\,,\,
\left(\matrix{1&0&0&0\cr0&0&0&1\cr0&1&0&0\cr0&0&1&0\cr}\right)_+\,,\nonumber\\
&&\quad\left(\matrix{0&0&1&0\cr0&1&0&0\cr0&0&0&1\cr1&0&0&0\cr}\right)_+\,,\,
\left(\matrix{0&0&0&1\cr0&1&0&0\cr1&0&0&0\cr0&0&1&0\cr}\right)_+\,,\,
\left(\matrix{0&1&0&0\cr0&0&0&1\cr0&0&1&0\cr1&0&0&0\cr}\right)_+\,,\,
\left(\matrix{0&0&0&1\cr1&0&0&0\cr0&0&1&0\cr0&1&0&0\cr}\right)_+\,,\nonumber\\
&&\quad\left(\matrix{0&1&0&0\cr0&0&1&0\cr1&0&0&0\cr0&0&0&1\cr}\right)_+\,,\,
\left(\matrix{0&0&1&0\cr1&0&0&0\cr0&1&0&0\cr0&0&0&1\cr}\right)_+\,,\,
\left(\matrix{0&1&0&0\cr0&0&1&0\cr0&0&0&1\cr1&0&0&0\cr}\right)_-\,,\,
\left(\matrix{0&1&0&0\cr0&0&0&1\cr1&0&0&0\cr0&0&1&0\cr}\right)_-\,,\nonumber\\
&&\quad\left.\left(\matrix{0&0&1&0\cr0&0&0&1\cr0&1&0&0\cr1&0&0&0\cr}\right)\,,\,
\left(\matrix{0&0&1&0\cr1&0&0&0\cr0&0&0&1\cr0&1&0&0\cr}\right)_-\,,\,
\left(\matrix{0&0&0&1\cr0&0&1&0\cr1&0&0&0\cr0&1&0&0\cr}\right)_-\,,\,
\left(\matrix{0&0&0&1\cr1&0&0&0\cr0&1&0&0\cr0&0&1&0\cr}\right)\right\}\,.
\label{071}
\end{eqnarray}

\noindent The discriminant is given by

\begin{eqnarray}
c_4({\bf a})&=&{1\over{4!}}\,\left[\left(\matrix{2&0&0&0\cr0&2&0&0\cr0&0&2&0\cr0&0&0&2\cr}
\right)-6\,\left(\matrix{2&0&0&0\cr0&2&0&0\cr0&0&1&1\cr0&0&1&1\cr}\right)+3\,\left(\matrix{
1&1&0&0\cr1&1&0&0\cr0&0&1&1\cr0&0&1&1\cr}\right)\right.\nonumber\\
&&\quad\quad\left.+8\,\left(\matrix{2&0&0&0\cr0&1&1&0\cr0&0&1&1\cr0&1&0&1\cr}\right)-6\,
\left(\matrix{1&1&0&0\cr0&1&1&0\cr0&0&1&1\cr1&0&0&1\cr}\right)\right]\nonumber\\
&=&{1\over{4!}}\,\left[\left<{\bf a}\right>^4-6\,\left<{\bf a}\right>^2\,\left<{\bf a}^2
\right>+3\,\left<{\bf a}^2\right>^2+8\,\left<{\bf a}\right>\,\left<{\bf a}^3\right>-6\,\left<{\bf a}^4\right>\right]\,.
\label{072}
\end{eqnarray}

The Cayley--Hamilton theorem is obtained by deriving the corresponding discriminant with respect to ${\bf I}$. In the language of semi--magic squares this means that one of the ${\bf I}$'s, one at a time, is not present. Therefore, the indices of that column are not contracted. We indicate that derivative by the same original semi--magic square and `resalt' the uncontracted column by boldface numbers. The resulting Cayley--Hamilton theorem is now written as

\begin{eqnarray}
\left(\matrix{2&{\bf 0}\cr0&{\bf 2}\cr}\right)-\left(\matrix{1&{\bf 1}\cr1&{\bf 1}\cr}\right)
-c_2({\bf a})\,[{\bf 2}]&\equiv&0\,,\nonumber\\
\nonumber\\
{1\over2}\,\left[\left(\matrix{2&0&{\bf 0}\cr0&2&{\bf 0}\cr0&0&{\bf 2}\cr}\right)
-\left(\matrix{1&1&{\bf 0}\cr1&1&{\bf 0}\cr0&0&{\bf 2}\cr}\right)\right]
-\left(\matrix{2&0&{\bf 0}\cr0&1&{\bf 1}\cr0&1&{\bf 1}\cr}\right)
+\left(\matrix{1&1&{\bf 0}\cr0&1&{\bf 1}\cr1&0&{\bf 1}\cr}\right)
-c_3({\bf a})\,[{\bf 2}]&\equiv&0\,,\nonumber\\
\nonumber\\
{1\over6}\,\left[
\left(\matrix{2&0&0&{\bf 0}\cr0&2&0&{\bf 0}\cr0&0&2&{\bf 0}\cr0&0&0&{\bf 2}\cr}\right)
-3\,\left(\matrix{2&0&0&{\bf 0}\cr0&1&1&{\bf 0}\cr0&1&1&{\bf 0}\cr0&0&0&{\bf 2}\cr}\right)
+2\,\left(\matrix{1&1&0&{\bf 0}\cr0&1&1&{\bf 0}\cr1&0&1&{\bf 0}\cr0&0&0&{\bf 2}\cr}\right)
\right]&&\nonumber\\
-{1\over2}\,\left[
\left(\matrix{2&0&0&{\bf 0}\cr0&2&0&{\bf 0}\cr0&0&1&{\bf 1}\cr0&0&1&{\bf 1}\cr}\right)
-\left(\matrix{1&1&0&{\bf 0}\cr1&1&0&{\bf 0}\cr0&0&1&{\bf 1}\cr0&0&1&{\bf 1}\cr}\right)
\right]&&\nonumber\\
+\left(\matrix{2&0&0&{\bf 0}\cr0&1&0&{\bf 1}\cr0&1&1&{\bf 0}\cr0&0&1&{\bf 1}\cr}\right)
-\left(\matrix{1&0&0&{\bf 1}\cr1&1&0&{\bf 0}\cr0&1&1&{\bf 0}\cr0&0&1&{\bf 1}\cr}\right)
-c_4({\bf a})\,[{\bf 2}]&\equiv&0\,,
\label{073}
\end{eqnarray}

\noindent where $[{\bf 2}]$ stands for the unit matrix ${\bf I}^{-1}$. Contracting with ${\bf I}$, $[2]=d$, we obtain (\ref{064}), (\ref{065}) and (\ref{069}).

\subsection{Graphical Construction of Invariants}

Each algebraic invariant is represented uniquely by a semi--magic square. On the other hand, semi--magic squares are obtained by considering all possible permutations of indices. However, for large values of $n$ this algorithm becomes unpractical. In order to avoid this difficulty we now develop a graphical algorithm for the construction and characterisation of algebraic invariants which allows to simplify this task. Let us represent the matrix ${\bf a}$ by a vertical grid with two boxes, namely

\begin{equation}
\put(-14,-14){\grid(14,28)(14,14)}
\label{074}
\end{equation}

\noindent The product of $n$ matrices is represented by

\begin{equation}
\put(-112,-14){\grid(42,28)(14,14)}\put(-63,-2){$\cdots$}\put(-42,-14){\grid(42,28)(14,14)}
\put(-106,-25){1}\put(-92,-25){2}\put(-8,-25){n}
\label{075}
\end{equation}

\noindent Each algebraic invariant is characterised by the way in which indices are contracted. We can always choose to keep fix the indices of the first row and look at how the indices in the second row are contracted with the indices in the first row. A void grid indicates that no permutation has been performed

\begin{equation}
\put(-84,-14){\grid(84,28)(14,14)}\,.
\label{076}
\end{equation}

\noindent A permutation of the indices $i$th and $j$th is indicated by

\begin{equation}
\put(-84,-14){\grid(84,28)(14,14)}
\put(-49,7){\circle{4}}\put(-21,7){\circle{4}}\put(-47,7){\line(1,0){24}}
\put(-50,-25){i}\put(-22,-25){j}
\label{077}
\end{equation}

\noindent A double permutation is indicated by

\begin{equation}
\put(-84,-14){\grid(84,28)(14,14)}
\put(-77,7){\circle{4}}\put(-63,7){\circle{4}}\put(-75,7){\line(1,0){10}}
\put(-49,7){\circle{4}}\put(-21,7){\circle{4}}\put(-47,7){\line(1,0){24}}
\label{078}
\end{equation}

\noindent A cyclic permutation is indicated by

\begin{equation}
\put(-84,-14){\grid(84,28)(14,14)}
\put(-77,7){\circle{4}}\put(-63,7){\circle{4}}\put(-49,7){\circle{4}}
\put(-75,7){\line(1,0){10}}\put(-61,7){\line(1,0){10}}
\label{079}
\end{equation}

\noindent In this case, however, it is necessary to take into account the sense in which the permutation is performed. There are 2 possibilities

\begin{equation}
\begin{array}{@{}l@{\hspace{91pt}}l@{}}
\put(-84,-14){\grid(84,28)(14,14)}
\put(-77,7){\circle{4}}\put(-63,7){\circle{4}}\put(-49,7){\circle{4}}
\put(-75,7){\vector(1,0){10}}\put(-61,7){\vector(1,0){10}}
&\put(-84,-14){\grid(84,28)(14,14)}
\put(-77,7){\circle{4}}\put(-63,7){\circle{4}}\put(-49,7){\circle{4}}
\put(-65,7){\vector(-1,0){10}}\put(-51,7){\vector(-1,0){10}}\cr
&
\end{array}
\label{080}
\end{equation}

\noindent When the sense of the permutation is irrelevant we use the right--oriented grid. The next possibility is

\begin{equation}
\put(-84,-14){\grid(84,28)(14,14)}
\put(-77,7){\circle{4}}\put(-63,7){\circle{4}}\put(-49,7){\circle{4}}\put(-35,7){\circle{4}}
\put(-75,7){\line(1,0){10}}\put(-61,7){\line(1,0){10}}\put(-47,7){\line(1,0){10}}
\label{081}
\end{equation}

\noindent In this case there are 6 possibles senses for the permutation, namely,

\begin{equation}
\begin{array}{@{}l@{\hspace{91pt}}l@{\hspace{91pt}}l@{}}
\put(-84,-14){\grid(84,28)(14,14)}
\put(-77,7){\circle{4}}\put(-63,7){\circle{4}}\put(-49,7){\circle{4}}\put(-35,7){\circle{4}}
\put(-75,7){\vector(1,0){10}}\put(-61,7){\vector(1,0){10}}\put(-47,7){\vector(1,0){10}}
&\put(-84,-14){\grid(84,28)(14,14)}
\put(-77,9){\circle{4}}\put(-63,5){\circle{4}}\put(-49,9){\circle{4}}\put(-35,7){\circle{4}}
\put(-75,9){\vector(1,0){24}}\put(-47,9){\vector(1,0){12}}\put(-35,5){\vector(-1,0){26}}
&\put(-84,-14){\grid(84,28)(14,14)}
\put(-77,11){\circle{4}}\put(-63,5){\circle{4}}\put(-49,3){\circle{4}}\put(-35,9){\circle{4}}
\put(-75,11){\vector(1,0){40}}\put(-35,7){\vector(-1,0){28}}\put(-63,3){\vector(1,0){12}}\cr
&&\cr
\put(-84,-14){\grid(84,28)(14,14)}
\put(-77,7){\circle{4}}\put(-63,7){\circle{4}}\put(-49,7){\circle{4}}\put(-35,7){\circle{4}}
\put(-65,7){\vector(-1,0){10}}\put(-51,7){\vector(-1,0){10}}\put(-37,7){\vector(-1,0){10}}
&\put(-84,-14){\grid(84,28)(14,14)}
\put(-77,9){\circle{4}}\put(-63,5){\circle{4}}\put(-49,9){\circle{4}}\put(-35,7){\circle{4}}
\put(-51,9){\vector(-1,0){24}}\put(-35,9){\vector(-1,0){12}}\put(-61,5){\vector(1,0){26}}
&\put(-84,-14){\grid(84,28)(14,14)}
\put(-77,11){\circle{4}}\put(-63,5){\circle{4}}\put(-49,3){\circle{4}}\put(-35,9){\circle{4}}
\put(-35,11){\vector(-1,0){40}}\put(-63,7){\vector(1,0){28}}\put(-51,3){\vector(-1,0){12}}\cr
&&
\end{array}
\label{082}
\end{equation}

\noindent As in the previous case, when the sense of the permutation is irrelevant we use only the right--oriented grid.

The semi--magic square corresponding to a given grid is obtained as follows. The number of empty boxes in each column corresponds to the diagonal entries in the semi--magic square. The lines correspond to the off--diagonal terms. For example

\begin{equation}
\put(-42,-14){\grid(42,28)(14,14)}
\put(-21,7){\circle{4}}\put(-7,7){\circle{4}}\put(-19,7){\line(1,0){10}}
\sim\left(\matrix{2&0&0\cr0&1&1\cr0&1&1\cr}\right)\,.
\label{083}
\end{equation}

\noindent The multiplicity is the number of possible ways in which the given permutation can be performed over the $n$ boxes (indices). The parity is given by the number of lines for the permutation. For example

\begin{equation}
\put(-42,-14){\grid(42,28)(14,14)}
\put(-21,7){\circle{4}}\put(-7,7){\circle{4}}\put(-19,7){\line(1,0){10}}
=-3\,\left(\matrix{2&0&0\cr0&1&1\cr0&1&1\cr}\right)\,.
\label{084}
\end{equation}

Let us perform the explicit construction of the semi--magic squares and discriminants for the first values of $n$. For $n=2$ the result is

\begin{equation}
\begin{array}{@{}l@{\hspace{35pt}}l@{}}
\put(-28,-14){\grid(28,28)(14,14)}=\left(\matrix{2&0\cr0&2\cr}\right)\,,
&\put(-28,-14){\grid(28,28)(14,14)}
\put(-21,7){\circle{4}}\put(-7,7){\circle{4}}\put(-19,7){\line(1,0){10}}
=-\left(\matrix{1&1\cr1&1\cr}\right)\,.\cr
&
\end{array}
\label{085}
\end{equation}

\noindent Therefore, we obtain (\ref{064}). For $n=3$ the result is

\begin{equation}
\begin{array}{@{}l@{\hspace{49pt}}l@{}}
\put(-42,-14){\grid(42,28)(14,14)}=\left(\matrix{2&0&0\cr0&2&0\cr0&0&2\cr}\right)\,,
&\put(-42,-14){\grid(42,28)(14,14)}
\put(-21,7){\circle{4}}\put(-7,7){\circle{4}}\put(-19,7){\line(1,0){10}}
=-3\,\left(\matrix{2&0&0\cr0&1&1\cr0&1&1\cr}\right)\,,\cr
&\cr
\put(-42,-14){\grid(42,28)(14,14)}
\put(-35,7){\circle{4}}\put(-21,7){\circle{4}}\put(-7,7){\circle{4}}
\put(-33,7){\vector(1,0){10}}\put(-19,7){\vector(1,0){10}}
=2\,\left(\matrix{1&1&0\cr0&1&1\cr1&0&1\cr}\right)\,.&\cr
&
\end{array}
\label{086}
\end{equation}

\noindent Therefore, we obtain (\ref{066}). For $n=4$ the result is

\begin{equation}
\begin{array}{@{}l@{\hspace{63pt}}l@{}}
\put(-56,-14){\grid(56,28)(14,14)}
=\left(\matrix{2&0&0&0\cr0&2&0&0\cr0&0&2&0\cr0&0&0&2\cr}\right)\,,
&\put(-56,-14){\grid(56,28)(14,14)}
\put(-21,7){\circle{4}}\put(-7,7){\circle{4}}\put(-19,7){\line(1,0){10}}
=-6\,\left(\matrix{2&0&0&0\cr0&2&0&0\cr0&0&1&1\cr0&0&1&1\cr}\right)\,,\cr
&\cr
\put(-56,-14){\grid(56,28)(14,14)}
\put(-49,7){\circle{4}}\put(-35,7){\circle{4}}\put(-47,7){\line(1,0){10}}
\put(-21,7){\circle{4}}\put(-7,7){\circle{4}}\put(-19,7){\line(1,0){10}}
=3\,\left(\matrix{1&1&0&0\cr1&1&0&0\cr0&0&1&1\cr0&0&1&1\cr}\right)\,,
&\put(-56,-14){\grid(56,28)(14,14)}
\put(-35,7){\circle{4}}\put(-21,7){\circle{4}}\put(-7,7){\circle{4}}
\put(-33,7){\vector(1,0){10}}\put(-19,7){\vector(1,0){10}}
=8\,\left(\matrix{2&0&0&0\cr0&1&1&0\cr0&0&1&1\cr0&1&0&1\cr}\right)\,,\cr
&\cr
\put(-56,-14){\grid(56,28)(14,14)}
\put(-49,7){\circle{4}}\put(-35,7){\circle{4}}\put(-21,7){\circle{4}}\put(-7,7){\circle{4}}
\put(-47,7){\vector(1,0){10}}\put(-33,7){\vector(1,0){10}}\put(-19,7){\vector(1,0){10}}
=-6\,\left(\matrix{1&1&0&0\cr0&1&1&0\cr0&0&1&1\cr1&0&0&1\cr}\right)\,.&\cr
&
\end{array}
\label{087}
\end{equation}

\noindent Therefore, we obtain (\ref{068}).

The number of semi--magic squares depends on the number of possible permutations. For large values of $n$ and $r$ this counting becomes quite involved. Therefore, it is advisable to have an easy recipe to obtain the correct counting.

For each order we have a different number of possible permutations $P_n$. We can represent them as

\begin{eqnarray}
P_2&=&{\bf 0}-{\bf 1}\,,\nonumber\\
P_3&=&{\bf 0}-3\cdot{\bf 1}+2\cdot{\bf 2}\,,\nonumber\\
P_4&=&{\bf 0}-6\cdot{\bf 1}+3\cdot{\bf 1}^2+8\cdot{\bf 2}-6\cdot{\bf 3}\,,\nonumber\\
P_5&=&{\bf 0}-10\cdot{\bf 1}+15\cdot{\bf 1}^2+20\cdot{\bf 2}-20\cdot({\bf 2}\,{\bf 1})-30\cdot
{\bf 3}+24\cdot{\bf 4}\,,\nonumber\\
P_6&=&{\bf 0}-15\cdot{\bf 1}+45\cdot{\bf 1}^2-15\cdot{\bf 1}^3+40\cdot{\bf 2}-120\cdot({\bf 2}\,
{\bf 1})\nonumber\\
&&+40\cdot{\bf 2}^2-90\cdot{\bf 3}+90\cdot({\bf 3}\,{\bf 1})+144\cdot{\bf 4}-120\cdot{\bf 5}\,.
\label{088}
\end{eqnarray}

\noindent The number of terms involving some given permutations is given by the coefficients in (\ref{088}). This can be easily verified in the graphical construction above.

\section{Hypermatrices. The Fourth--Rank Case}

There is not a natural multiplication operation for hypermatrices in the sense that the product of two hypermatrices be again a hypermatrix of the same rank. Therefore, the construction of algebraic invariants must be performed in a different way which, by the way, gives a clue for the definition of amultiplication operation for hypermatrices. We now turn to the construction of algebraic invariants for fourth--rank matrices. This algorithm can be easily extended to matrices of an arbitrary even--rank $r$; for the odd--rank case see section 4.

Our construction is based on alternating products. To this purpose let us consider a fourth--rank matrix ${\bf A}$ with components $A_{ijkl}$ and a fourth--rank matrix ${\bf\Delta}$ with components $\Delta^{ijkl}$ as a fourth--rank generalization of the unit matrix. For practical purposes the discriminants are better represented by semi--magic squares of rank 4. They are constructed using the graphical algorithm of the section above. 

\subsection{Alternating Products and Discriminants}

Let us now introduce a fourth--rank unit matrix $\Delta^{ijkl}$. In order to fix the ideas we can consider a unit matrix defined by the relations

\begin{equation}
\Delta^{ijkl}=\cases{1\,,&if $i=j=k=l$,\cr0\,,&otherwise.\cr}\,.
\label{089}
\end{equation}

\noindent However, other definitions are possible and they are currently under study \cite{Ta02}. In a way similar to (\ref{040}) we define

\begin{equation}
Q_s^{i_1j_1k_1l_1\cdots i_s j_s k_s l_s}={1\over{s!}}\,\Delta^{|[i_1j_1k_1l_1}\,\cdots\,
\Delta^{i_s j_s k_s l_s]|}\,.
\label{090}
\end{equation}

\noindent For the first values of $s$ we obtain

\begin{eqnarray}
Q_1^{ijkl}&=&\Delta^{ijkl}\,,\nonumber\\
Q_2^{i_1j_1k_1l_1i_2j_2k_2l_2}&=&{1\over2}\,[\Delta^{i_1j_1k_1l_1}\,\Delta^{i_2j_2k_2l_2}
\nonumber\\
&&-(\Delta^{i_1j_1k_1l_2}\,\Delta^{i_2j_2k_2l_1}+\Delta^{i_1j_1k_2l_1}\,\Delta^{i_2j_2k_1l_2}
\nonumber\\
&&+\Delta^{i_1j_2k_1l_1}\,\Delta^{i_2j_1k_2l_2}+\Delta^{i_2j_1k_1l_1}\,\Delta^{i_1j_2k_2l_2})
\nonumber\\
&&+(\Delta^{i_1j_1k_2l_2}\,\Delta^{i_2j_2k_1l_1}+\Delta^{i_1j_2k_1l_2}\,\Delta^{i_2j_1
k_2l_1}+\Delta^{i_2j_1k_1l_2}\,\Delta^{i_1j_2k_2l_1})]\,,
\label{091}
\end{eqnarray}

\noindent etc. Then, the discriminants are defined by

\begin{equation}
C_s({\bf A})=Q_s^{i_1j_1k_1l_1\cdots i_s j_s k_s l_s}\,A_{i_1j_1k_1l_1}\,\cdots\,A_{i_s j_s k_s l_s}\,.
\label{092}
\end{equation}

Let us now consider a mixed second--rank matrix ${\bf U}$ with components ${U^i}_j$ and its inverse ${\bf U}^{-1}$ with components ${(U^{-1})^i}_j$. Then we define a similarity transformation for fourth--rank covariant and contravariant matrices as

\begin{eqnarray}
{\bf A}'&=&[{\bf A},\,{\bf U},\,{\bf U},{\bf U},\,{\bf U}]\,,\nonumber\\
{\bf B}'&=&[{\bf B},\,{\bf U}^{-1},\,{\bf U}^{-1},{\bf U}^{-1},\,{\bf U}^{-1}]\,,
\label{093}
\end{eqnarray}

\noindent defined by the components

\begin{eqnarray}
(A')_{i_1i_2i_3i_4}&=&A_{j_1j_2j_3j_4}\,{U^{j_1}}_{i_1}\,{U^{j_2}}_{i_2}\,{U^{j_3}}_{i_3}\,{U^{j_4}}_{i_4}\,,\nonumber\\
(B')^{i_1i_2i_3i_4}&=&B^{j_1j_2j_3j_4}\,{(U^{-1})^{i_1}}_{j_1}\,{(U^{-1})^{i_2}}_{j_2}\,{(U^{-1}
)^{i_3}}_{j_3}\,{(U^{-1})^{i_4}}_{j_4}\,.
\label{094}
\end{eqnarray}

\noindent Then, the discriminants are invariants under this kind of similarity transformations.

The determinant for a higher--rank matrix can be defined in complete analogy with the definition for ordinary matrices. Therefore, in analogy with (\ref{045}) we define

\begin{equation}
A=\det({\bf A})=C_d({\bf A})\,.
\label{095}
\end{equation}

\noindent Let us observe that a relation similar to (\ref{043}) holds for fourth--rank symbols, namely,

\begin{equation}
Q_d^{i_1\cdots i_d j_1\cdots j_d k_1\cdots k_d l_1\cdots l_d}
={1\over{d!}}\,\Delta^{|[i_1j_1k_1l_1}\,\cdots\,\Delta^{i_d j_d k_d l_d]|}
={1\over{d!}}\,\epsilon^{i_1\cdots i_d}\,\cdots\,\epsilon^{l_1\cdots l_d}\,.
\label{096}
\end{equation}

\noindent Therefore

\begin{equation}
A=\det({\bf A})={1\over{d!}}\,\epsilon^{i_1\cdots i_d}\,\cdots\,\epsilon^{l_1\cdots l_d}\,A_{i
_1j_1k_1l_1}\,\cdots\,A_{i_d j_d k_d l_d}\,.
\label{097}
\end{equation}

\noindent In analogy with (\ref{047}) we define

\begin{equation}
A^{ijkl}={1\over A}\,{{\partial A}\over{\partial A_{ijkl}}}\,.
\label{098}
\end{equation}

\noindent Then

\begin{equation}
A^{ijkl}={1\over{(d-1)!}}\,{1\over A}\,\epsilon^{i i_1\cdots i_{(d-1)}}\,\cdots\,\epsilon^{l l
_1\cdots l_{(d-1)}}\,A_{i_1j_1k_1l_1}\,\cdots\,A_{i_{(d-1)}j_{(d-1)}k_{(d-1)}l_{(d-1)}}\,.
\label{099}
\end{equation}

\noindent This matrix satisfies

\begin{equation}
A^{i k_1k_2k_3}\,A_{j k_1k_2k_3}=\delta^i_j\,.
\label{100}
\end{equation}

\noindent The definitions (\ref{094}) and (\ref{095}) were used in previous works \cite{Ta93,TRMC,TR,TU} concerning the applications of fourth--rank geometry to the formulation of an alternative theory for the gravitational field.

Let us observe that $C_d({\bf A})$ is symmetric under the interchange of ${\bf A}$ and ${\bf\Delta}$. Therefore, in analogy with (\ref{030}) the inverse of the matrix ${\bf\Delta}$ is defined by

\begin{equation}
{\bf\Delta}^{-1}={1\over{C_d({\bf A})}}\,{{\partial C_d({\bf A})}\over{\partial{\bf\Delta}}}\,.
\label{101}
\end{equation}

\noindent In terms of components

\begin{equation}
\Delta_{ijkl}={1\over{C_d({\bf A})}}\,{{\partial C_d({\bf A})}\over{\partial\Delta^{ijkl}}}\,.
\label{102}
\end{equation}

\noindent This matrix satisfies

\begin{equation}
{\bf\Delta}^{-1}\,\cdot\,{\bf\Delta}={\bf\Delta}\,\cdot\,{\bf\Delta}^{-1}={\bf I}\,.
\label{103}
\end{equation}

\noindent In terms of components

\begin{equation}
\Delta_{ik_1k_2k_3}\,\Delta^{jk_1k_2k_3}=\delta^j_i\,.
\label{104}
\end{equation}

\noindent The eq. (\ref{096}) can be rewritten as

\begin{equation}
{{\partial C_d({\bf A})}\over{\partial{\bf\Delta}}}-C_d({\bf A})\,{\bf\Delta}\equiv0\,.
\label{105}
\end{equation}

\noindent This is the statement of the Cayley--Hamilton theorem for hypermatrices. 

\bigskip

As an example of the relation above let us consider the simple case $d=2$. The determinant (\ref{094}) is then given by

\begin{eqnarray}
A&=&A_{1111}\,A_{2222}-(A_{1112}\,A_{2221}+A_{1121}\,A_{2212}+A_{1211}\,A_{2122}+A_{
2111}\,A_{1222})\nonumber\\
&&+(A_{1122}\,A_{2211}+A_{1212}\,A_{2121}+A_{2112}\,A_{1221})\,.
\label{106}
\end{eqnarray}

\noindent The components of the matrix $A^{ijkl}$ are given by

\begin{equation}
\begin{array}{@{}llll@{}}
A^{1111}={1\over A}\,A_{2222}\,,&A^{1112}=-{1\over A}\,A_{2221}\,,
&A^{1121}=-{1\over A}\,A_{2212}\,,&A^{1211}=-{1\over A}\,A_{2122}\,,\cr
&\cr
A^{2111}=-{1\over A}\,A_{1222}\,,&A^{1122}={1\over A}\,A_{2211}\,,
&A^{1212}={1\over A}\,A_{2121}\,,&A^{2112}={1\over A}\,A_{1221}\,,\cr
&&&
\end{array}
\label{107}
\end{equation}

\noindent and similar expressions for the other components. In order to check the validity of eq. (\ref{097}) let us consider the cases $(11)$ and $(12)$. We can then verify taht

\begin{eqnarray}
A^{1ijk}\,A_{1ijk}&=&1\,,\nonumber\\
A^{1ijk}\,A_{2ijk}&=&0\,,
\label{108}
\end{eqnarray}

\noindent and similar relations for the other indices.

\subsection{Semi--Magic Squares}

The algebraic invariants which can be constructed in this case are given by the semi--magic squares of rank $4$. Their number is $H_1(4)=1$, $H_2(4)=5$, $H_3(4)=120$, $H_4(4)=7558$. As for the second--rank case we must take care only of the representatives for each equivalence class. The number of equivalence classes $p_4(n)$ is given by the generating function

\begin{equation}
\sum_{n=0}^\infty\,p_4(n)\,x^n=\prod_{n=1}^\infty\,{1\over{(1-x^n)^{n!}}}\,,
\label{109}
\end{equation}

\noindent For the first values of $n$ $p_4(n)$ is given by

\begin{equation}
p_4(n)=\{1,\,1,\,3,\,9,\,36,\,\cdots\}\,.
\label{110}
\end{equation}

The representatives for each equivalence class are

\begin{eqnarray}
K_{4,1}&=&\{(4)\}\,,\nonumber\\
\nonumber\\
K_{4,2}&=&\left\{\left(\matrix{4&0\cr0&4\cr}\right)\,,\,\left(\matrix{3&1\cr1&3\cr}\right)\,,\,
\left(\matrix{2&2\cr2&2\cr}\right)\right\}\,,\nonumber\\
\nonumber\\
K_{4,3}&=&\left\{\left(\matrix{4&0&0\cr0&4&0\cr0&0&4\cr}\right)\,,\,
\left(\matrix{4&0&0\cr0&3&1\cr0&1&3\cr}\right)\,,\,
\left(\matrix{4&0&0\cr0&2&2\cr0&2&2\cr}\right)\,,\,
\left(\matrix{3&1&0\cr0&3&1\cr1&0&3\cr}\right)\,,\,
\left(\matrix{3&1&0\cr1&2&1\cr0&1&3\cr}\right)\right.\,,\nonumber\\
&&\quad\left.\left(\matrix{3&0&1\cr0&2&2\cr1&2&1\cr}\right)\,,
\left(\matrix{2&2&0\cr0&2&2\cr2&0&2\cr}\right)\,,\,
\left(\matrix{0&2&2\cr2&1&1\cr2&1&1\cr}\right)\,,\,
\left(\matrix{2&1&1\cr1&2&1\cr1&1&2\cr}\right)\right\}\,,\nonumber\\
\nonumber\\
K_{4,4}&=&\left\{\left(\matrix{4&0&0&0\cr0&4&0&0\cr0&0&4&0\cr0&0&0&4\cr}\right)\,,\,
\left(\matrix{4&0&0&0\cr0&4&0&0\cr0&0&3&1\cr0&0&1&3\cr}\right)\,,\,
\left(\matrix{4&0&0&0\cr0&4&0&0\cr0&0&2&2\cr0&0&2&2\cr}\right)\,,\,
\left(\matrix{4&0&0&0\cr0&3&1&0\cr0&0&3&1\cr0&1&0&3\cr}\right)\right.\,,\nonumber\\
&&\quad\left(\matrix{4&0&0&0\cr0&3&1&0\cr0&1&2&1\cr0&0&1&3\cr}\right)\,,\,
\left(\matrix{4&0&0&0\cr0&3&0&1\cr0&0&2&2\cr0&1&2&1\cr}\right)\,,\,
\left(\matrix{4&0&0&0\cr0&2&2&0\cr0&0&2&2\cr0&2&0&2\cr}\right)\,,\,
\left(\matrix{4&0&0&0\cr0&0&2&2\cr0&2&1&1\cr0&2&1&1\cr}\right)\,,\nonumber\\
&&\quad\left(\matrix{4&0&0&0\cr0&2&1&1\cr0&1&2&1\cr0&1&1&2\cr}\right)\,,\,
\left(\matrix{3&1&0&0\cr0&3&1&0\cr0&0&3&1\cr1&0&0&3\cr}\right)\,,\,
\left(\matrix{3&1&0&0\cr1&3&0&0\cr0&0&3&1\cr0&0&1&3\cr}\right)\,,\,
\left(\matrix{3&0&0&1\cr0&3&0&1\cr1&0&3&0\cr0&1&1&2\cr}\right)\,,\nonumber\\
&&\quad\left(\matrix{3&0&0&1\cr0&3&0&1\cr0&0&3&1\cr1&1&1&1\cr}\right)\,,\,
\left(\matrix{3&1&0&0\cr1&3&0&0\cr0&0&2&2\cr0&0&2&2\cr}\right)\,,\,
\left(\matrix{3&0&0&1\cr0&3&0&1\cr0&0&2&2\cr1&1&2&0\cr}\right)\,,\,
\left(\matrix{3&1&0&0\cr0&3&0&1\cr1&0&2&1\cr0&0&2&2\cr}\right)\,,\nonumber\\
&&\quad\left(\matrix{3&0&0&1\cr0&3&1&0\cr0&1&2&1\cr1&0&1&2\cr}\right)\,,\,
\left(\matrix{3&0&0&1\cr0&3&1&0\cr1&0&2&1\cr0&1&1&2\cr}\right)\,,\,
\left(\matrix{3&0&0&1\cr0&3&0&1\cr0&1&2&1\cr1&0&2&1\cr}\right)_{\bf a}\,,\,
\left(\matrix{3&0&0&1\cr0&3&1&0\cr0&0&2&2\cr1&1&1&1\cr}\right)_{\bf a}\,,\nonumber\\
&&\quad\left(\matrix{3&0&0&1\cr0&2&2&0\cr0&2&0&2\cr1&0&2&1\cr}\right)\,,\,
\left(\matrix{3&0&1&0\cr0&2&0&2\cr1&0&2&1\cr0&2&1&1\cr}\right)\,,\,
\left(\matrix{3&1&0&0\cr0&2&0&2\cr1&0&2&1\cr0&1&2&1\cr}\right)\,,\,
\left(\matrix{3&1&0&0\cr0&2&1&1\cr0&1&2&1\cr1&0&1&2\cr}\right)\,,\nonumber\\
&&\quad\left(\matrix{3&1&0&0\cr1&1&1&1\cr0&2&1&1\cr0&0&2&2\cr}\right)_{\bf b}\,,\,
\left(\matrix{3&1&0&0\cr1&1&2&0\cr0&1&1&2\cr0&1&1&2\cr}\right)_{\bf b}\,,\,
\left(\matrix{3&0&0&1\cr0&2&1&1\cr0&1&2&1\cr1&1&1&1\cr}\right)\,,\,
\left(\matrix{2&2&0&0\cr0&2&2&0\cr0&0&2&2\cr2&0&0&2\cr}\right)\,,\nonumber\\
&&\quad\left(\matrix{2&2&0&0\cr2&2&0&0\cr0&0&2&2\cr0&0&2&2\cr}\right)\,,\,
\left(\matrix{2&2&0&0\cr2&1&1&0\cr0&1&1&2\cr0&0&2&2\cr}\right)_{\bf c}\,,\,
\left(\matrix{2&1&0&1\cr0&2&2&0\cr1&0&2&1\cr1&1&0&2\cr}\right)_{\bf c}\,,\,
\left(\matrix{2&0&1&1\cr0&2&1&1\cr1&1&2&0\cr1&1&0&2\cr}\right)_{\bf d}\,,\,\nonumber\\
&&\quad\left(\matrix{2&1&1&0\cr0&2&1&1\cr1&0&2&1\cr1&1&0&2\cr}\right)_{\bf d}\,,\,
\left(\matrix{2&1&1&0\cr1&2&1&0\cr1&0&1&2\cr0&1&1&2\cr}\right)\,,\,
\left(\matrix{2&1&1&0\cr1&2&0&1\cr1&1&1&1\cr0&0&2&2\cr}\right)\,,\,
\left(\matrix{2&2&0&0\cr0&0&2&2\cr1&1&1&1\cr1&1&1&1\cr}\right)\,,\nonumber\\
&&\quad\left.\left(\matrix{2&0&1&1\cr2&0&1&1\cr0&2&1&1\cr0&2&1&1\cr}\right)\,,\,
\left(\matrix{2&1&0&1\cr0&2&1&1\cr1&0&2&1\cr1&1&1&1\cr}\right)\,,\,
\left(\matrix{2&0&1&1\cr0&2&1&1\cr1&1&1&1\cr1&1&1&1\cr}\right)\,,\,
\left(\matrix{1&1&1&1\cr1&1&1&1\cr1&1&1&1\cr1&1&1&1\cr}\right)\right\}\,.\nonumber\\
&&
\label{111}
\end{eqnarray}

\noindent Let us observe that in (\ref{108}) there are 40 magic squares of order 4 rather than the 36 implied by (\ref{106}); the couples $(19,20)$, $(25,26)$, $(34,35)$ and $(36,37)$, labelled by ${\bf a}$, ${\bf b}$, ${\bf c}$ and ${\bf c}$, respectively, have the same connectivity structure but, since rows and columns corresponds to different matrices, they represent different invariants.

\subsection{Construction of Discriminants}

Each semi--magic square represents one and only one algebraic invariant. However, in this case, the corresponding invariant can no longer be represented by mean of traces, as was done for ordinary matrices. The semi--magic squares of order 1 and 2 in (\ref{106}) correspond to the following invariants

\begin{eqnarray}
(4)&=&\Delta^{ijkl}\,A_{ijkl}\,,\nonumber\\
\nonumber\\
\left(\matrix{4&0\cr0&4\cr}\right)&=&\left(\Delta^{ijkl}\,A_{ijkl}\right)^2\,,\nonumber\\
\left(\matrix{3&1\cr1&3\cr}\right)&=&\Delta^{i_1j_1k_1l_1}\,A_{l_1i_2j_2k_2}\,
\Delta^{i_2j_2k_2l_2}\,A_{l_2i_1j_1k_1}\,,\nonumber\\
\left(\matrix{2&2\cr2&2\cr}\right)&=&\Delta^{i_1j_1k_1l_1}\,A_{k_1l_1i_2j_2}\,
\Delta^{i_2j_2k_2l_2}\,A_{k_2l_2i_1j_1}\,.
\label{112}
\end{eqnarray}

\noindent It is obvious from the expressions above that semi--magic squares are more practical for representing algebraic invariants.

The corresponding discriminants are linear combinations of the monomial algebraic invariants (semi--magic squares) of the same order. In order to determine the coefficients of this linear combination we proceed in a way similar to that for ordinary matrices. The matrices can be contracted according to the allowed number of possible permutations. The possible permutations are the same as described in sec. 3.2. However, this time they must be summed three times. It is obvious that the number of terms which must be computed growth very fast, as $(n!)^{r-1}$. Therefore, even when this algorithm provides a direct answer, a more practical way to evaluate the coefficients is necessary. Let us therefore consider the graphical algorithm developed in sec. 2.6. 

For $n=1$ there is only one possibility, namely

\begin{equation}
C_1({\bf A})=(4)\,.
\label{113}
\end{equation}

For $n=2$ we have

\begin{equation}
\begin{array}{@{}l@{\hspace{35pt}}l@{}}
\put(-28,-28){\grid(28,56)(14,14)}=\left(\matrix{4&0\cr0&4\cr}\right)\,,
&\put(-28,-28){\grid(28,56)(14,14)}
\put(-21,21){\circle{4}}\put(-7,21){\circle{4}}\put(-19,21){\line(1,0){10}}
=-3\,\left(\matrix{3&1\cr1&3\cr}\right)\,,\cr
&\cr
\put(-28,-28){\grid(28,56)(14,14)}
\put(-21,21){\circle{4}}\put(-7,21){\circle{4}}\put(-19,21){\line(1,0){10}}
\put(-21,7){\circle{4}}\put(-7,7){\circle{4}}\put(-19,7){\line(1,0){10}}
=3\,\left(\matrix{2&2\cr2&2\cr}\right)\,,
&\put(-28,-28){\grid(28,56)(14,14)}
\put(-21,21){\circle{4}}\put(-7,21){\circle{4}}\put(-19,21){\line(1,0){10}}
\put(-21,7){\circle{4}}\put(-7,7){\circle{4}}\put(-19,7){\line(1,0){10}}
\put(-21,-7){\circle{4}}\put(-7,-7){\circle{4}}\put(-19,-7){\line(1,0){10}}
=-\left(\matrix{1&3\cr3&1\cr}\right)\,.\cr
&
\end{array}
\label{114}
\end{equation}

\noindent The coefficients in this expressions are obtained from

\begin{equation}
P_2^3=({\bf 0}-{\bf 1})^3={\bf 0}\cdot{\bf 0}\cdot{\bf 0}-3\cdot{\bf 0}\cdot{\bf 0}\cdot{\bf 1}+3\cdot{\bf 0}\cdot{\bf 1}\cdot{\bf 1}-{\bf 1}\cdot{\bf 1}\cdot{\bf 1}\,.
\label{115}
\end{equation}

\noindent The resulting discriminant is

\begin{equation}
C_2({\bf A})={1\over2}\,\left[\left(\matrix{4&0\cr0&4\cr}\right)-4\,\left(
\matrix{3&1\cr1&3\cr}\right)+3\,\left(\matrix{2&2\cr2&2\cr}\right)\right]\,.
\label{116}
\end{equation}

For $n=3$ the grids are given in Appendix A. We can verify that the coefficientes are given by

\begin{eqnarray}
P_3^3&=&({\bf 0}-3\cdot{\bf 1}+2\cdot{\bf 2})^3\nonumber\\
&=&{\bf 0}\cdot{\bf 0}\cdot{\bf 0}-9\cdot{\bf 0}\cdot{\bf 0}\cdot{\bf 1}+27\cdot{\bf 0}\cdot{\bf 1}\cdot{\bf 1}+6\cdot{\bf 0}\cdot{\bf 0}\cdot{\bf 2}-27\cdot{\bf 1}\cdot{\bf 1}\cdot{\bf 1}\nonumber\\
&&-36\cdot{\bf 0}\cdot{\bf 1}\cdot{\bf 2}+54\cdot{\bf 1}\cdot{\bf 1}\cdot{\bf 2}+12\cdot{\bf 0}
\cdot{\bf 2}\cdot{\bf 2}-36\cdot{\bf 1}\cdot{\bf 2}\cdot{\bf 2}+8\cdot{\bf 2}\cdot{\bf 2}\cdot{
\bf 2}\,.
\label{117}
\end{eqnarray}

\noindent The discriminant is

\begin{eqnarray}
C_3({\bf A})&=&{1\over6}\,\left[\left(\matrix{4&0&0\cr0&4&0\cr0&0&4\cr}\right)
-12\,\left(\matrix{4&0&0\cr0&3&1\cr0&1&3\cr}\right)
+9\,\left(\matrix{4&0&0\cr0&2&2\cr0&2&2\cr}\right)\right.\nonumber\\
&&\quad\quad+8\,\left(\matrix{3&1&0\cr0&3&1\cr1&0&3\cr}\right)
+36\,\left(\matrix{3&1&0\cr1&2&1\cr0&1&3\cr}\right)
-72\,\left(\matrix{3&1&0\cr1&1&2\cr0&2&2\cr}\right)\nonumber\\
&&\quad\quad+6\,\left.\left(\matrix{2&2&0\cr0&2&2\cr2&0&2\cr}\right)
+36\,\left(\matrix{0&2&2\cr2&1&1\cr2&1&1\cr}\right)
-12\,\left(\matrix{2&1&1\cr1&2&1\cr1&1&2\cr}\right)\right]\,,
\label{118}
\end{eqnarray}

For $n=4$ the grids are given in Appendix B. Now we do not verify that the coefficientes in the relation 

\begin{eqnarray}
P_4^3&=&({\bf 0}-6\cdot{\bf 1}+3\cdot{\bf 1}^2+8\cdot{\bf 2}-6\cdot{\bf 3})^3\nonumber\\
&=&{\bf 0}\cdot{\bf 0}\cdot{\bf 0}
-18\cdot{\bf 0}\cdot{\bf 0}\cdot{\bf 1}
+108\cdot{\bf 0}\cdot{\bf 1}\cdot{\bf 1}
+9\cdot{\bf 0}\cdot{\bf 0}\cdot{\bf 1}^2
+24\cdot{\bf 0}\cdot{\bf 0}\cdot{\bf 2}\nonumber\\
&&-216\cdot{\bf 1}\cdot{\bf 1}\cdot{\bf 1}
-108\cdot{\bf 0}\cdot{\bf 1}\cdot{\bf 1}^2
-288\cdot{\bf 0}\cdot{\bf 1}\cdot{\bf 2}
-18\cdot{\bf 0}\cdot{\bf 0}\cdot{\bf 3}\nonumber\\
&&+324\cdot{\bf 1}\cdot{\bf 1}\cdot{\bf 1}^2
+864\cdot{\bf 1}\cdot{\bf 1}\cdot{\bf 2}
+108\cdot{\bf 0}\cdot{\bf 1}\cdot{\bf 3}
+27\cdot{\bf 0}\cdot{\bf 1}^2\cdot{\bf 1}^2\nonumber\\
&&+144\cdot{\bf 0}\cdot{\bf 1}^2\cdot{\bf 2}
+192\cdot{\bf 0}\cdot{\bf 2}\cdot{\bf 2}
-648\cdot{\bf 1}\cdot{\bf 1}\cdot{\bf 3}
-162\cdot{\bf 1}\cdot{\bf 1}^2\cdot{\bf 1}^2\nonumber\\
&&-864\cdot{\bf 1}\cdot{\bf 1}^2\cdot{\bf 2}
-1152\cdot{\bf 1}\cdot{\bf 2}\cdot{\bf 2}
-54\cdot{\bf 0}\cdot{\bf 1}^2\cdot{\bf 3}
-288\cdot{\bf 0}\cdot{\bf 2}\cdot{\bf 3}\nonumber\\
&&+648\cdot{\bf 1}\cdot{\bf 1}^2\cdot{\bf 3}
+1728\cdot{\bf 1}\cdot{\bf 2}\cdot{\bf 3}
+27\cdot{\bf 1}^2\cdot{\bf 1}^2\cdot{\bf 1}^2
+216\cdot{\bf 1}^2\cdot{\bf 1}^2\cdot{\bf 2}\nonumber\\
&&+576\cdot{\bf 1}^2\cdot{\bf 2}\cdot{\bf 2}
+512\cdot{\bf 2}\cdot{\bf 2}\cdot{\bf 2}
+108\cdot{\bf 0}\cdot{\bf 3}\cdot{\bf 3}
-648\cdot{\bf 1}\cdot{\bf 3}\cdot{\bf 3}\nonumber\\
&&-162\cdot{\bf 1}^2\cdot{\bf 1}^2\cdot{\bf 3}
-864\cdot{\bf 1}^2\cdot{\bf 2}\cdot{\bf 3}
-1152\cdot{\bf 2}\cdot{\bf 2}\cdot{\bf 3}
+324\cdot{\bf 1}^2\cdot{\bf 3}\cdot{\bf 3}\nonumber\\
&&+864\cdot{\bf 2}\cdot{\bf 3}\cdot{\bf 3}
-216\cdot{\bf 3}\cdot{\bf 3}\cdot{\bf 3}\,.
\label{119}
\end{eqnarray}

\noindent corresponds to that given in the Appendix B. Instead, we use them to check the correctness of the coefficients there. The discriminant is

\begin{eqnarray}
C_4({\bf A})&=&{1\over{24}}\,\left[\left(\matrix{4&0&0&0\cr0&4&0&0\cr0&0&4&0\cr0&0&0&4\cr}
\right)-24\,\left(\matrix{4&0&0&0\cr0&4&0&0\cr0&0&3&1\cr0&0&1&3\cr}\right)
+18\left(\matrix{4&0&0&0\cr0&4&0&0\cr0&0&2&2\cr0&0&2&2\cr}\right)\right.\nonumber\\
&&\quad\quad+32\,\left(\matrix{4&0&0&0\cr0&3&1&0\cr0&0&3&1\cr0&1&0&3\cr}\right)
+144\,\left(\matrix{4&0&0&0\cr0&3&1&0\cr0&1&2&1\cr0&0&1&3\cr}\right)
-288\,\left(\matrix{4&0&0&0\cr0&3&0&1\cr0&0&2&2\cr0&1&2&1\cr}\right)\nonumber\\
&&\quad\quad+24\,\left(\matrix{4&0&0&0\cr0&2&2&0\cr0&0&2&2\cr0&2&0&2\cr}\right)
+144\,\left(\matrix{4&0&0&0\cr0&0&2&2\cr0&2&1&1\cr0&2&1&1\cr}\right)
-48\,\left(\matrix{4&0&0&0\cr0&2&1&1\cr0&1&2&1\cr0&1&1&2\cr}\right)\nonumber\\
&&\quad\quad-24\,\left(\matrix{3&1&0&0\cr0&3&1&0\cr0&0&3&1\cr1&0&0&3\cr}\right)
+48\,\left(\matrix{3&1&0&0\cr1&3&0&0\cr0&0&3&1\cr0&0&1&3\cr}\right)
-288\,\left(\matrix{3&0&0&1\cr0&3&0&1\cr1&0&3&0\cr0&1&1&2\cr}\right)\nonumber\\
&&\quad\quad-96\,\left(\matrix{3&0&0&1\cr0&3&0&1\cr0&0&3&1\cr1&1&1&1\cr}\right)
-72\,\left(\matrix{3&1&0&0\cr1&3&0&0\cr0&0&2&2\cr0&0&2&2\cr}\right)
+144\,\left(\matrix{3&0&0&1\cr0&3&0&1\cr0&0&2&2\cr1&1&2&0\cr}\right)\nonumber\\
&&\quad\quad+288\,\left(\matrix{3&1&0&0\cr0&3&0&1\cr1&0&2&1\cr0&0&2&2\cr}\right)
-432\,\left(\matrix{3&0&0&1\cr0&3&1&0\cr0&1&2&1\cr1&0&1&2\cr}\right)
+288\,\left(\matrix{3&0&0&1\cr0&3&1&0\cr1&0&2&1\cr0&1&1&2\cr}\right)\nonumber\\
&&\quad\quad+288\,\left(\matrix{3&0&0&1\cr0&3&0&1\cr0&1&2&1\cr1&0&2&1\cr}\right)
+288\,\left(\matrix{3&0&0&1\cr0&3&1&0\cr0&0&2&2\cr1&1&1&1\cr}\right)
-288\,\left(\matrix{3&0&0&1\cr0&2&2&0\cr0&2&0&2\cr1&0&2&1\cr}\right)\nonumber\\
&&\quad\quad+864\,\left(\matrix{3&0&1&0\cr0&2&0&2\cr1&0&2&1\cr0&2&1&1\cr}\right)
-576\,\left(\matrix{3&1&0&0\cr0&2&0&2\cr1&0&2&1\cr0&1&2&1\cr}\right)
+144\,\left(\matrix{3&1&0&0\cr0&2&1&1\cr0&1&2&1\cr1&0&1&2\cr}\right)\nonumber\\
&&\quad\quad-576\,\left(\matrix{3&1&0&0\cr1&1&1&1\cr0&2&1&1\cr0&0&2&2\cr}\right)
-576\,\left(\matrix{3&1&0&0\cr1&1&2&0\cr0&1&1&2\cr0&1&1&2\cr}\right)
+288\,\left(\matrix{3&0&0&1\cr0&2&1&1\cr0&1&2&1\cr1&1&1&1\cr}\right)\nonumber\\
&&\quad\quad+18\,\left(\matrix{2&2&0&0\cr0&2&2&0\cr0&0&2&2\cr2&0&0&2\cr}\right)
+27\,\left(\matrix{2&2&0&0\cr2&2&0&0\cr0&0&2&2\cr0&0&2&2\cr}\right)
-144\,\left(\matrix{2&2&0&0\cr2&1&1&0\cr0&1&1&2\cr0&0&2&2\cr}\right)\nonumber\\
&&\quad\quad+480\,\left(\matrix{2&1&0&1\cr0&2&2&0\cr1&0&2&1\cr1&1&0&2\cr}\right)
+288\,\left(\matrix{2&0&1&1\cr0&2&1&1\cr1&1&2&0\cr1&1&0&2\cr}\right)
-504\,\left(\matrix{2&1&1&0\cr0&2&1&1\cr1&0&2&1\cr1&1&0&2\cr}\right)\nonumber\\
&&\quad\quad-288\,\left(\matrix{2&1&1&0\cr1&2&1&0\cr1&0&1&2\cr0&1&1&2\cr}\right)
-288\,\left(\matrix{2&1&1&0\cr1&2&0&1\cr1&1&1&1\cr0&0&2&2\cr}\right)
+144\,\left(\matrix{2&2&0&0\cr0&0&2&2\cr1&1&1&1\cr1&1&1&1\cr}\right)\nonumber\\
&&\quad\quad+144\,\left(\matrix{2&0&1&1\cr2&0&1&1\cr0&2&1&1\cr0&2&1&1\cr}\right)
+528\,\left(\matrix{2&1&0&1\cr0&2&1&1\cr1&0&2&1\cr1&1&1&1\cr}\right)
-144\,\left(\matrix{2&0&1&1\cr0&2&1&1\cr1&1&1&1\cr1&1&1&1\cr}\right)\nonumber\\
&&\quad\quad\left.+24\,\left(\matrix{1&1&1&1\cr1&1&1&1\cr1&1&1&1\cr1&1&1&1\cr}\right)\right]\,.
\label{120}
\end{eqnarray}

\subsection{The Cayley--Hamilton Theorem}

Let us write the Cayley--Hamilton theorem in terms of almost--magic rectangles. For the first values of $d$ we obtain

\begin{eqnarray}
\left(\matrix{4&{\bf 0}\cr0&{\bf 4}\cr}\right)-4\,\left(\matrix{3&{\bf 1}\cr1&{\bf 3}\cr}\right)
+3\,\left(\matrix{2&{\bf 2}\cr2&{\bf 2}\cr}\right)-C_2({\bf A})\,[{\bf 4}]&\equiv&0\,,
\nonumber\\
\nonumber\\
{1\over2}\,\left[\left(\matrix{4&0&{\bf 0}\cr0&4&{\bf 0}\cr0&0&{\bf 4}\cr}\right)
-4\,\left(\matrix{3&1&{\bf 0}\cr1&3&{\bf 0}\cr0&0&{\bf 4}\cr}\right)
+3\,\left(\matrix{2&2&{\bf 0}\cr2&2&{\bf 0}\cr0&0&{\bf 4}\cr}\right)\right]
-4\,\left(\matrix{4&0&{\bf 0}\cr0&3&{\bf 1}\cr0&1&{\bf 3}\cr}\right)&&\nonumber\\
+3\,\left(\matrix{4&0&{\bf 0}\cr0&2&{\bf 2}\cr0&2&{\bf 2}\cr}\right)
+4\,\left(\matrix{3&1&{\bf 0}\cr0&3&{\bf 1}\cr1&0&{\bf 3}\cr}\right)
+6\,\left(\matrix{3&0&{\bf 1}\cr0&3&{\bf 1}\cr1&1&{\bf 2}\cr}\right)
+12\,\left(\matrix{3&1&{\bf 0}\cr1&2&{\bf 1}\cr0&1&{\bf 3}\cr}\right)&&\nonumber\\
-12\,\left(\matrix{3&1&{\bf 0}\cr1&1&{\bf 2}\cr0&2&{\bf 2}\cr}\right)
-12\,\left(\matrix{3&0&{\bf 1}\cr1&2&{\bf 1}\cr0&2&{\bf 2}\cr}\right)
+3\,\left(\matrix{2&2&{\bf 0}\cr2&0&{\bf 2}\cr0&2&{\bf 2}\cr}\right)
-12\,\left(\matrix{2&2&{\bf 0}\cr2&1&{\bf 1}\cr0&1&{\bf 3}\cr}\right)&&\nonumber\\
+6\,\left(\matrix{2&2&{\bf 0}\cr1&1&{\bf 2}\cr1&1&{\bf 2}\cr}\right)
+12\,\left(\matrix{2&1&{\bf 1}\cr2&1&{\bf 1}\cr0&2&{\bf 2}\cr}\right)
-6\,\left(\matrix{2&1&{\bf 1}\cr1&2&{\bf 1}\cr1&1&{\bf 2}\cr}\right)
-C_3({\bf A})\,[{\bf 4}]&\equiv&0\,,
\label{121}
\end{eqnarray}

\noindent where $[{\bf 4}]$ stands for the unit matrix ${\bf\Delta}^{-1}$. Contracting with ${\bf\Delta}$, $[4]=d$, we obtain (\ref{115}), (\ref{116}) and (\ref{117}), respectively. Therefore, the above is the Cayley--Hamilton theorem for fourth--rank matrices. The expression of the Cayley--Hamilton theorem for $d=4$ is exhibited in the Appendix C.

\bigskip

In order to ilustrate our result let us consider the simple case of a completely symmetric fourth--rank matrix $G_{ijkl}$ in two dimensions, $i,j,k,l=1,2$. Then, the determinant, according to (\ref{112}) and (\ref{115}), is given by

\begin{equation}
C_2({\bf G})=G_{1111}\,G_{2222}-4\,G_{1112}\,G_{1222}+3\,G_{1122}^2\,.
\label{122}
\end{equation}

\noindent In terms of ${\bf G}$'s and ${\bf\Delta}$'s the almost magic rectangles in (\ref{118}) are given by

\begin{eqnarray}
\left(\matrix{4&{\bf 0}\cr0&{\bf 4}\cr}\right)_{abcd}&=&(\Delta^{ijkl}\,G_{ijkl})\,G_{abcd}\,,
\nonumber\\
\left(\matrix{3&{\bf 1}\cr1&{\bf 3}\cr}\right)_{abcd}&=&G_{(a|ijk}\,\Delta^{ijkl}\,G_{l|bcd)}\,,
\nonumber\\
\left(\matrix{2&{\bf 2}\cr2&{\bf 2}\cr}\right)_{abcd}&=&G_{(ab|ij}\,\Delta^{ijkl}\,G_{kl|cd)}\,.
\label{123}
\end{eqnarray}

\noindent In detail we have

\begin{eqnarray}
\left(\matrix{4&{\bf 0}\cr0&{\bf 4}\cr}\right)_{1111}&=&(G_{1111}+G_{2222})\,G_{1111}\,,
\nonumber\\
\left(\matrix{4&{\bf 0}\cr0&{\bf 4}\cr}\right)_{1112}&=&(G_{1111}+G_{2222})\,G_{1112}\,,
\nonumber\\
\left(\matrix{4&{\bf 0}\cr0&{\bf 4}\cr}\right)_{1122}&=&(G_{1111}+G_{2222})\,G_{1122}\,,
\nonumber\\
\nonumber\\
\left(\matrix{3&{\bf 1}\cr1&{\bf 3}\cr}\right)_{1111}&=&G_{1111}^2+G_{1112}\,G_{1222}\,,
\nonumber\\
\left(\matrix{3&{\bf 1}\cr1&{\bf 3}\cr}\right)_{1112}&=&{1\over4}\,(4\,G_{1111}\,G_{1112}
+G_{1112}\,G_{2222}+3\,G_{1122}\,G_{1222})\,,\nonumber\\
\left(\matrix{3&{\bf 1}\cr1&{\bf 3}\cr}\right)_{1122}&=&{1\over2}\,[(G_{1111}+G_{2222})\,
G_{1122}+G_{1112}^2+G_{1222}^2]\,,\nonumber\\
\nonumber\\
\left(\matrix{2&{\bf 2}\cr2&{\bf 2}\cr}\right)_{1111}&=&G_{1111}^2+G_{1122}^2\,,\nonumber\\
\left(\matrix{2&{\bf 2}\cr2&{\bf 2}\cr}\right)_{1112}&=&G_{1111}\,G_{1112}+G_{1122}\,G_{1222}\,,
\nonumber\\
\left(\matrix{2&{\bf 2}\cr2&{\bf 2}\cr}\right)_{1122}&=&{1\over3}\,[(G_{1111}+G_{2222})\,
G_{1122}+2\,G_{1112}^2+2\,G_{1222}^2]\,,
\label{124}
\end{eqnarray}

\noindent and similar expressions for the other indices. Then we have

\begin{eqnarray}
\left(\matrix{4&{\bf 0}\cr0&{\bf 4}\cr}\right)_{1111}-4\,\left(\matrix{3&{\bf 1}\cr1&{\bf 3}\cr
}\right)_{1111}+3\,\left(\matrix{2&{\bf 2}\cr2&{\bf 2}\cr}\right)_{1111}-C_2({\bf A})\,[{\bf 4}
]_{1111}&\equiv&0\,,\nonumber\\
\left(\matrix{4&{\bf 0}\cr0&{\bf 4}\cr}\right)_{1112}-4\,\left(\matrix{3&{\bf 1}\cr1&{\bf 3}\cr
}\right)_{1112}+3\,\left(\matrix{2&{\bf 2}\cr2&{\bf 2}\cr}\right)_{1112}-C_2({\bf A})\,[{\bf 4}
]_{1112}&\equiv&0\,,\nonumber\\
\left(\matrix{4&{\bf 0}\cr0&{\bf 4}\cr}\right)_{1122}-4\,\left(\matrix{3&{\bf 1}\cr1&{\bf 3}\cr
}\right)_{1122}+3\,\left(\matrix{2&{\bf 2}\cr2&{\bf 2}\cr}\right)_{1122}-C_2({\bf A})\,[{\bf 4}
]_{1122}&\equiv&0\,,
\label{125}
\end{eqnarray}

\noindent and similar results for the other indices.

\subsection{Generalised Multiplication and Ring Structure}

Now we introduce a definition for the multiplication of fourth--rank matrices which allows to give a ring structure to the space of fourth--rank matrices. Such as we defined determinants by means of an inverse analogy, we define the product of fourth--rank hypermatrices with a similar relation. To this purpose, in analogy with (\ref{033}), we define

\begin{eqnarray}
T_1({\bf A})&=&C_1({\bf A})\,,\nonumber\\
T_2({\bf A})&=&{1\over2}\,C_1^2({\bf A})-C_2({\bf A})\,,\nonumber\\
T_3({\bf A})&=&{1\over3}\,C_1^3({\bf A})-C_1({\bf A})\,C_2({\bf A})+C_3({\bf A})\,,\nonumber\\
T_4({\bf A})&=&{1\over4}\,C_1^4({\bf A})-C_1^2({\bf A})\,C_2({\bf A})+C_1({\bf A})\,C_3({\bf A})
+{1\over2}\,C_2^2({\bf A})-C_4({\bf A})\,.
\label{126}
\end{eqnarray}

\noindent Now, we invert the relation (\ref{035}) to define powers of fourth--rank matrices as

\begin{equation}
({\bf A}^s)={{\partial T_s({\bf A})}\over{\partial{\bf I}}}\,.
\label{127}
\end{equation}

For the first values of $s$ we obtain

\begin{eqnarray}
({\bf A})&=&({\bf 4})\,,\nonumber\\
({\bf A}^2)&=&4\,\left(\matrix{3&{\bf 1}\cr1&{\bf 3}\cr}\right)-3\,\left(\matrix{2&{\bf 2}\cr2&{
\bf 2}\cr}\right)\,,\nonumber\\
\nonumber\\
({\bf A}^3)&=&4\,\left(\matrix{3&1&{\bf 0}\cr0&3&{\bf 1}\cr1&0&{\bf 3}\cr}\right)
+6\,\left(\matrix{3&0&{\bf 1}\cr0&3&{\bf 1}\cr1&1&{\bf 2}\cr}\right)
+12\,\left(\matrix{3&1&{\bf 0}\cr1&2&{\bf 1}\cr0&1&{\bf 3}\cr}\right)
-12\,\left(\matrix{3&1&{\bf 0}\cr1&1&{\bf 2}\cr0&2&{\bf 2}\cr}\right)\nonumber\\
&&-12\,\left(\matrix{3&0&{\bf 1}\cr1&2&{\bf 1}\cr0&2&{\bf 2}\cr}\right)
+3\,\left(\matrix{2&2&{\bf 0}\cr2&0&{\bf 2}\cr0&2&{\bf 2}\cr}\right)
-12\,\left(\matrix{2&2&{\bf 0}\cr2&1&{\bf 1}\cr0&1&{\bf 3}\cr}\right)
+6\,\left(\matrix{2&2&{\bf 0}\cr1&1&{\bf 2}\cr1&1&{\bf 2}\cr}\right)\nonumber\\
&&+12\,\left(\matrix{2&1&{\bf 1}\cr2&1&{\bf 1}\cr0&2&{\bf 2}\cr}\right)
-6\,\left(\matrix{2&1&{\bf 1}\cr1&2&{\bf 1}\cr1&1&{\bf 2}\cr}\right)\,.
\label{128}
\end{eqnarray}

\noindent The expression for $({\bf A}^4)$ is displayed in Appendix D.

Now, the discriminants can be written as

\begin{eqnarray}
C_0({\bf A})&=&1\,,\nonumber\\
C_1({\bf A})&=&\left[{\bf A}\right]\,,\nonumber\\
C_2({\bf A})&=&{1\over2}\,\left[\left[{\bf A}\right]^2-\left[{\bf A}^2\right]\right]\,,
\nonumber\\
C_3({\bf A})&=&{1\over{3!}}\,\left[\left[{\bf A}\right]^3-3\,\left[{\bf A}\right]\,\left[{\bf A
}^2\right]+2\,\left[{\bf A}^3\right]\right]\,,\nonumber\\
C_4({\bf A})&=&{1\over{4!}}\,\left[\left[{\bf A}\right]^4-6\,\left[{\bf A}\right]^2\,\left[{\bf A}^2\right]+8\,\left[{\bf A}\right]\,\left[{\bf A}^3\right]+3\,\left[{\bf A}^2\right]^2-6\,
\left[{\bf A}^4\right]\right]\,,
\label{129}
\end{eqnarray}

\noindent where 

\begin{equation}
[{\bf A}]={\rm Trace}(({\bf A}))={\bf\Delta}\,\cdot\,({\bf A})\,.
\label{130}
\end{equation}

\noindent The Cayley--Hamilton theorem can now be rewritten as

\begin{eqnarray}
({\bf A})-C_1({\bf A})\,{\bf I}&\equiv&0\,,\nonumber\\
({\bf A}^2)-C_1({\bf A})\,({\bf A})+C_2({\bf A})\,{\bf I}&\equiv&0\,,\nonumber\\
({\bf A}^3)-C_1({\bf A})\,({\bf A}^2)+C_2({\bf A})\,({\bf A})-C_3({\bf A})\,{\bf I}&\equiv&0\,,
\nonumber\\
({\bf A}^4)-C_1({\bf A})\,({\bf A}^3)+C_2({\bf A})\,({\bf A}^2)-C_3({\bf A})\,({\bf A})+C_4({\bf A})\,{\bf I}&\equiv&0\,.
\label{131}
\end{eqnarray}

\noindent which is formally exactly the same as (\ref{033}).

Since ${\bf I}$ is a fourth--rank matrix, the multiplication defined above is not a binary operation. In fact, we can define binary, trinary and tetranary products as

\begin{eqnarray}
\left<{\bf A},\,{\bf A}\right>_7&=&{\bf A}\,{\bf I}\,{\bf A}\,,\nonumber\\
\left<{\bf A},\,{\bf A},\,{\bf A}\right>_{36}&=&{\bf A}\,{\bf I}\,{\bf A}\,{\bf I}\,{\bf A}\,,
\nonumber\\
\left<{\bf A},\,{\bf A},\,{\bf A},\,{\bf A}\right>_{217}&=&{\bf A}\,{\bf I}\,{\bf A}\,{\bf I}\,
{\bf A}\,{\bf I}\,{\bf A}\,.
\label{132}
\end{eqnarray}

\noindent The subscript denotes the number of terms involved in each expression. Then, the products defined above are given by

\begin{eqnarray}
({\bf A}^2)&=&\left<{\bf A},\,{\bf A}\right>\,,\nonumber\\
({\bf A}^3)&=&\left<{\bf A},\,{\bf A},\,{\bf A}\right>+\left<{\bf A}^2,\,{\bf A}\right>\,,
\nonumber\\
({\bf A}^4)&=&\left<{\bf A},\,{\bf A},\,{\bf A},\,{\bf A}\right>+\left<{\bf A}^2,\,{\bf A},\,{\bf A}\right>+\left<{\bf A}^3,\,{\bf A}\right>+\left<{\bf A}^2,\,{\bf A}^2\right>\,.
\label{133}
\end{eqnarray}

Let us finally observe that

\begin{equation}
({\bf A}^p)\,\cdot\,({\bf A}^q)\not=({\bf A}^{p+q})\,.
\label{134}
\end{equation}

\section{The Third--Rank Case}

In this section we study the third--rank case as a representative of odd--rank hypermatrices; all other odd--rank cases can be dealt with in a similar form. 

Let us start by considering a third--rank matrix ${\bf A}$ with components $A_{ijk}$ and a third--rank matrix ${\bf\Delta}$ with components $\Delta^{ijk}$ as a third--rank generalization of the unit matrix. For practical purposes the discriminants are better represented by semi--magic squares of rank 3. They are constructed using the graphical algorithm of the section above. 

Let us introduce a third--rank unit matrix $\Delta^{ijk}$. In order to fix the ideas we can consider a unit matrix defined by the relations

\begin{equation}
\Delta^{ijk}=\cases{1\,,&if $i=j=k$,\cr0\,,&otherwise.\cr}\,.
\label{401}
\end{equation}

\noindent However, other definitions are possible and they are currently under study \cite{Ta02}. 

Let us therefore proceed to the construction of discriminants for third--rank matrices. For $n=1$ there is only one possibility, namely

\begin{equation}
C_1({\bf A})=(3)\,.
\label{402}
\end{equation}

For $n=2$ we have

\begin{equation}
\begin{array}{@{}l@{\hspace{35pt}}l@{\hspace{35pt}}l@{}}
\put(-28,-17){\grid(28,42)(14,14)}=\left(\matrix{3&0\cr0&3\cr}\right)\,,
&\put(-28,-17){\grid(28,42)(14,14)}
\put(-21,18){\circle{4}}\put(-7,18){\circle{4}}\put(-19,18){\line(1,0){10}}
=-2\,\left(\matrix{2&1\cr1&2\cr}\right)\,,
&\put(-28,-17){\grid(28,42)(14,14)}
\put(-21,18){\circle{4}}\put(-7,18){\circle{4}}\put(-19,18){\line(1,0){10}}
\put(-21,4){\circle{4}}\put(-7,4){\circle{4}}\put(-19,4){\line(1,0){10}}
=\left(\matrix{1&2\cr2&1\cr}\right)\,,\cr
&&
\end{array}
\label{403}
\end{equation}

\noindent The coefficients in (\ref{403}) are obtained from

\begin{equation}
P_2^2=({\bf 0}-{\bf 1})^2={\bf 0}\cdot{\bf 0}-2\cdot{\bf 0}\cdot{\bf 1}+{\bf 1}\cdot{\bf 1}\,.
\label{404}
\end{equation}

\noindent The corresponding discriminant is given by

\begin{equation}
C_2({\bf A})={1\over2}\,\left[\left(\matrix{3&0\cr0&3\cr}\right)-\left(\matrix{2&1\cr1&2\cr}
\right)\right]\,.
\label{405}
\end{equation}

For $n=3$ the corresponding grids are given by

\begin{equation}
\begin{array}{@{}l@{\hspace{49pt}}l@{}}
{\put(-42,-17){\grid(42,42)(14,14)}}=\left(\matrix{3&0&0\cr0&3&0\cr0&0&3\cr}\right)\,,
&\put(-42,-17){\grid(42,42)(14,14)}
\put(-21,18){\circle{4}}\put(-7,18){\circle{4}}\put(-19,18){\line(1,0){10}}
=-6\,\left(\matrix{3&0&0\cr0&2&1\cr0&1&2\cr}\right)\,,\cr
&\cr
\put(-42,-17){\grid(42,42)(14,14)}
\put(-21,18){\circle{4}}\put(-7,18){\circle{4}}\put(-19,18){\line(1,0){10}}
\put(-21,4){\circle{4}}\put(-7,4){\circle{4}}\put(-19,4){\line(1,0){10}}
=3\,\left(\matrix{3&0&0\cr0&1&2\cr0&2&1\cr}\right)\,,
&\put(-42,-17){\grid(42,42)(14,14)}
\put(-21,18){\circle{4}}\put(-7,18){\circle{4}}\put(-19,18){\line(1,0){10}}
\put(-35,4){\circle{4}}\put(-21,4){\circle{4}}\put(-33,4){\line(1,0){10}}
=6\,\left(\matrix{2&1&0\cr1&1&1\cr0&1&2\cr}\right)\,,\cr
&\cr
\put(-42,-17){\grid(42,42)(14,14)}
\put(-35,18){\circle{4}}\put(-21,18){\circle{4}}\put(-7,18){\circle{4}}
\put(-33,18){\vector(1,0){10}}\put(-19,18){\vector(1,0){10}}
=4\,\left(\matrix{2&1&0\cr0&2&1\cr1&0&2\cr}\right)\,,
&{\put(-42,-17){\grid(42,42)(14,14)}
\put(-35,18){\circle{4}}\put(-21,18){\circle{4}}\put(-7,18){\circle{4}}
\put(-33,18){\vector(1,0){10}}\put(-19,18){\vector(1,0){10}}
\put(-21,4){\circle{4}}\put(-7,4){\circle{4}}\put(-19,4){\line(1,0){10}}
=-12\,\left(\matrix{2&1&0\cr0&1&2\cr1&1&1\cr}\right)}\,,\cr
&\cr
{\put(-42,-17){\grid(42,42)(14,14)}
\put(-35,18){\circle{4}}\put(-21,18){\circle{4}}\put(-7,18){\circle{4}}
\put(-33,18){\vector(1,0){10}}\put(-19,18){\vector(1,0){10}}
\put(-35,4){\circle{4}}\put(-21,4){\circle{4}}\put(-7,4){\circle{4}}
\put(-33,4){\vector(1,0){10}}\put(-19,4){\vector(1,0){10}}
=2\,\left(\matrix{1&2&0\cr0&1&2\cr2&0&1\cr}\right)}\,,
&{\put(-42,-17){\grid(42,42)(14,14)}
\put(-35,18){\circle{4}}\put(-21,18){\circle{4}}\put(-7,18){\circle{4}}
\put(-33,18){\vector(1,0){10}}\put(-19,18){\vector(1,0){10}}
\put(-35,4){\circle{4}}\put(-21,4){\circle{4}}\put(-7,4){\circle{4}}
\put(-23,4){\vector(-1,0){10}}\put(-9,4){\vector(-1,0){10}}
=2\,\left(\matrix{1&1&1\cr1&1&1\cr1&1&1\cr}\right)}\,.\cr
&
\end{array}
\label{406}
\end{equation}

\noindent The coefficients are obtained from

\begin{eqnarray}
P_3^2&=&({\bf 0}-3\cdot{\bf 1}+2\cdot{\bf 2})^2\nonumber\\
&=&{\bf 0}\cdot{\bf 0}-6\cdot{\bf 0}\cdot{\bf 1}+9\cdot{\bf 1}\cdot{\bf 1}+4\cdot{\bf 0}\cdot{
\bf 2}-12\cdot{\bf 1}\cdot{\bf 2}+4\cdot{\bf 2}\cdot{\bf 2}\,.
\label{407}
\end{eqnarray}

\noindent Then, the discriminant is given by

\begin{eqnarray}
C_3({\bf A})&=&{1\over6}\,\left[\left(\matrix{3&0&0\cr0&3&0\cr0&0&3\cr}\right)
-3\,\left(\matrix{3&0&0\cr0&2&1\cr0&1&2\cr}\right)
+6\,\left(\matrix{2&1&0\cr0&2&1\cr1&0&2\cr}\right)\right.\nonumber\\
&&\quad\left.-6\,\left(\matrix{2&1&0\cr1&1&1\cr0&1&2\cr}\right)
+2\,\left(\matrix{1&1&1\cr1&1&1\cr1&1&1\cr}\right)\right]\,.
\label{408}
\end{eqnarray}

For $n=4$ the grids are given in the Appendix C. The coefficients are obtained from

\begin{eqnarray}
P_4^2&=&({\bf 0}-6\cdot{\bf 1}+3\cdot{\bf 1}^2+8\cdot{\bf 2}-6\cdot{\bf 3})^2\nonumber\\
&=&{\bf 0}\cdot{\bf 0}
+\cdot{\bf 0}\cdot{\bf 1}
+\cdot{\bf 1}\cdot{\bf 1}
+\cdot{\bf 0}\cdot{\bf 1}^2
+\cdot{\bf 0}\cdot{\bf 2}\nonumber\\
&&+\cdot{\bf 1}\cdot{\bf 1}^2
+\cdot{\bf 1}\cdot{\bf 2}
+\cdot{\bf 0}\cdot{\bf 3}
+\cdot{\bf 1}\cdot{\bf 3}
+\cdot{\bf 1}^2\cdot{\bf 1}^2\nonumber\\
&&+\cdot{\bf 1}^2\cdot{\bf 2}
+\cdot{\bf 2}\cdot{\bf 2}
+\cdot{\bf 1}^2\cdot{\bf 3}
+\cdot{\bf 2}\cdot{\bf 3}
+\cdot{\bf 3}\cdot{\bf 3}\,.
\label{409}
\end{eqnarray}

\noindent The discriminant is then given by

\begin{eqnarray}
C_4({\bf a})&=&{1\over{24}}\,\left[
\left(\matrix{3&0&0&0\cr0&3&0&0\cr0&0&3&0\cr0&0&0&3\cr}\right)
-6\,\left(\matrix{3&0&0&0\cr0&3&0&0\cr0&0&2&1\cr0&0&1&2\cr}\right)
+24\,\left(\matrix{3&0&0&0\cr0&2&1&0\cr0&0&2&1\cr0&1&0&2\cr}\right)\right.\,,\nonumber\\
&&\quad-24\,\left(\matrix{3&0&0&0\cr0&2&1&0\cr0&1&1&1\cr0&0&1&2\cr}\right)
+8\,\left(\matrix{3&0&0&0\cr0&1&1&1\cr0&1&1&1\cr0&1&1&1\cr}\right)
-6\,\left(\matrix{2&1&0&0\cr0&2&1&0\cr0&0&2&1\cr1&0&0&2\cr}\right)\nonumber\\
&&\quad+3\,\left(\matrix{2&1&0&0\cr1&2&0&0\cr0&0&2&1\cr0&0&1&2\cr}\right)
-48\,\left(\matrix{2&0&0&1\cr0&2&0&1\cr1&0&2&0\cr0&1&1&1\cr}\right)
+24\,\left(\matrix{2&0&0&1\cr0&2&0&1\cr0&0&2&1\cr1&1&1&0\cr}\right)\nonumber\\
&&\quad\left.+36\,\left(\matrix{2&0&1&0\cr0&2&0&1\cr1&0&1&1\cr0&1&1&1\cr}\right)
-24\,\left(\matrix{2&1&0&0\cr1&0&1&1\cr0&1&1&1\cr0&1&1&1\cr}\right)
+12\,\left(\matrix{1&1&1&0\cr0&1&1&1\cr1&0&1&1\cr1&1&0&1\cr}\right)\right]\,.
\label{410}
\end{eqnarray}

Let us now consider the previous expressions for $d=2$ for a third--rank hypermatrix ${\bf a}$ with components $a_{ijk}$. Then, the relevant discriminants are

\begin{eqnarray}
C_1({\bf a})&=&a_{000}+a_{111}\,,\nonumber\\
C_2({\bf a})&=&a_{000}\,a_{111}-{1\over3}\,\left(a_{001}\,a_{110}+a_{010}\,a_{101}+a_{100}\,
a_{011}\right)\,.
\label{411}
\end{eqnarray}

\noindent According with (\ref{040}) the components of the inverse hypermatrix are given by

\begin{equation}
a^{ijk}={1\over{C_2({\bf a})}}\,{{\partial C_2({\bf a})}\over{\partial a_{ijk}}}\,.
\label{412}
\end{equation}

\noindent In detail we obtain

\begin{eqnarray}
a^{000}&=&{1\over{C_2({\bf a})}}\,a_{111}\,,\nonumber\\
a^{001}&=&-{1\over{3C_2({\bf a})}}\,a_{110}\,,\nonumber\\
a^{010}&=&-{1\over{3C_2({\bf a})}}\,a_{101}\,,\nonumber\\
a^{100}&=&-{1\over{3C_2({\bf a})}}\,a_{011}\,,
\label{413}
\end{eqnarray}

\noindent and similar relations for the other components. Then, we obtain

\begin{eqnarray}
a_{0ij}\,a^{0ij}&=&1\,,\nonumber\\
a_{0ij}\,a^{1ij}&\not=&0\,,
\label{414}
\end{eqnarray}

\noindent and similar relations for the other components. Therefore

\begin{equation}
a_{ikl}\,a^{jkl}\not=\delta^j_i\,.
\label{415}
\end{equation}

\noindent Therefore, even when $C_2({\bf a})$ is an invariant for a third--rank hypermatrix it does not have one of the essential properties of a determinant. 

In order to understand this result let us consider the relations similar to (\ref{040}). We define

\begin{equation}
Q_s^{i_1j_1k_1\cdots i_s j_s k_s}={1\over{s!}}\,\Delta^{|[i_1j_1k_1}\,\cdots\,\Delta^{i_s j_s k_s]|}\,.
\label{416}
\end{equation}

\noindent For the first values of $s$ we obtain

\begin{eqnarray}
Q_1^{ijk}&=&\Delta^{ijk}\,,\nonumber\\
Q_2^{i_1j_1k_1i_2j_2k_2}&=&{1\over2}\,[\Delta^{i_1j_1k_1}\,\Delta^{i_2j_2k_2}\nonumber\\
&&-(\Delta^{i_1j_1k_1}\,\Delta^{i_2j_2k_2}+\Delta^{i_1j_1k_2}\,\Delta^{i_2j_2k_1}\nonumber\\
&&+\Delta^{i_1j_2k_1}\,\Delta^{i_2j_1k_2}+\Delta^{i_2j_1k_1}\,\Delta^{i_1j_2k_2})\nonumber\\
&&+(\Delta^{i_1j_1k_2}\,\Delta^{i_2j_2k_1}+\Delta^{i_1j_2k_1}\,\Delta^{i_2j_1k_2}+\Delta^{i_2j_1
k_1}\,\Delta^{i_1j_2k_2})]\,,
\label{417}
\end{eqnarray}

\noindent etc. Then, the discriminants are defined by

\begin{equation}
C_s({\bf A})=Q_s^{i_1j_1k_1\cdots i_s j_s k_s}\,A_{i_1j_1k_1}\,\cdots\,A_{i_s j_s k_s}\,.
\label{418}
\end{equation}

Let us now consider a mixed second--rank matrix ${\bf U}$ with components ${U^i}_j$ and its inverse ${\bf U}^{-1}$ with components ${(U^{-1})^i}_j$. Then we define a similarity transformation for fourth--rank covariant and contravariant matrices as

\begin{eqnarray}
{\bf A}'&=&[{\bf A},\,{\bf U},\,{\bf U},{\bf U},\,{\bf U}]\,,\nonumber\\
{\bf B}'&=&[{\bf B},\,{\bf U}^{-1},\,{\bf U}^{-1},{\bf U}^{-1},\,{\bf U}^{-1}]\,,
\label{419}
\end{eqnarray}

\noindent defined by the components

\begin{eqnarray}
(A')_{i_1i_2i_3}&=&A_{j_1j_2j_3}\,{U^{j_1}}_{i_1}\,{U^{j_2}}_{i_2}\,{U^{j_3}}_{i_3}\,,
\nonumber\\
(B')^{i_1i_2i_3}&=&B^{j_1j_2j_3}\,{(U^{-1})^{i_1}}_{j_1}\,{(U^{-1})^{i_2}}_{j_2}\,{(U^{-1})^{i
_3}}_{j_3}\,.
\label{420}
\end{eqnarray}

\noindent Then, the discriminants are invariants under this kind of similarity transformations.

Let us observe that a relation similar to (\ref{043}) does not holds for third--rank symbols, namely,

\begin{equation}
Q_d^{i_1\cdots i_d j_1\cdots j_d k_1\cdots k_d}
={1\over{d!}}\,\Delta^{|[i_1j_1k_1}\,\cdots\,\Delta^{i_d j_d k_d]|}
\not={1\over{d!}}\,\epsilon^{i_1\cdots i_d}\,\cdots\,\epsilon^{k_1\cdots k_d}\,.
\label{421}
\end{equation}

\noindent The determinant for a higher--rank matrix could be defined in complete analogy with the definition for ordinary matrices. Therefore, $C_d({\bf A})\not=\det({\bf A})$. In fact

\begin{equation}
C_d({\bf A})=Q_d^{i_1\cdots i_d j_1\cdots j_d k_1\cdots k_d}\,\,A_{i_1j_1k_1}\,\cdots\,
A_{i_d j_d k_d}
={1\over{d!}}\,\Delta^{|[i_1j_1k_1}\,\cdots\,\Delta^{i_d j_d k_d]|}\,\,A_{i_1j_1k_1}\,\cdots\,
A_{i_d j_d k_d}\,.
\label{422}
\end{equation}

\noindent However

\begin{equation}
A=\det({\bf A})={1\over{d!}}\,\epsilon^{i_1\cdots i_d}\,\cdots\,\epsilon^{k_1\cdots k_d}\,A_{i
_1j_1k_1}\,\cdots\,A_{i_d j_d k_d}\equiv0\,.
\label{423}
\end{equation}

\noindent This result holds for any odd--rank hypermatrix in odd dimension. In fact

\begin{equation}
\det({\bf A})=\underbrace{\epsilon^{i_{11}\cdots i_{1d}}\,\cdots\,\epsilon^{i_{r1}\cdots i_{rd}}
}_{r\,{\rm times}}\,\underbrace{A_{i_{11}\cdots i_{r1}}\,\cdots\,A_{i_{1d}\cdots i_{rd}}}_{d\,{
\rm times}}\equiv0\,.
\label{424}
\end{equation}

\noindent This result is due to the odd number of $\epsilon$'s which add all contributions to zero. This result can be easily verified for the simple case at hand, $r=3$, $d=2$. We have

\begin{equation}
\epsilon^{i_1i_2}\,\epsilon^{j_1j_2}\,\epsilon^{k_1k_2}\,A_{i_1j_1k_1}\,A_{i_2j_2k_2}\equiv0\,.
\label{425}
\end{equation}

In order to obtain some indication as to the correct way to define a determinant for odd--rank matrices, let us consider the simple case of a completely symmetric third--rank matrix ${\bf a}$ with components $a_{ijk}$ in dimension $d$. Then, let us look for a solution $a^{ijk}$ to the equation

\begin{equation}
a^{ik_1k_2}\,a_{jk_1k_2}=\delta^i_j\,,
\label{151}
\end{equation}

\noindent which defines an inverse matrix in a way similar to (\ref{095}). The number of unknowns is $d(d+1)(d+2)/6$ while the number of equations (\ref{118}) is $d^2$. This algebraic system of equations is underdetermined, except for $d=2$. In this last case the solution is given by

\begin{eqnarray}
a^{000}&=&{1\over a}\,\left(a_{000}\,a_{111}^2+2\,a_{011}^3-3\,a_{001}\,a_{011}\,a_{111}
\right)\,,\nonumber\\
a^{001}&=&{1\over a}\,\left(2\,a_{111}\,a_{001}^2-a_{000}\,a_{011}\,a_{111}-a_{001}\,a_{011}^2
\right)\,,
\label{152}
\end{eqnarray}

\noindent and similar relations for $a^{111}$ and $a^{011}$, where

\begin{equation}
a={1\over{18}}\,\left[a_{000}^2\,a_{111}^2-6\,a_{000}\,a_{001}\,a_{011}\,a_{111}+4\,a_{000}\,
a_{011}^3+4\,a_{111}\,a_{001}^3-3\,a_{001}^2\,a_{011}^2\right]\,.
\label{153}
\end{equation}

\noindent Let us observe that

\begin{eqnarray}
a^{000}&=&{1\over a}\,{{\partial a}\over{\partial a_{000}}}\,,\nonumber\\
a^{001}&=&{1\over3}\,{1\over a}\,{{\partial a}\over{\partial a_{001}}}\,,
\label{154}
\end{eqnarray}

\noindent or

\begin{equation}
a^{ijk}={1\over a}\,{{\partial a}\over{\partial a_{ijk}}}\,.
\label{155}
\end{equation}

\noindent Therefore, the role of the determinant, in a way similar to (\ref{047}) and (\ref{100}), is played by $a$. 

Let us observe that for any matrix ${\bf A}$ the product of $\epsilon$'s and ${\bf A}$'s have sense only for an expression which is at least quadratic in ${\bf A}$'s and ${\bf A}$'s must be of even rank. On the other hand, $a$ is quartic in ${\bf a}$'s and in order to relate $a$ with some determinant it must be with the determinant of some hypermatrix ${\bf A}$ which is quadratic in ${\bf a}$'s, therefore a sixth--rank matrix. Let us therefore consider a sixthh-rank matrix ${\bf A}$ with components $A_{i_1i_2i_3i_4i_5i_6}$. Since the rank is even we can use an extension of the definition (\ref{092}) to sixth--rank matrices. For $d=2$ we obtain

\begin{equation}
A={1\over2}\,\epsilon^{i_1j_1}\,\cdots\,\epsilon^{i_6j_6}\,A_{i_1\cdots i_6}\,A_{j_1\cdots j_6}
\,.
\label{156}
\end{equation}

Let us therefore construct discriminants for rank 6 hypermatrices. For $n=2$ the corresponding grids are

\begin{equation}
\begin{array}{@{}l@{\hspace{35pt}}l@{\hspace{35pt}}l@{}}
{\put(-28,-42){\grid(28,84)(14,14)}}=\left(\matrix{6&0\cr0&6\cr}\right)\,,
&{\put(-28,-42){\grid(28,84)(14,14)}}
\put(-21,35){\circle{4}}\put(-7,35){\circle{4}}\put(-19,35){\line(1,0){10}}
=-5\,\left(\matrix{5&1\cr1&5\cr}\right)\,,
&{\put(-28,-42){\grid(28,84)(14,14)}}
\put(-21,35){\circle{4}}\put(-7,35){\circle{4}}\put(-19,35){\line(1,0){10}}
\put(-21,21){\circle{4}}\put(-7,21){\circle{4}}\put(-19,21){\line(1,0){10}}
=10\,\left(\matrix{4&2\cr2&4\cr}\right)\,,\cr
&&\cr
{\put(-28,-42){\grid(28,84)(14,14)}}
\put(-21,35){\circle{4}}\put(-7,35){\circle{4}}\put(-19,35){\line(1,0){10}}
\put(-21,21){\circle{4}}\put(-7,21){\circle{4}}\put(-19,21){\line(1,0){10}}
\put(-21,7){\circle{4}}\put(-7,7){\circle{4}}\put(-19,7){\line(1,0){10}}
=-10\,\left(\matrix{3&3\cr3&3\cr}\right)\,,
&{\put(-28,-42){\grid(28,84)(14,14)}}
\put(-21,35){\circle{4}}\put(-7,35){\circle{4}}\put(-19,35){\line(1,0){10}}
\put(-21,21){\circle{4}}\put(-7,21){\circle{4}}\put(-19,21){\line(1,0){10}}
\put(-21,7){\circle{4}}\put(-7,7){\circle{4}}\put(-19,7){\line(1,0){10}}
\put(-21,-7){\circle{4}}\put(-7,-7){\circle{4}}\put(-19,-7){\line(1,0){10}}
=5\,\left(\matrix{2&4\cr4&2\cr}\right)\,,
&{\put(-28,-42){\grid(28,84)(14,14)}}
\put(-21,35){\circle{4}}\put(-7,35){\circle{4}}\put(-19,35){\line(1,0){10}}
\put(-21,21){\circle{4}}\put(-7,21){\circle{4}}\put(-19,21){\line(1,0){10}}
\put(-21,7){\circle{4}}\put(-7,7){\circle{4}}\put(-19,7){\line(1,0){10}}
\put(-21,-7){\circle{4}}\put(-7,-7){\circle{4}}\put(-19,-7){\line(1,0){10}}
\put(-21,-21){\circle{4}}\put(-7,-21){\circle{4}}\put(-19,-21){\line(1,0){10}}
=-\left(\matrix{1&5\cr5&1\cr}\right)\,,\cr
&&
\end{array}
\label{157}
\end{equation}

\noindent The coefficientes are obtained from

\begin{eqnarray}
P_2^5&=&({\bf 0}-{\bf 1})^5\cr
&=&{\bf 0}\cdot{\bf 0}\cdot{\bf 0}\cdot{\bf 0}\cdot{\bf 0}
-5\cdot{\bf 0}\cdot{\bf 0}\cdot{\bf 0}\cdot{\bf 0}\cdot{\bf 1}
+10\cdot{\bf 0}\cdot{\bf 0}\cdot{\bf 0}\cdot{\bf 1}\cdot{\bf 1}\nonumber\\
&&-10\cdot{\bf 0}\cdot{\bf 0}\cdot{\bf 1}\cdot{\bf 1}\cdot{\bf 1}
+5\cdot{\bf 0}\cdot{\bf 1}\cdot{\bf 1}\cdot{\bf 1}\cdot{\bf 1}
-{\bf 1}\cdot{\bf 1}\cdot{\bf 1}\cdot{\bf 1}\cdot{\bf 1}\,.
\label{158}
\end{eqnarray}

\noindent Then, the discriminant is given by

\begin{equation}
C_2({\bf A})={1\over2}\,\left[\left(\matrix{6&0\cr0&6\cr}\right)-6\,\left(\matrix{5&1\cr1&5\cr}
\right)+15\,\left(\matrix{4&2\cr2&4\cr}\right)-10\,\left(\matrix{3&3\cr3&3\cr}\right)\right]\,.
\label{159}
\end{equation}

For $n=3$ the grids and semi--magic squares are given in Appendix D. The coefficients are obtained from

\begin{eqnarray}
P_3^5&=&({\bf 0}-3\cdot{\bf 1}+2\cdot{\bf 2})^5\cr
&=&{\bf 0}\cdot{\bf 0}\cdot{\bf 0}\cdot{\bf 0}\cdot{\bf 0}
-15\cdot{\bf 0}\cdot{\bf 0}\cdot{\bf 0}\cdot{\bf 0}\cdot{\bf 1}
+90\cdot{\bf 0}\cdot{\bf 0}\cdot{\bf 0}\cdot{\bf 1}\cdot{\bf 1}\nonumber\\
&&+10\cdot{\bf 0}\cdot{\bf 0}\cdot{\bf 0}\cdot{\bf 0}\cdot{\bf 2}
-270\cdot{\bf 0}\cdot{\bf 0}\cdot{\bf 1}\cdot{\bf 1}\cdot{\bf 1}
-120\cdot{\bf 0}\cdot{\bf 0}\cdot{\bf 0}\cdot{\bf 1}\cdot{\bf 2}\nonumber\\
&&+405\cdot{\bf 0}\cdot{\bf 1}\cdot{\bf 1}\cdot{\bf 1}\cdot{\bf 1}
+540\cdot{\bf 0}\cdot{\bf 0}\cdot{\bf 1}\cdot{\bf 1}\cdot{\bf 2}
+40\cdot{\bf 0}\cdot{\bf 0}\cdot{\bf 0}\cdot{\bf 2}\cdot{\bf 2}\nonumber\\
&&-243\cdot{\bf 1}\cdot{\bf 1}\cdot{\bf 1}\cdot{\bf 1}\cdot{\bf 1}
-1080\cdot{\bf 0}\cdot{\bf 1}\cdot{\bf 1}\cdot{\bf 1}\cdot{\bf 2}
-360\cdot{\bf 0}\cdot{\bf 0}\cdot{\bf 1}\cdot{\bf 2}\cdot{\bf 2}\nonumber\\
&&+810\cdot{\bf 1}\cdot{\bf 1}\cdot{\bf 1}\cdot{\bf 1}\cdot{\bf 2}
+1080\cdot{\bf 0}\cdot{\bf 1}\cdot{\bf 1}\cdot{\bf 2}\cdot{\bf 2}
+80\cdot{\bf 0}\cdot{\bf 0}\cdot{\bf 2}\cdot{\bf 2}\cdot{\bf 2}\nonumber\\
&&-1080\cdot{\bf 1}\cdot{\bf 1}\cdot{\bf 1}\cdot{\bf 2}\cdot{\bf 2}
-480\cdot{\bf 0}\cdot{\bf 1}\cdot{\bf 2}\cdot{\bf 2}\cdot{\bf 2}
+720\cdot{\bf 1}\cdot{\bf 1}\cdot{\bf 2}\cdot{\bf 2}\cdot{\bf 2}\nonumber\\
&&+80\cdot{\bf 0}\cdot{\bf 2}\cdot{\bf 2}\cdot{\bf 2}\cdot{\bf 2}
-240\cdot{\bf 1}\cdot{\bf 2}\cdot{\bf 2}\cdot{\bf 2}\cdot{\bf 2}
+32\cdot{\bf 2}\cdot{\bf 2}\cdot{\bf 2}\cdot{\bf 2}\cdot{\bf 2}\,.
\label{160}
\end{eqnarray}

\noindent The discriminant is given by

\begin{eqnarray}
C_3({\bf A})&=&{1\over6}\,\left[\left(\matrix{6&0&0\cr0&6&0\cr0&0&6\cr}\right)
-18\,\left(\matrix{6&0&0\cr0&5&1\cr0&1&5\cr}\right)
+45\,\left(\matrix{6&0&0\cr0&4&2\cr0&2&4\cr}\right)\right.\nonumber\\
&&\quad\quad-30\,\left(\matrix{6&0&0\cr0&3&3\cr0&3&3\cr}\right)
+12\,\left(\matrix{5&1&0\cr0&5&1\cr1&0&5\cr}\right)
+90\,\left(\matrix{5&0&1\cr0&5&1\cr1&1&4\cr}\right)\nonumber\\
&&\quad\quad-180\,\left(\matrix{5&0&1\cr0&2&4\cr1&4&1\cr}\right)
-360\,\left(\matrix{5&0&1\cr0&4&2\cr1&2&3\cr}\right)
+360\,\left(\matrix{5&0&1\cr0&3&3\cr1&3&2\cr}\right)\nonumber\\
&&\quad\quad+30\,\left(\matrix{4&2&0\cr0&4&2\cr2&0&4\cr}\right)
-90\,\left(\matrix{4&1&1\cr1&4&1\cr1&1&4\cr}\right)
+720\,\left(\matrix{1&4&1\cr4&0&2\cr1&2&3\cr}\right)\nonumber\\
&&\quad\quad+270\,\left(\matrix{4&2&0\cr2&2&2\cr0&2&4\cr}\right)
-360\,\left(\matrix{4&0&2\cr0&3&3\cr2&3&1\cr}\right)
+180\,\left(\matrix{4&1&1\cr1&3&2\cr1&2&3\cr}\right)\nonumber\\
&&\quad\quad-540\,\left(\matrix{4&2&0\cr1&2&3\cr1&2&3\cr}\right)
-540\,\left(\matrix{4&1&1\cr2&2&2\cr0&3&3\cr}\right)
+20\,\left(\matrix{3&3&0\cr0&3&3\cr3&0&3\cr}\right)\nonumber\\
&&\quad\quad+540\,\left(\matrix{0&3&3\cr3&2&1\cr3&1&2\cr}\right)
-600\,\left(\matrix{3&2&1\cr1&3&2\cr2&1&3\cr}\right)
+540\,\left(\matrix{3&1&2\cr1&3&2\cr2&2&2\cr}\right)\nonumber\\
&&\quad\quad\left.-90\,\left(\matrix{2&2&2\cr2&2&2\cr2&2&2\cr}\right)\right]\,.
\label{161}
\end{eqnarray}

Let us now return to the two--dimensional case. Then we have

\begin{eqnarray}
A&=&{1\over2}\,\left[A_{000000}\,A_{111111}-(A_{000001}\,A_{111110}+A_{000010}\,A_{111101}
+A_{000100}\,A_{111011}\right.\nonumber\\
&&+A_{001000}\,A_{110111}+A_{010000}\,A_{101111}+A_{100000}\,A_{011111})+(A_{000011}\,A_{111100}\nonumber\\
&&+A_{000101}\,A_{111010}+A_{001001}\,A_{110110}+A_{010001}\,A_{101110}+A_{100001}\,A_{011110}
\nonumber\\
&&+A_{000110}\,A_{111001}+A_{001010}\,A_{110101}+A_{010010}\,A_{101101}+A_{100010}\,A_{011101}
\nonumber\\
&&+A_{001100}\,A_{110011}+A_{010100}\,A_{101011}+A_{100100}\,A_{011011}+A_{011000}\,A_{100111}
\nonumber\\
&&+A_{101000}\,A_{010111}+A_{110000}\,A_{001111})-(A_{000111}\,A_{111000}+A_{001011}\,A_{110100}
\nonumber\\
&&+A_{010011}\,A_{101100}+A_{100011}\,A_{011100}+A_{001101}\,A_{110010}+A_{010101}\,A_{101010}
\nonumber\\
&&+\left.A_{100101}\,A_{011010}+A_{011001}\,A_{100110}+A_{101001}\,A_{010110}+A_{110001}\,A_{001110})\right]\,.
\label{162}
\end{eqnarray}

\noindent Let us consider a third--rank matrix ${\bf a}$ with components $a_{ijk}$. Then, we define the sixth--rank matrix

\begin{eqnarray}
A_{i_1j_1k_1i_2j_2k_2}&=&a_{i_1j_1k_1}\,\otimes\,a_{i_2j_2k_2}\nonumber\\
&=&{1\over4}\,(a_{i_1j_1k_1}\,a_{i_2j_2k_2}+a_{i_1j_1k_2}\,a_{i_2j_2k_1}+a_{i_1j_2k_1}\,
a_{i_2j_1k_2}+a_{i_2j_1k_1}\,a_{i_1j_2k_2})\,.
\label{163}
\end{eqnarray}

\noindent For $d=2$, ${\bf A}$ has 64 components. We exhibit 32 of them; the other 32 components are obtained by interchange of $0$'s and $1$'s. We obtain

\begin{eqnarray}
A_{000000}&=&a_{000}^2\,,\nonumber\\
A_{000001}&=&A_{000010}=A_{000100}=A_{001000}=A_{010000}=A_{100000}=a_{000}\,a_{001}\,,
\nonumber\\
A_{000011}&=&A_{010001}=A_{001010}=A_{011000}={1\over2}\,(a_{000}\,a_{011}+a_{001}\,a_{010})\,,
\nonumber\\
A_{000101}&=&A_{100001}=A_{001100}=A_{101000}={1\over2}\,(a_{000}\,a_{101}+a_{001}\,a_{100})\,,
\nonumber\\
A_{000110}&=&A_{100010}=A_{010100}=A_{110000}={1\over2}\,(a_{000}\,a_{110}+a_{010}\,a_{100})\,,
\nonumber\\
A_{001001}&=&a_{001}^2\,,\nonumber\\
A_{010010}&=&a_{010}^2\,,\nonumber\\
A_{100100}&=&a_{100}^2\,,\nonumber\\
A_{000111}&=&A_{100011}=A_{010101}=A_{011100}=A_{110001}\nonumber\\
&=&{1\over4}\,(a_{000}\,a_{111}+a_{001}\,a_{110}+a_{010}\,a_{101}+a_{100}\,a_{011})\,,
\nonumber\\
A_{001011}&=&A_{011001}=a_{001}\,a_{011}\,,\nonumber\\
A_{001101}&=&A_{101001}=a_{001}\,a_{101}\,,\nonumber\\
A_{001110}&=&A_{110001}=a_{001}\,a_{110}\,.
\label{164}
\end{eqnarray}

\noindent The determinant, using (\ref{126}), is given by

\begin{equation}
A={3\over4}\,C\,,
\label{165}
\end{equation}

\noindent where

\begin{eqnarray}
C&=&(a_{000}^2\,a_{111}^2+a_{001}^2\,a_{110}^2+a_{010}^2\,a_{101}^2+a_{100}^2\,a_{011}^2)
\nonumber\\
&&-2\,[a_{000}\,a_{111}\,(a_{001}\,a_{110}+a_{010}\,a_{101}+a_{100}\,a_{011})\nonumber\\
&&+a_{001}\,a_{010}\,a_{101}\,a_{110}+a_{001}\,a_{011}\,a_{110}\,a_{100}+a_{010}\,a_{011}\,a_{101}\,a_{100}]\nonumber\\
&&+4\,(a_{000}\,a_{011}\,a_{101}\,a_{110}+a_{001}\,a_{010}\,a_{100}\,a_{111})\,,
\label{166}
\end{eqnarray}

\noindent is the determinant introduced by Cayley \cite{Ca}. 

Let su mention that there is a further algorithm to construct the Cayley hyperdeterminant. Let us consider the symmetric matrix

\begin{equation}
g_{ij}=a_{ik_1k_2}\,a_{jl_1l_2}\,\epsilon^{k_1l_1}\,\epsilon^{k_2l_2}\,.
\label{167}
\end{equation}

\noindent Then we have

\begin{eqnarray}
g_{00}&=&2\,(a_{000}\,a_{011}-a_{001}\,a_{010})\,,\nonumber\\
g_{01}&=&a_{000}\,a_{111}-a_{001}\,a_{110}-a_{010}\,a_{101}+a_{011}\,a_{100}\,,\nonumber\\
g_{11}&=&2\,(a_{111}\,a_{100}-a_{110}\,a_{101})\,.
\label{168}
\end{eqnarray}

\noindent The determinant of this matrix is given by

\begin{eqnarray}
g&=&\det({\bf g})=g_{00}\,g_{11}\,-g_{01}^2\nonumber\\
&=&4\,(a_{000}\,a_{011}-a_{001}\,a_{010})\,(a_{111}\,a_{100}-a_{110}\,a_{101})\nonumber\\
&&-(a_{000}\,a_{111}-a_{001}\,a_{110}-a_{010}\,a_{101}+a_{011}\,a_{100})^2=-C\,.
\label{169}
\end{eqnarray}

\noindent This algorithm appeared in \cite{Cr}.

\section{Concluding Remarks}

We have developed an algorithm to construct algebraic invariants for hypermatrices. We constructed hyperdeterminants and exhibit an extension of the Cayley--Hamilton theorem to hypermatrices.

These algebraic invariants were considered by Cayley \cite{Ca}; see \cite{GKZ1,GKZ2} for an updated account. 

Higher--rank tensors look similar to hypermatrices and the results presented here are a first step for the construction of algebraic invariants for higher--rank tensors. Higher--rank tensors appear in several contexts such as in Finsler geometry \cite{As,Ru}, fourth--rank gravity \cite{Ta93,TRMC,TR,TU}, dual models for higher spin gauge fields \cite{Hu1,Hu2,Hu3}. 

\section*{Acknowledgements}

This work was partially done at the {\bf Abdus Salam} International Centre for Theoretical Physics, Trieste. This work was partially supported by Direcci\'on de Investigaci\'on -- Bogot\'a, Universidad Nacional de Colombia. The author thanks D. G. Glynn for several useful comments and criticisms. Several calculations were kindly and patiently checked by Mrs. Viviana Dionicio.

\newpage
\appendix

\section{Partitions and Discriminants}

A partition is a way of writing an integer $n$ as a sumof positive integers where the order of the addends is not significant. The partitions of a number $n$ correspond to the set of solutions $(m_1,\cdots,m_n)$ to the Diophantine equation

\begin{equation}
1\,m_1+2\,m_2+\cdots+n\,m_n=n\,.
\label{viv1}
\end{equation}

\noindent The number of solutions to eq. (\ref{viv1}) is given by $p(n)$. This number is given by the generating function \cite{Gu1,Gu2}

\begin{equation}
\sum_{n=0}^\infty\,p(n)\,x^n=\prod_{n=1}^\infty\,{1\over{(1-x^n)}}\,.
\label{viv2}
\end{equation}

\noindent For the first values of $n$ $p(n)$ is given by

\begin{equation}
p(n)=\{1,\,1,\,2,\,3,\,5,\,7,\,11,\,15,\,22,\,30,\,\cdots\}\,.
\label{viv3}
\end{equation}

\noindent For the first values of $n$ the solutions to (\ref{viv1}) are

\begin{eqnarray}
M_1&=&\{(1)\}\,,\nonumber\\
M_2&=&\{(2,\,0),\,(0,\,1)\}\,,\nonumber\\
M_3&=&\{(3,\,0,\,0),(1,\,1,\,0),(0,\,0,\,1)\}\,,\nonumber\\
M_4&=&\{(4,\,0,\,0,\,0),\,(2,\,1,\,0,\,0),\,(0,\,2,\,0,\,0),\,(1,\,0,\,1,\,0),\,(0,\,0,\,0,\,1)
\}\,,\nonumber\\
M_5&=&\{(5,\,0,\,0,\,0,\,0),\,(3,\,1,\,0,\,0,\,0),\,(2,\,0,\,1,\,0,\,0),\,(1,\,2,\,0,\,0,\,0)
\nonumber\\
&&\,(1,\,0,\,0,\,1,\,0),\,(0,\,1,\,1,\,0,\,0),\,(0,\,0,\,0,\,0,\,1)\}\,,\nonumber\\
M_6&=&\{(6,\,0,\,0,\,0,\,0,\,0),\,(4,\,1,\,0,\,0,\,0,\,0),\,(3,\,0,\,1,\,0,\,0,\,0),\,(2,\,2,\,0,\,0,\,0,\,0),\nonumber\\
&&\,(2,\,0,\,0,\,1,\,0,\,0),\,(1,\,1,\,1,\,0,\,0,\,0),\,(1,\,0,\,0,\,0,\,1,\,0),\,(0,\,3,\,0,\,0,\,0,\,0),\nonumber\\
&&\,(0,\,0,\,2,\,0,\,0,\,0),\,(0,\,1,\,0,\,1,\,0,\,0),\,(0,\,0,\,0,\,0,\,0,\,1)\}\,.
\label{viv4}
\end{eqnarray}

Using this result we can now construct discriminants by

\begin{equation}
c_n({\bf a})=\sum_{k=1}^{p(n)}\,\prod_{j=1}^n\,{{(-1)^{(j-1)\cdot m_{kj}}}\over{j^{m_{kj}}m_{kj}
!}}\,\left<{\bf a}^j\right>^{m_{kj}}\,.
\label{viv5}
\end{equation}

Permutations are obtained in a similar way

\begin{equation}
P_n=n!\,\sum_{k=1}^{p(n)}\,\prod_{j=1}^n\,{{(-1)^{(j-1)\cdot m_{kj}}}\over{j^{m_{kj}}m_{kj}!}}\,
({\bf j-1})^{m_{kj}}\,.
\label{viv6}
\end{equation}

\section{Grids for $r=4$, $n=3$}

\begin{equation}
\begin{array}{@{}l@{\hspace{49pt}}l@{}}
{\put(-42,-28){\grid(42,56)(14,14)}}=\left(\matrix{4&0&0\cr0&4&0\cr0&0&4\cr}\right)\,,
&\put(-42,-28){\grid(42,56)(14,14)}
\put(-21,21){\circle{4}}\put(-7,21){\circle{4}}\put(-19,21){\line(1,0){10}}
=-9\,\left(\matrix{4&0&0\cr0&3&1\cr0&1&3\cr}\right)\,,\cr
&\cr
\put(-42,-28){\grid(42,56)(14,14)}
\put(-21,21){\circle{4}}\put(-7,21){\circle{4}}\put(-19,21){\line(1,0){10}}
\put(-21,7){\circle{4}}\put(-7,7){\circle{4}}\put(-19,7){\line(1,0){10}}
=9\,\left(\matrix{4&0&0\cr0&2&2\cr0&2&2\cr}\right)\,,
&\put(-42,-28){\grid(42,56)(14,14)}
\put(-21,21){\circle{4}}\put(-7,21){\circle{4}}\put(-19,21){\line(1,0){10}}
\put(-35,7){\circle{4}}\put(-21,7){\circle{4}}\put(-33,7){\line(1,0){10}}
=18\,\left(\matrix{3&1&0\cr1&2&1\cr0&1&3\cr}\right)\,,\cr
&\cr
\put(-42,-28){\grid(42,56)(14,14)}
\put(-35,21){\circle{4}}\put(-21,21){\circle{4}}\put(-7,21){\circle{4}}
\put(-33,21){\vector(1,0){10}}\put(-19,21){\vector(1,0){10}}
=6\,\left(\matrix{3&1&0\cr0&3&1\cr1&0&3\cr}\right)\,,
&\put(-42,-28){\grid(42,56)(14,14)}
\put(-21,21){\circle{4}}\put(-7,21){\circle{4}}\put(-19,21){\line(1,0){10}}
\put(-21,7){\circle{4}}\put(-7,7){\circle{4}}\put(-19,7){\line(1,0){10}}
\put(-21,-7){\circle{4}}\put(-7,-7){\circle{4}}\put(-19,-7){\line(1,0){10}}
=-3\,\left(\matrix{4&0&0\cr0&1&3\cr0&3&1\cr}\right)\,,\cr
&\cr
{\put(-42,-28){\grid(42,56)(14,14)}
\put(-21,21){\circle{4}}\put(-7,21){\circle{4}}\put(-19,21){\line(1,0){10}}
\put(-21,7){\circle{4}}\put(-7,7){\circle{4}}\put(-19,7){\line(1,0){10}}
\put(-35,-7){\circle{4}}\put(-21,-7){\circle{4}}\put(-33,-7){\line(1,0){10}}
=-18\,\left(\matrix{3&1&0\cr1&1&2\cr0&2&2\cr}\right)}\,,
&{\put(-42,-28){\grid(42,56)(14,14)}
\put(-21,21){\circle{4}}\put(-7,21){\circle{4}}\put(-19,21){\line(1,0){10}}
\put(-35,7){\circle{4}}\put(-21,7){\circle{4}}\put(-33,7){\line(1,0){10}}
\put(-35,-7){\circle{4}}\put(-7,-7){\circle{4}}\put(-33,-7){\line(1,0){24}}
=-6\,\left(\matrix{2&1&1\cr1&2&1\cr1&1&2\cr}\right)}\,,\cr
&\cr
\end{array}
\label{A1a}
\end{equation}

\newpage
\section{Grids for $r=4$, $n=4$}

$$
\begin{array}{@{}l@{\hspace{63pt}}l@{}}
{\put(-56,-28){\grid(56,56)(14,14)}}
=\left(\matrix{4&0&0&0\cr0&4&0&0\cr0&0&4&0\cr0&0&0&4\cr}\right)\,,
&\put(-56,-28){\grid(56,56)(14,14)}
\put(-21,21){\circle{4}}\put(-7,21){\circle{4}}\put(-19,21){\line(1,0){10}}
=-18\,\left(\matrix{4&0&0&0\cr0&4&0&0\cr0&0&3&1\cr0&0&1&3\cr}\right)\,,\cr
&\cr
{\put(-56,-28){\grid(56,56)(14,14)}
\put(-21,21){\circle{4}}\put(-7,21){\circle{4}}\put(-19,21){\line(1,0){10}}
\put(-21,7){\circle{4}}\put(-7,7){\circle{4}}\put(-19,7){\line(1,0){10}}}
=18\,\left(\matrix{4&0&0&0\cr0&4&0&0\cr0&0&2&2\cr0&0&2&2\cr}\right)\,,
&{\put(-56,-28){\grid(56,56)(14,14)}
\put(-21,21){\circle{4}}\put(-7,21){\circle{4}}\put(-19,21){\line(1,0){10}}
\put(-35,7){\circle{4}}\put(-21,7){\circle{4}}\put(-33,7){\line(1,0){10}}}
=72\,\left(\matrix{4&0&0&0\cr0&3&1&0\cr0&1&2&1\cr0&0&1&3\cr}\right)\,,\cr
&\cr
{\put(-56,-28){\grid(56,56)(14,14)}
\put(-21,21){\circle{4}}\put(-7,21){\circle{4}}\put(-19,21){\line(1,0){10}}
\put(-49,7){\circle{4}}\put(-35,7){\circle{4}}\put(-47,7){\line(1,0){10}}}
=18\,\left(\matrix{3&1&0&0\cr1&3&0&0\cr0&0&3&1\cr0&0&1&3\cr}\right)\,,
&\put(-56,-28){\grid(56,56)(14,14)}
\put(-21,21){\circle{4}}\put(-7,21){\circle{4}}\put(-19,21){\line(1,0){10}}
\put(-49,21){\circle{4}}\put(-35,21){\circle{4}}\put(-47,21){\line(1,0){10}}
=9\,\left(\matrix{3&1&0&0\cr1&3&0&0\cr0&0&3&1\cr0&0&1&3\cr}\right)\,,\cr
&\cr
\put(-56,-28){\grid(56,56)(14,14)}
\put(-35,21){\circle{4}}\put(-21,21){\circle{4}}\put(-7,21){\circle{4}}
\put(-33,21){\vector(1,0){10}}\put(-19,21){\vector(1,0){10}}
=24\,\left(\matrix{4&0&0&0\cr0&3&1&0\cr0&0&3&1\cr0&1&0&3\cr}\right)\,,
&\put(-56,-28){\grid(56,56)(14,14)}
\put(-21,21){\circle{4}}\put(-7,21){\circle{4}}\put(-19,21){\line(1,0){10}}
\put(-21,7){\circle{4}}\put(-7,7){\circle{4}}\put(-19,7){\line(1,0){10}}
\put(-21,-7){\circle{4}}\put(-7,-7){\circle{4}}\put(-19,-7){\line(1,0){10}}
=-6\,\left(\matrix{4&0&0&0\cr0&4&0&0\cr0&0&1&3\cr0&0&3&1\cr}\right)\,,\cr
&\cr
\end{array}
\label{B1a}
$$

$$
\begin{array}{@{}l@{\hspace{63pt}}l@{}}
\put(-56,-28){\grid(56,56)(14,14)}
\put(-35,21){\circle{4}}\put(-21,21){\circle{4}}\put(-7,21){\circle{4}}
\put(-33,21){\vector(1,0){10}}\put(-19,21){\vector(1,0){10}}
\put(-49,7){\circle{4}}\put(-35,7){\circle{4}}\put(-47,7){\line(1,0){10}}
=-144\,\left(\matrix{3&1&0&0\cr1&2&1&0\cr0&0&3&1\cr0&1&0&3\cr}\right)\,,
&\put(-56,-28){\grid(56,56)(14,14)}
\put(-49,21){\circle{4}}\put(-35,21){\circle{4}}\put(-21,21){\circle{4}}\put(-7,21){\circle{4}}
\put(-47,21){\vector(1,0){10}}\put(-33,21){\vector(1,0){10}}\put(-19,21){\vector(1,0){10}}
=-18\,\left(\matrix{3&1&0&0\cr0&3&1&0\cr0&0&3&1\cr1&0&0&3\cr}\right)\,,\cr
&\cr
\put(-56,-28){\grid(56,56)(14,14)}
\put(-49,21){\circle{4}}\put(-35,21){\circle{4}}\put(-47,21){\line(1,0){10}}
\put(-21,21){\circle{4}}\put(-7,21){\circle{4}}\put(-19,21){\line(1,0){10}}
\put(-21,7){\circle{4}}\put(-7,7){\circle{4}}\put(-19,7){\line(1,0){10}}
\put(-21,-7){\circle{4}}\put(-7,-7){\circle{4}}\put(-19,-7){\line(1,0){10}}
=18\,\left(\matrix{3&1&0&0\cr1&3&0&0\cr0&0&1&3\cr0&0&3&1\cr}\right)\,,
&\put(-56,-28){\grid(56,56)(14,14)}
\put(-49,21){\circle{4}}\put(-35,21){\circle{4}}\put(-47,21){\line(1,0){10}}
\put(-21,21){\circle{4}}\put(-7,21){\circle{4}}\put(-19,21){\line(1,0){10}}
\put(-21,7){\circle{4}}\put(-7,7){\circle{4}}\put(-19,7){\line(1,0){10}}
\put(-35,-7){\circle{4}}\put(-21,-7){\circle{4}}\put(-33,-7){\line(1,0){10}}
=144\,\left(\matrix{3&1&0&0\cr1&2&1&0\cr0&1&1&2\cr0&0&2&2\cr}\right)\,,\cr
&\cr
\end{array}
\label{B1b}
$$

$$
\begin{array}{@{}l@{\hspace{63pt}}l@{}}
\put(-56,-28){\grid(56,56)(14,14)}
\put(-35,21){\circle{4}}\put(-21,21){\circle{4}}\put(-7,21){\circle{4}}
\put(-33,21){\vector(1,0){10}}\put(-19,21){\vector(1,0){10}}
\put(-35,7){\circle{4}}\put(-21,7){\circle{4}}\put(-7,7){\circle{4}}
\put(-33,7){\vector(1,0){10}}\put(-19,7){\vector(1,0){10}}
=24\,\left(\matrix{4&0&0&0\cr0&2&2&0\cr0&0&2&2\cr0&2&0&2\cr}\right)\,,
&\put(-56,-28){\grid(56,56)(14,14)}
\put(-35,21){\circle{4}}\put(-21,21){\circle{4}}\put(-7,21){\circle{4}}
\put(-33,21){\vector(1,0){10}}\put(-19,21){\vector(1,0){10}}
\put(-35,7){\circle{4}}\put(-21,7){\circle{4}}\put(-7,7){\circle{4}}
\put(-23,7){\vector(-1,0){10}}\put(-9,7){\vector(-1,0){10}}
=24\,\left(\matrix{4&0&0&0\cr0&2&1&1\cr0&1&2&1\cr0&1&1&2\cr}\right)\,,\cr
&\cr
\put(-56,-28){\grid(56,56)(14,14)}
\put(-35,21){\circle{4}}\put(-21,21){\circle{4}}\put(-7,21){\circle{4}}
\put(-33,21){\vector(1,0){10}}\put(-19,21){\vector(1,0){10}}
\put(-49,7){\circle{4}}\put(-35,7){\circle{4}}\put(-21,7){\circle{4}}
\put(-47,7){\vector(1,0){10}}\put(-33,7){\vector(1,0){10}}
=72\,\left(\matrix{3&1&0&0\cr0&2&2&0\cr1&0&2&1\cr0&1&0&3\cr}\right)\,,
&\put(-56,-28){\grid(56,56)(14,14)}
\put(-35,21){\circle{4}}\put(-21,21){\circle{4}}\put(-7,21){\circle{4}}
\put(-33,21){\vector(1,0){10}}\put(-19,21){\vector(1,0){10}}
\put(-49,7){\circle{4}}\put(-35,7){\circle{4}}\put(-21,7){\circle{4}}
\put(-37,7){\vector(-1,0){10}}\put(-23,7){\vector(-1,0){10}}
=72\,\left(\matrix{3&0&1&0\cr1&2&1&0\cr0&1&2&1\cr0&1&0&3\cr}\right)\,,\cr
&\cr
\end{array}
\label{B1c}
$$

$$
\begin{array}{@{}l@{\hspace{63pt}}l@{}}
\put(-56,-28){\grid(56,56)(14,14)}
\put(-35,21){\circle{4}}\put(-21,21){\circle{4}}\put(-7,21){\circle{4}}
\put(-33,21){\vector(1,0){10}}\put(-19,21){\vector(1,0){10}}
\put(-35,7){\circle{4}}\put(-21,7){\circle{4}}\put(-7,7){\circle{4}}
\put(-33,7){\vector(1,0){10}}\put(-19,7){\vector(1,0){10}}
\put(-49,-7){\circle{4}}\put(-35,-7){\circle{4}}\put(-47,-7){\line(1,0){10}}
=-72\,\left(\matrix{3&1&0&0\cr1&1&2&0\cr0&0&2&2\cr0&2&0&2\cr}\right)\,,
&\put(-56,-28){\grid(56,56)(14,14)}
\put(-35,21){\circle{4}}\put(-21,21){\circle{4}}\put(-7,21){\circle{4}}
\put(-33,21){\vector(1,0){10}}\put(-19,21){\vector(1,0){10}}
\put(-35,7){\circle{4}}\put(-21,7){\circle{4}}\put(-7,7){\circle{4}}
\put(-23,7){\vector(-1,0){10}}\put(-9,7){\vector(-1,0){10}}
\put(-49,-7){\circle{4}}\put(-35,-7){\circle{4}}\put(-47,-7){\line(1,0){10}}
=-72\,\left(\matrix{3&1&0&0\cr1&1&1&1\cr0&1&2&1\cr0&1&1&2\cr}\right)\,,\cr
&\cr
\put(-56,-28){\grid(56,56)(14,14)}
\put(-35,21){\circle{4}}\put(-21,21){\circle{4}}\put(-7,21){\circle{4}}
\put(-33,21){\vector(1,0){10}}\put(-19,21){\vector(1,0){10}}
\put(-49,7){\circle{4}}\put(-35,7){\circle{4}}\put(-21,7){\circle{4}}
\put(-47,7){\vector(1,0){10}}\put(-33,7){\vector(1,0){10}}
\put(-35,-7){\circle{4}}\put(-21,-7){\circle{4}}\put(-33,-7){\line(1,0){10}}
=-72\,\left(\matrix{3&1&0&0\cr0&1&3&0\cr1&1&1&1\cr0&1&0&3\cr}\right)\,,
&\put(-56,-28){\grid(56,56)(14,14)}
\put(-35,21){\circle{4}}\put(-21,21){\circle{4}}\put(-7,21){\circle{4}}
\put(-33,21){\vector(1,0){10}}\put(-19,21){\vector(1,0){10}}
\put(-49,7){\circle{4}}\put(-35,7){\circle{4}}\put(-21,7){\circle{4}}
\put(-37,7){\vector(-1,0){10}}\put(-23,7){\vector(-1,0){10}}
\put(-35,-7){\circle{4}}\put(-21,-7){\circle{4}}\put(-33,-7){\line(1,0){10}}
=-72\,\left(\matrix{3&0&1&0\cr1&1&2&0\cr0&2&1&1\cr0&1&0&3\cr}\right)\,,\cr
&\cr
\end{array}
\label{B1d}
$$

$$
\begin{array}{@{}l@{\hspace{63pt}}l@{}}
\put(-56,-28){\grid(56,56)(14,14)}
\put(-49,21){\circle{4}}\put(-35,21){\circle{4}}\put(-21,21){\circle{4}}\put(-7,21){\circle{4}}
\put(-47,21){\vector(1,0){10}}\put(-33,21){\vector(1,0){10}}\put(-19,21){\vector(1,0){10}}
\put(-49,9){\circle{4}}\put(-21,9){\circle{4}}\put(-47,9){\line(1,0){24}}
\put(-35,5){\circle{4}}\put(-7,5){\circle{4}}\put(-33,5){\line(1,0){24}}
=-36\,\left(\matrix{2&1&1&0\cr0&2&1&1\cr1&0&2&1\cr1&1&0&2\cr}\right)\,,
&\put(-56,-28){\grid(56,56)(14,14)}
\put(-49,21){\circle{4}}\put(-35,21){\circle{4}}\put(-21,21){\circle{4}}\put(-7,21){\circle{4}}
\put(-47,21){\vector(1,0){10}}\put(-33,21){\vector(1,0){10}}\put(-19,21){\vector(1,0){10}}
\put(-35,7){\circle{4}}\put(-21,7){\circle{4}}\put(-7,7){\circle{4}}
\put(-33,7){\vector(1,0){10}}\put(-19,7){\vector(1,0){10}}
=-144\,\left(\matrix{3&1&0&0\cr0&2&2&0\cr0&0&2&2\cr1&1&0&2\cr}\right)\,,\cr
&\cr
\put(-56,-28){\grid(56,56)(14,14)}
\put(-49,21){\circle{4}}\put(-35,21){\circle{4}}\put(-21,21){\circle{4}}\put(-7,21){\circle{4}}
\put(-47,21){\vector(1,0){10}}\put(-33,21){\vector(1,0){10}}\put(-19,21){\vector(1,0){10}}
\put(-35,7){\circle{4}}\put(-21,7){\circle{4}}\put(-7,7){\circle{4}}
\put(-23,7){\vector(-1,0){10}}\put(-9,7){\vector(-1,0){10}}
=-144\,\left(\matrix{3&1&0&0\cr0&2&1&1\cr0&1&2&1\cr1&0&1&2\cr}\right)\,,
&\put(-56,-28){\grid(56,56)(14,14)}
\put(-49,21){\circle{4}}\put(-35,21){\circle{4}}\put(-47,21){\line(1,0){10}}
\put(-21,21){\circle{4}}\put(-7,21){\circle{4}}\put(-19,21){\line(1,0){10}}
\put(-49,7){\circle{4}}\put(-35,7){\circle{4}}\put(-47,7){\line(1,0){10}}
\put(-21,7){\circle{4}}\put(-7,7){\circle{4}}\put(-19,7){\line(1,0){10}}
\put(-49,-7){\circle{4}}\put(-35,-7){\circle{4}}\put(-47,-7){\line(1,0){10}}
\put(-21,-7){\circle{4}}\put(-7,-7){\circle{4}}\put(-19,-7){\line(1,0){10}}
=3\,\left(\matrix{1&3&0&0\cr3&1&0&0\cr0&0&1&3\cr0&0&3&1\cr}\right)\,,\cr
&\cr
\end{array}
\label{B1e}
$$

$$
\begin{array}{@{}l@{\hspace{63pt}}l@{}}
\put(-56,-28){\grid(56,56)(14,14)}
\put(-35,21){\circle{4}}\put(-21,21){\circle{4}}\put(-7,21){\circle{4}}
\put(-33,21){\vector(1,0){10}}\put(-19,21){\vector(1,0){10}}
\put(-35,7){\circle{4}}\put(-21,7){\circle{4}}\put(-7,7){\circle{4}}
\put(-33,7){\vector(1,0){10}}\put(-19,7){\vector(1,0){10}}
\put(-35,-7){\circle{4}}\put(-21,-7){\circle{4}}\put(-7,-7){\circle{4}}
\put(-23,-7){\vector(-1,0){10}}\put(-9,-7){\vector(-1,0){10}}
=24\,\left(\matrix{4&0&0&0\cr0&1&2&1\cr0&1&1&2\cr0&2&1&1\cr}\right)\,,
&\put(-56,-28){\grid(56,56)(14,14)}
\put(-35,21){\circle{4}}\put(-21,21){\circle{4}}\put(-7,21){\circle{4}}
\put(-33,21){\vector(1,0){10}}\put(-19,21){\vector(1,0){10}}
\put(-35,7){\circle{4}}\put(-21,7){\circle{4}}\put(-7,7){\circle{4}}
\put(-33,7){\vector(1,0){10}}\put(-19,7){\vector(1,0){10}}
\put(-49,-7){\circle{4}}\put(-35,-7){\circle{4}}\put(-21,-7){\circle{4}}
\put(-47,-7){\vector(1,0){10}}\put(-33,-7){\vector(1,0){10}}
=72\,\left(\matrix{3&1&0&0\cr0&1&3&0\cr1&0&1&2\cr0&2&0&2\cr}\right)\,,\cr
&\cr
\end{array}
\label{B1f}
$$

$$
\begin{array}{@{}l@{\hspace{63pt}}l@{}}
\put(-56,-28){\grid(56,56)(14,14)}
\put(-49,21){\circle{4}}\put(-35,21){\circle{4}}\put(-21,21){\circle{4}}\put(-7,21){\circle{4}}
\put(-47,21){\vector(1,0){10}}\put(-33,21){\vector(1,0){10}}\put(-19,21){\vector(1,0){10}}
\put(-35,7){\circle{4}}\put(-21,7){\circle{4}}\put(-7,7){\circle{4}}
\put(-23,7){\vector(-1,0){10}}\put(-9,7){\vector(-1,0){10}}
\put(-35,-7){\circle{4}}\put(-7,-7){\circle{4}}\put(-33,-7){\line(1,0){24}}
=144\,\left(\matrix{3&1&0&0\cr0&1&1&2\cr0&1&2&1\cr1&1&1&1\cr}\right)\,,
&\put(-56,-28){\grid(56,56)(14,14)}
\put(-49,21){\circle{4}}\put(-35,21){\circle{4}}\put(-21,21){\circle{4}}\put(-7,21){\circle{4}}
\put(-47,21){\vector(1,0){10}}\put(-33,21){\vector(1,0){10}}\put(-19,21){\vector(1,0){10}}
\put(-35,7){\circle{4}}\put(-21,7){\circle{4}}\put(-7,7){\circle{4}}
\put(-33,7){\vector(1,0){10}}\put(-19,7){\vector(1,0){10}}
\put(-49,-7){\circle{4}}\put(-21,-7){\circle{4}}\put(-47,-7){\line(1,0){24}}
=144\,\left(\matrix{2&1&1&0\cr0&2&2&0\cr1&0&1&2\cr1&1&0&2\cr}\right)\,,\cr
&\cr
\end{array}
\label{B1g}
$$

$$
\begin{array}{@{}l@{\hspace{63pt}}l@{}}
\put(-56,-28){\grid(56,56)(14,14)}
\put(-49,21){\circle{4}}\put(-35,21){\circle{4}}\put(-21,21){\circle{4}}\put(-7,21){\circle{4}}
\put(-47,21){\vector(1,0){10}}\put(-33,21){\vector(1,0){10}}\put(-19,21){\vector(1,0){10}}
\put(-35,7){\circle{4}}\put(-21,7){\circle{4}}\put(-7,7){\circle{4}}
\put(-23,7){\vector(-1,0){10}}\put(-9,7){\vector(-1,0){10}}
\put(-35,-5){\circle{4}}\put(-49,-9){\circle{4}}\put(-21,-5){\circle{4}}\put(-7,-9){\circle{4}}
\put(-33,-5){\line(1,0){10}}\put(-47,-9){\line(1,0){38}}
=-144\,\left(\matrix{2&1&0&1\cr0&1&2&1\cr0&2&1&1\cr2&0&1&1\cr}\right)\,,
&\put(-56,-28){\grid(56,56)(14,14)}
\put(-49,21){\circle{4}}\put(-35,21){\circle{4}}\put(-21,21){\circle{4}}\put(-7,21){\circle{4}}
\put(-47,21){\vector(1,0){10}}\put(-33,21){\vector(1,0){10}}\put(-19,21){\vector(1,0){10}}
\put(-35,7){\circle{4}}\put(-21,7){\circle{4}}\put(-7,7){\circle{4}}
\put(-33,7){\vector(1,0){10}}\put(-19,7){\vector(1,0){10}}
\put(-35,-7){\circle{4}}\put(-21,-7){\circle{4}}\put(-7,-7){\circle{4}}
\put(-33,-7){\vector(1,0){10}}\put(-19,-7){\vector(1,0){10}}
=-72\,\left(\matrix{3&1&0&0\cr0&1&3&0\cr0&0&1&3\cr1&2&0&1\cr}\right)\,,\cr
&\cr
\end{array}
\label{B1h}
$$

$$
\begin{array}{@{}l@{\hspace{63pt}}l@{}}
\put(-56,-28){\grid(56,56)(14,14)}
\put(-49,21){\circle{4}}\put(-35,21){\circle{4}}\put(-21,21){\circle{4}}\put(-7,21){\circle{4}}
\put(-47,21){\vector(1,0){10}}\put(-33,21){\vector(1,0){10}}\put(-19,21){\vector(1,0){10}}
\put(-49,9){\circle{4}}\put(-35,5){\circle{4}}\put(-21,9){\circle{4}}\put(-7,7){\circle{4}}
\put(-47,9){\vector(1,0){24}}\put(-19,9){\vector(1,0){12}}\put(-7,5){\vector(-1,0){26}}
\put(-35,-7){\circle{4}}\put(-7,-7){\circle{4}}\put(-33,-7){\line(1,0){24}}
=-72\,\left(\matrix{2&1&1&0\cr1&1&1&1\cr0&0&2&2\cr1&2&0&1\cr}\right)\,,
&\put(-56,-28){\grid(56,56)(14,14)}
\put(-49,21){\circle{4}}\put(-35,21){\circle{4}}\put(-21,21){\circle{4}}\put(-7,21){\circle{4}}
\put(-47,21){\vector(1,0){10}}\put(-33,21){\vector(1,0){10}}\put(-19,21){\vector(1,0){10}}
\put(-49,9){\circle{4}}\put(-35,5){\circle{4}}\put(-21,9){\circle{4}}\put(-7,7){\circle{4}}
\put(-47,9){\vector(1,0){24}}\put(-19,9){\vector(1,0){12}}\put(-7,5){\vector(-1,0){26}}
\put(-49,-7){\circle{4}}\put(-21,-7){\circle{4}}\put(-47,-7){\line(1,0){24}}
=-72\,\left(\matrix{1&1&2&0\cr1&2&1&0\cr1&0&1&2\cr1&1&0&2\cr}\right)\,,\cr
&\cr
\end{array}
\label{B1i}
$$

\begin{equation}
\begin{array}{@{}l@{\hspace{63pt}}l@{}}
\put(-56,-28){\grid(56,56)(14,14)}
\put(-49,21){\circle{4}}\put(-35,21){\circle{4}}\put(-21,21){\circle{4}}\put(-7,21){\circle{4}}
\put(-47,21){\vector(1,0){10}}\put(-33,21){\vector(1,0){10}}\put(-19,21){\vector(1,0){10}}
\put(-49,9){\circle{4}}\put(-35,5){\circle{4}}\put(-21,9){\circle{4}}\put(-7,7){\circle{4}}
\put(-47,9){\vector(1,0){24}}\put(-19,9){\vector(1,0){12}}\put(-7,5){\vector(-1,0){26}}
\put(-49,-7){\circle{4}}\put(-35,-7){\circle{4}}\put(-21,-7){\circle{4}}
\put(-47,-7){\vector(1,0){10}}\put(-33,-7){\vector(1,0){10}}
=72\,\left(\matrix{1&2&1&0\cr1&1&2&0\cr1&0&1&2\cr1&1&0&2\cr}\right)\,,
&\put(-56,-28){\grid(56,56)(14,14)}
\put(-49,21){\circle{4}}\put(-35,21){\circle{4}}\put(-21,21){\circle{4}}\put(-7,21){\circle{4}}
\put(-47,21){\vector(1,0){10}}\put(-33,21){\vector(1,0){10}}\put(-19,21){\vector(1,0){10}}
\put(-49,9){\circle{4}}\put(-35,5){\circle{4}}\put(-21,9){\circle{4}}\put(-7,7){\circle{4}}
\put(-47,9){\vector(1,0){24}}\put(-19,9){\vector(1,0){12}}\put(-7,5){\vector(-1,0){26}}
\put(-49,-7){\circle{4}}\put(-35,-7){\circle{4}}\put(-21,-7){\circle{4}}
\put(-37,-7){\vector(-1,0){10}}\put(-23,-7){\vector(-1,0){10}}
=72\,\left(\matrix{1&1&2&0\cr2&1&1&0\cr0&1&1&2\cr1&1&0&2\cr}\right)\,,\cr
&\cr
\end{array}
\label{B1j}
\end{equation}

\section{The Cayley--Hamilton Theorem for $r=4$, $d=4$}

\begin{eqnarray}
{1\over6}\,\left[
\left(\matrix{4&0&0&{\bf 0}\cr0&4&0&{\bf 0}\cr0&0&4&{\bf 0}\cr0&0&0&{\bf 4}\cr}\right)
-12\,\left(\matrix{4&0&0&{\bf 0}\cr0&3&1&{\bf 0}\cr0&1&3&{\bf 0}\cr0&0&0&{\bf 4}\cr}\right)
+9\,\left(\matrix{4&0&0&{\bf 0}\cr0&2&2&{\bf 0}\cr0&2&2&{\bf 0}\cr0&0&0&{\bf 4}\cr}\right)
\right.&&\nonumber\\
+8\,\left(\matrix{3&1&0&{\bf 0}\cr0&3&1&{\bf 0}\cr1&0&3&{\bf 0}\cr0&0&0&{\bf 4}\cr}\right)
+36\,\left(\matrix{3&1&0&{\bf 0}\cr1&2&1&{\bf 0}\cr0&1&3&{\bf 0}\cr0&0&0&{\bf 4}\cr}\right)
-72\,\left(\matrix{3&1&0&{\bf 0}\cr1&1&2&{\bf 0}\cr0&2&2&{\bf 0}\cr0&0&0&{\bf 4}\cr}\right)
&&\nonumber\\
\left.+6\,\left(\matrix{2&2&0&{\bf 0}\cr0&2&2&{\bf 0}\cr2&0&2&{\bf 0}\cr0&0&0&{\bf 4}\cr}\right)
+36\,\left(\matrix{0&2&2&{\bf 0}\cr2&1&1&{\bf 0}\cr2&1&1&{\bf 0}\cr0&0&0&{\bf 4}\cr}\right)
-12\,\left(\matrix{2&1&1&{\bf 0}\cr1&2&1&{\bf 0}\cr1&1&2&{\bf 0}\cr0&0&0&{\bf 4}\cr}\right)
\right]&&\nonumber\\
-2\,\left(\matrix{4&0&0&{\bf 0}\cr0&4&0&{\bf 0}\cr0&0&3&{\bf 1}\cr0&0&1&{\bf 3}\cr}\right)
+{3\over2}\,\left(\matrix{4&0&0&{\bf 0}\cr0&4&0&{\bf 0}\cr0&0&2&{\bf 2}\cr0&0&2&{\bf 2}\cr}\right)
+4\,\left(\matrix{4&0&0&{\bf 0}\cr0&3&1&{\bf 0}\cr0&0&3&{\bf 1}\cr0&1&0&{\bf 3}\cr}\right)
&&\nonumber\\
+6\,\left(\matrix{4&0&0&{\bf 0}\cr0&3&0&{\bf 1}\cr0&0&3&{\bf 1}\cr0&1&1&{\bf 2}\cr}\right)
+12\,\left(\matrix{4&0&0&{\bf 0}\cr0&3&1&{\bf 0}\cr0&1&2&{\bf 1}\cr0&0&1&{\bf 3}\cr}\right)
-12\,\left(\matrix{4&0&0&{\bf 0}\cr0&3&1&{\bf 0}\cr0&0&2&{\bf 2}\cr0&1&1&{\bf 2}\cr}\right)
&&\nonumber\\
-12\,\left(\matrix{4&0&0&{\bf 0}\cr0&3&0&{\bf 1}\cr0&0&2&{\bf 2}\cr0&1&2&{\bf 1}\cr}\right)
-12\,\left(\matrix{4&0&0&{\bf 0}\cr0&2&2&{\bf 0}\cr0&2&1&{\bf 1}\cr0&0&1&{\bf 3}\cr}\right)
+3\,\left(\matrix{4&0&0&{\bf 0}\cr0&2&2&{\bf 0}\cr0&2&0&{\bf 2}\cr0&0&2&{\bf 2}\cr}\right)
&&\nonumber\\
+6\,\left(\matrix{4&0&0&{\bf 0}\cr0&2&2&{\bf 0}\cr0&1&1&{\bf 2}\cr0&1&1&{\bf 2}\cr}\right)
+12\,\left(\matrix{4&0&0&{\bf 0}\cr0&2&1&{\bf 1}\cr0&2&1&{\bf 1}\cr0&0&2&{\bf 2}\cr}\right)
-6\,\left(\matrix{4&0&0&{\bf 0}\cr0&2&1&{\bf 1}\cr0&1&2&{\bf 1}\cr0&1&1&{\bf 2}\cr}\right)
&&\nonumber\\
-4\,\left(\matrix{3&1&0&{\bf 0}\cr1&3&0&{\bf 0}\cr0&0&3&{\bf 1}\cr0&0&1&{\bf 3}\cr}\right)
+8\,\left(\matrix{3&1&0&{\bf 0}\cr0&3&1&{\bf 0}\cr0&0&3&{\bf 1}\cr1&0&0&{\bf 3}\cr}\right)
-12\,\left(\matrix{3&1&0&{\bf 0}\cr0&3&0&{\bf 1}\cr0&0&3&{\bf 1}\cr1&0&1&{\bf 2}\cr}\right)
&&\nonumber\\
-12\,\left(\matrix{3&0&1&{\bf 0}\cr0&3&1&{\bf 0}\cr1&0&0&{\bf 3}\cr0&1&2&{\bf 1}\cr}\right)
-12\,\left(\matrix{3&0&1&{\bf 0}\cr1&3&0&{\bf 0}\cr0&1&2&{\bf 1}\cr0&0&1&{\bf 2}\cr}\right)
-12\,\left(\matrix{3&0&1&{\bf 0}\cr0&3&0&{\bf 1}\cr1&1&2&{\bf 0}\cr0&0&1&{\bf 2}\cr}\right)
&&\nonumber\\
-4\,\left(\matrix{3&0&0&{\bf 1}\cr0&3&0&{\bf 1}\cr0&0&3&{\bf 1}\cr1&1&1&{\bf 1}\cr}\right)
-12\,\left(\matrix{3&0&1&{\bf 0}\cr0&3&1&{\bf 0}\cr1&1&1&{\bf 1}\cr0&0&1&{\bf 3}\cr}\right)
-6\,\left(\matrix{3&1&0&{\bf 0}\cr1&3&0&{\bf 0}\cr0&0&2&{\bf 2}\cr0&0&2&{\bf 2}\cr}\right)
&&\nonumber\\
-6\,\left(\matrix{3&0&0&{\bf 1}\cr0&2&2&{\bf 0}\cr0&2&2&{\bf 0}\cr1&0&0&{\bf 3}\cr}\right)
+6\,\left(\matrix{3&0&0&{\bf 1}\cr0&3&0&{\bf 1}\cr0&0&2&{\bf 2}\cr1&1&2&{\bf 0}\cr}\right)
+6\,\left(\matrix{3&0&1&{\bf 0}\cr0&3&1&{\bf 0}\cr0&0&2&{\bf 2}\cr1&1&0&{\bf 2}\cr}\right)
&&\nonumber\\
+12\,\left(\matrix{3&0&1&{\bf 0}\cr0&0&1&{\bf 3}\cr0&2&2&{\bf 2}\cr1&2&0&{\bf 1}\cr}\right)
+12\,\left(\matrix{3&1&0&{\bf 0}\cr0&3&0&{\bf 1}\cr1&0&2&{\bf 1}\cr0&0&2&{\bf 2}\cr}\right)
+12\,\left(\matrix{3&1&0&{\bf 0}\cr0&3&1&{\bf 0}\cr1&0&1&{\bf 2}\cr0&0&2&{\bf 2}\cr}\right)
&&\nonumber\\
+12\,\left(\matrix{3&0&0&{\bf 1}\cr0&0&1&{\bf 3}\cr1&2&1&{\bf 0}\cr0&2&2&{\bf 0}\cr}\right)
+12\,\left(\matrix{3&0&1&{\bf 0}\cr0&2&2&{\bf 0}\cr0&2&1&{\bf 1}\cr1&0&0&{\bf 3}\cr}\right)
-36\,\left(\matrix{3&0&0&{\bf 1}\cr0&3&1&{\bf 0}\cr0&1&2&{\bf 1}\cr1&0&1&{\bf 2}\cr}\right)
&&\nonumber\\
-36\,\left(\matrix{3&0&1&{\bf 0}\cr0&1&0&{\bf 3}\cr0&2&1&{\bf 1}\cr1&1&2&{\bf 0}\cr}\right)
+12\,\left(\matrix{3&0&0&{\bf 1}\cr0&3&1&{\bf 0}\cr1&0&2&{\bf 1}\cr0&1&1&{\bf 2}\cr}\right)
+12\,\left(\matrix{3&0&1&{\bf 0}\cr0&3&0&{\bf 1}\cr1&0&1&{\bf 2}\cr0&1&2&{\bf 1}\cr}\right)
&&\nonumber\\
+24\,\left(\matrix{3&0&1&{\bf 0}\cr0&1&0&{\bf 3}\cr1&2&1&{\bf 0}\cr0&1&2&{\bf 1}\cr}\right)
+12\,\left(\matrix{3&0&0&{\bf 1}\cr0&3&0&{\bf 1}\cr0&1&2&{\bf 1}\cr1&0&2&{\bf 1}\cr}\right)
+12\,\left(\matrix{3&0&1&{\bf 0}\cr0&3&1&{\bf 0}\cr0&1&1&{\bf 2}\cr1&0&1&{\bf 2}\cr}\right)
&&\nonumber\\
+24\,\left(\matrix{3&0&1&{\bf 0}\cr1&2&1&{\bf 0}\cr0&2&1&{\bf 1}\cr0&0&1&{\bf 3}\cr}\right)
+24\,\left(\matrix{3&0&0&{\bf 1}\cr0&3&1&{\bf 0}\cr0&0&2&{\bf 2}\cr1&1&1&{\bf 1}\cr}\right)
+24\,\left(\matrix{3&0&1&{\bf 0}\cr0&1&0&{\bf 3}\cr0&2&2&{\bf 0}\cr1&1&1&{\bf 1}\cr}\right)
&&\nonumber\\
-12\,\left(\matrix{3&0&0&{\bf 1}\cr0&2&2&{\bf 0}\cr0&2&0&{\bf 2}\cr1&0&2&{\bf 1}\cr}\right)
-12\,\left(\matrix{3&0&1&{\bf 0}\cr0&2&0&{\bf 2}\cr0&2&2&{\bf 0}\cr1&0&1&{\bf 2}\cr}\right)
-12\,\left(\matrix{3&0&1&{\bf 0}\cr0&2&0&{\bf 2}\cr0&0&2&{\bf 2}\cr1&2&1&{\bf 0}\cr}\right)
&&\nonumber\\
-12\,\left(\matrix{2&2&0&{\bf 0}\cr2&0&2&{\bf 0}\cr0&2&1&{\bf 1}\cr0&0&1&{\bf 3}\cr}\right)
+36\,\left(\matrix{3&0&1&{\bf 0}\cr0&2&0&{\bf 2}\cr0&2&1&{\bf 1}\cr1&0&2&{\bf 1}\cr}\right)
+36\,\left(\matrix{3&0&0&{\bf 1}\cr0&2&2&{\bf 0}\cr0&2&1&{\bf 1}\cr1&0&1&{\bf 2}\cr}\right)
&&\nonumber\\
+36\,\left(\matrix{3&1&0&{\bf 0}\cr1&2&1&{\bf 0}\cr0&0&2&{\bf 2}\cr0&1&1&{\bf 2}\cr}\right)
+36\,\left(\matrix{2&0&2&{\bf 0}\cr0&2&1&{\bf 1}\cr2&1&1&{\bf 0}\cr0&1&0&{\bf 3}\cr}\right)
-24\,\left(\matrix{3&1&0&{\bf 0}\cr0&2&0&{\bf 2}\cr1&0&2&{\bf 1}\cr0&1&2&{\bf 1}\cr}\right)
&&\nonumber\\
-24\,\left(\matrix{3&1&0&{\bf 0}\cr0&2&2&{\bf 0}\cr1&2&1&{\bf 0}\cr0&2&1&{\bf 1}\cr}\right)
-24\,\left(\matrix{3&0&0&{\bf 1}\cr0&0&2&{\bf 2}\cr1&2&1&{\bf 0}\cr0&2&1&{\bf 1}\cr}\right)
-24\,\left(\matrix{2&0&2&{\bf 0}\cr1&2&1&{\bf 0}\cr0&2&1&{\bf 1}\cr1&0&0&{\bf 3}\cr}\right)
&&\nonumber\\
+18\,\left(\matrix{3&1&0&{\bf 0}\cr0&2&1&{\bf 1}\cr0&1&2&{\bf 1}\cr1&0&1&{\bf 2}\cr}\right)
+18\,\left(\matrix{3&1&0&{\bf 0}\cr0&2&1&{\bf 1}\cr1&1&2&{\bf 0}\cr0&1&1&{\bf 2}\cr}\right)
+18\,\left(\matrix{3&0&0&{\bf 1}\cr1&1&2&{\bf 0}\cr0&2&1&{\bf 1}\cr0&1&1&{\bf 2}\cr}\right)
&&\nonumber\\
+18\,\left(\matrix{2&1&1&{\bf 0}\cr1&2&1&{\bf 0}\cr0&1&2&{\bf 1}\cr1&0&0&{\bf 3}\cr}\right)
-48\,\left(\matrix{3&1&0&{\bf 0}\cr1&1&1&{\bf 1}\cr0&2&1&{\bf 1}\cr0&0&2&{\bf 2}\cr}\right)
-24\,\left(\matrix{3&0&0&{\bf 1}\cr1&1&1&{\bf 1}\cr0&1&1&{\bf 2}\cr0&2&2&{\bf 0}\cr}\right)
&&\nonumber\\
-24\,\left(\matrix{2&2&0&{\bf 0}\cr1&1&2&{\bf 0}\cr1&1&1&{\bf 1}\cr0&0&1&{\bf 3}\cr}\right)
-24\,\left(\matrix{3&1&0&{\bf 0}\cr1&1&2&{\bf 0}\cr0&1&1&{\bf 2}\cr0&1&1&{\bf 2}\cr}\right)
-24\,\left(\matrix{3&1&0&{\bf 0}\cr1&1&0&{\bf 2}\cr0&1&2&{\bf 1}\cr0&1&2&{\bf 1}\cr}\right)
&&\nonumber\\
-24\,\left(\matrix{3&0&0&{\bf 1}\cr1&2&0&{\bf 1}\cr0&1&2&{\bf 1}\cr0&1&2&{\bf 1}\cr}\right)
-24\,\left(\matrix{2&1&1&{\bf 0}\cr2&1&1&{\bf 0}\cr0&2&1&{\bf 1}\cr0&0&1&{\bf 3}\cr}\right)
+12\,\left(\matrix{3&0&0&{\bf 1}\cr0&2&1&{\bf 1}\cr0&1&2&{\bf 1}\cr1&1&1&{\bf 1}\cr}\right)
&&\nonumber\\
+24\,\left(\matrix{3&0&1&{\bf 0}\cr0&2&1&{\bf 1}\cr0&1&1&{\bf 2}\cr1&1&1&{\bf 1}\cr}\right)
+3\,\left(\matrix{2&0&0&{\bf 2}\cr2&2&0&{\bf 0}\cr0&2&2&{\bf 0}\cr0&0&2&{\bf 2}\cr}\right)
+12\,\left(\matrix{2&1&1&{\bf 0}\cr1&2&1&{\bf 0}\cr1&1&1&{\bf 1}\cr0&0&1&{\bf 3}\cr}\right)
&&\nonumber\\
+{9\over2}\,\left(\matrix{2&2&0&{\bf 0}\cr2&2&0&{\bf 0}\cr0&0&2&{\bf 2}\cr0&0&2&{\bf 2}\cr}\right)
-12\,\left(\matrix{2&2&0&{\bf 0}\cr2&1&1&{\bf 0}\cr0&1&1&{\bf 2}\cr0&0&2&{\bf 2}\cr}\right)
-12\,\left(\matrix{2&2&0&{\bf 0}\cr2&1&0&{\bf 1}\cr0&1&2&{\bf 1}\cr0&0&2&{\bf 2}\cr}\right)
&&\nonumber\\
+20\,\left(\matrix{2&1&0&{\bf 1}\cr0&2&2&{\bf 0}\cr1&0&2&{\bf 1}\cr1&1&0&{\bf 2}\cr}\right)
+20\,\left(\matrix{2&1&1&{\bf 0}\cr0&2&0&{\bf 2}\cr1&0&1&{\bf 2}\cr1&1&2&{\bf 0}\cr}\right)
+20\,\left(\matrix{2&0&1&{\bf 1}\cr0&2&0&{\bf 2}\cr1&2&1&{\bf 0}\cr1&0&2&{\bf 1}\cr}\right)
&&\nonumber\\
+20\,\left(\matrix{2&2&0&{\bf 0}\cr0&2&1&{\bf 1}\cr1&0&2&{\bf 1}\cr1&0&1&{\bf 2}\cr}\right)
+48\,\left(\matrix{2&0&1&{\bf 1}\cr0&2&1&{\bf 1}\cr1&1&2&{\bf 0}\cr1&1&0&{\bf 2}\cr}\right)
-84\,\left(\matrix{2&1&1&{\bf 0}\cr0&2&1&{\bf 1}\cr1&0&2&{\bf 1}\cr1&1&0&{\bf 2}\cr}\right)
&&\nonumber\\
-12\,\left(\matrix{2&1&1&{\bf 0}\cr1&2&1&{\bf 0}\cr1&0&1&{\bf 2}\cr0&1&1&{\bf 2}\cr}\right)
-12\,\left(\matrix{2&1&0&{\bf 1}\cr1&2&0&{\bf 1}\cr1&0&2&{\bf 1}\cr0&1&2&{\bf 1}\cr}\right)
-24\,\left(\matrix{2&1&0&{\bf 1}\cr1&1&0&{\bf 2}\cr1&1&2&{\bf 0}\cr0&1&2&{\bf 1}\cr}\right)
&&\nonumber\\
-24\,\left(\matrix{2&1&1&{\bf 0}\cr1&2&0&{\bf 1}\cr1&1&1&{\bf 1}\cr0&0&2&{\bf 2}\cr}\right)
-24\,\left(\matrix{2&1&0&{\bf 1}\cr1&0&1&{\bf 2}\cr1&1&1&{\bf 1}\cr0&2&2&{\bf 0}\cr}\right)
+24\,\left(\matrix{2&2&0&{\bf 0}\cr0&0&2&{\bf 2}\cr1&1&1&{\bf 1}\cr1&1&1&{\bf 1}\cr}\right)
&&\nonumber\\
+12\,\left(\matrix{2&0&1&{\bf 1}\cr2&0&1&{\bf 1}\cr0&2&1&{\bf 1}\cr0&2&1&{\bf 1}\cr}\right)
+12\,\left(\matrix{2&1&1&{\bf 0}\cr2&1&1&{\bf 0}\cr0&1&1&{\bf 2}\cr0&1&1&{\bf 2}\cr}\right)
+22\,\left(\matrix{2&1&0&{\bf 1}\cr0&2&1&{\bf 1}\cr1&0&2&{\bf 1}\cr1&1&1&{\bf 1}\cr}\right)
&&\nonumber\\
+66\,\left(\matrix{2&1&1&{\bf 0}\cr0&2&1&{\bf 1}\cr1&0&1&{\bf 2}\cr1&1&1&{\bf 1}\cr}\right)
-24\,\left(\matrix{2&0&1&{\bf 1}\cr0&2&1&{\bf 1}\cr1&1&1&{\bf 1}\cr1&1&1&{\bf 1}\cr}\right)
-24\,\left(\matrix{2&1&1&{\bf 0}\cr0&1&1&{\bf 2}\cr1&1&1&{\bf 1}\cr1&1&1&{\bf 1}\cr}\right)
&&\nonumber\\
+4\,\left(\matrix{1&1&1&{\bf 1}\cr1&1&1&{\bf 1}\cr1&1&1&{\bf 1}\cr1&1&1&{\bf 1}\cr}\right)
-C_4({\bf A})\,[{\bf 4}]&\equiv&0\,.
\label{C1a}
\end{eqnarray}

\section{Generalized Product for $r=4$, $d=4$}

\begin{eqnarray}
({\bf A}^4)&=&8\,\left(\matrix{3&1&0&{\bf 0}\cr0&3&1&{\bf 0}\cr0&0&3&{\bf 1}\cr1&0&0&{\bf 3}\cr}\right)
-12\,\left(\matrix{3&1&0&{\bf 0}\cr0&3&0&{\bf 1}\cr0&0&3&{\bf 1}\cr1&0&1&{\bf 2}\cr}\right)
-12\,\left(\matrix{3&0&1&{\bf 0}\cr0&3&1&{\bf 0}\cr1&0&0&{\bf 3}\cr0&1&2&{\bf 1}\cr}\right)
\nonumber\\
&&-12\,\left(\matrix{3&0&1&{\bf 0}\cr1&3&0&{\bf 0}\cr0&1&2&{\bf 1}\cr0&0&1&{\bf 2}\cr}\right)
-12\,\left(\matrix{3&0&1&{\bf 0}\cr0&3&0&{\bf 1}\cr1&1&2&{\bf 0}\cr0&0&1&{\bf 2}\cr}\right)
-4\,\left(\matrix{3&0&0&{\bf 1}\cr0&3&0&{\bf 1}\cr0&0&3&{\bf 1}\cr1&1&1&{\bf 1}\cr}\right)
\nonumber\\
&&-12\,\left(\matrix{3&0&1&{\bf 0}\cr0&3&1&{\bf 0}\cr1&1&1&{\bf 1}\cr0&0&1&{\bf 3}\cr}\right)
-6\,\left(\matrix{3&0&0&{\bf 1}\cr0&2&2&{\bf 0}\cr0&2&2&{\bf 0}\cr1&0&0&{\bf 3}\cr}\right)
+6\,\left(\matrix{3&0&0&{\bf 1}\cr0&3&0&{\bf 1}\cr0&0&2&{\bf 2}\cr1&1&2&{\bf 0}\cr}\right)
\nonumber\\
&&+6\,\left(\matrix{3&0&1&{\bf 0}\cr0&3&1&{\bf 0}\cr0&0&2&{\bf 2}\cr1&1&0&{\bf 2}\cr}\right)
+12\,\left(\matrix{3&0&1&{\bf 0}\cr0&0&1&{\bf 3}\cr0&2&2&{\bf 2}\cr1&2&0&{\bf 1}\cr}\right)
+12\,\left(\matrix{3&1&0&{\bf 0}\cr0&3&0&{\bf 1}\cr1&0&2&{\bf 1}\cr0&0&2&{\bf 2}\cr}\right)
\nonumber\\
&&+12\,\left(\matrix{3&1&0&{\bf 0}\cr0&3&1&{\bf 0}\cr1&0&1&{\bf 2}\cr0&0&2&{\bf 2}\cr}\right)
+12\,\left(\matrix{3&0&0&{\bf 1}\cr0&0&1&{\bf 3}\cr1&2&1&{\bf 0}\cr0&2&2&{\bf 0}\cr}\right)
+12\,\left(\matrix{3&0&1&{\bf 0}\cr0&2&2&{\bf 0}\cr0&2&1&{\bf 1}\cr1&0&0&{\bf 3}\cr}\right)
\nonumber\\
&&-36\,\left(\matrix{3&0&0&{\bf 1}\cr0&3&1&{\bf 0}\cr0&1&2&{\bf 1}\cr1&0&1&{\bf 2}\cr}\right)
-36\,\left(\matrix{3&0&1&{\bf 0}\cr0&1&0&{\bf 3}\cr0&2&1&{\bf 1}\cr1&1&2&{\bf 0}\cr}\right)
+12\,\left(\matrix{3&0&0&{\bf 1}\cr0&3&1&{\bf 0}\cr1&0&2&{\bf 1}\cr0&1&1&{\bf 2}\cr}\right)
\nonumber\\
&&+12\,\left(\matrix{3&0&1&{\bf 0}\cr0&3&0&{\bf 1}\cr1&0&1&{\bf 2}\cr0&1&2&{\bf 1}\cr}\right)
+24\,\left(\matrix{3&0&1&{\bf 0}\cr0&1&0&{\bf 3}\cr1&2&1&{\bf 0}\cr0&1&2&{\bf 1}\cr}\right)
+12\,\left(\matrix{3&0&0&{\bf 1}\cr0&3&0&{\bf 1}\cr0&1&2&{\bf 1}\cr1&0&2&{\bf 1}\cr}\right)
\nonumber\\
&&+12\,\left(\matrix{3&0&1&{\bf 0}\cr0&3&1&{\bf 0}\cr0&1&1&{\bf 2}\cr1&0&1&{\bf 2}\cr}\right)
+24\,\left(\matrix{3&0&1&{\bf 0}\cr1&2&1&{\bf 0}\cr0&2&1&{\bf 1}\cr0&0&1&{\bf 3}\cr}\right)
+24\,\left(\matrix{3&0&0&{\bf 1}\cr0&3&1&{\bf 0}\cr0&0&2&{\bf 2}\cr1&1&1&{\bf 1}\cr}\right)
\nonumber\\
&&+24\,\left(\matrix{3&0&1&{\bf 0}\cr0&1&0&{\bf 3}\cr0&2&2&{\bf 0}\cr1&1&1&{\bf 1}\cr}\right)
-12\,\left(\matrix{3&0&0&{\bf 1}\cr0&2&2&{\bf 0}\cr0&2&0&{\bf 2}\cr1&0&2&{\bf 1}\cr}\right)
-12\,\left(\matrix{3&0&1&{\bf 0}\cr0&2&0&{\bf 2}\cr0&2&2&{\bf 0}\cr1&0&1&{\bf 2}\cr}\right)
\nonumber\\
&&-12\,\left(\matrix{3&0&1&{\bf 0}\cr0&2&0&{\bf 2}\cr0&0&2&{\bf 2}\cr1&2&1&{\bf 0}\cr}\right)
-12\,\left(\matrix{2&2&0&{\bf 0}\cr2&0&2&{\bf 0}\cr0&2&1&{\bf 1}\cr0&0&1&{\bf 3}\cr}\right)
+36\,\left(\matrix{3&0&1&{\bf 0}\cr0&2&0&{\bf 2}\cr0&2&1&{\bf 1}\cr1&0&2&{\bf 1}\cr}\right)
\nonumber\\
&&+36\,\left(\matrix{3&0&0&{\bf 1}\cr0&2&2&{\bf 0}\cr0&2&1&{\bf 1}\cr1&0&1&{\bf 2}\cr}\right)
+36\,\left(\matrix{3&1&0&{\bf 0}\cr1&2&1&{\bf 0}\cr0&0&2&{\bf 2}\cr0&1&1&{\bf 2}\cr}\right)
+36\,\left(\matrix{2&0&2&{\bf 0}\cr0&2&1&{\bf 1}\cr2&1&1&{\bf 0}\cr0&1&0&{\bf 3}\cr}\right)
\nonumber\\
&&-24\,\left(\matrix{3&1&0&{\bf 0}\cr0&2&0&{\bf 2}\cr1&0&2&{\bf 1}\cr0&1&2&{\bf 1}\cr}\right)
-24\,\left(\matrix{3&1&0&{\bf 0}\cr0&2&2&{\bf 0}\cr1&2&1&{\bf 0}\cr0&2&1&{\bf 1}\cr}\right)
-24\,\left(\matrix{3&0&0&{\bf 1}\cr0&0&2&{\bf 2}\cr1&2&1&{\bf 0}\cr0&2&1&{\bf 1}\cr}\right)
\nonumber\\
&&-24\,\left(\matrix{2&0&2&{\bf 0}\cr1&2&1&{\bf 0}\cr0&2&1&{\bf 1}\cr1&0&0&{\bf 3}\cr}\right)
+18\,\left(\matrix{3&1&0&{\bf 0}\cr0&2&1&{\bf 1}\cr0&1&2&{\bf 1}\cr1&0&1&{\bf 2}\cr}\right)
+18\,\left(\matrix{3&1&0&{\bf 0}\cr0&2&1&{\bf 1}\cr1&1&2&{\bf 0}\cr0&1&1&{\bf 2}\cr}\right)
\nonumber\\
&&+18\,\left(\matrix{3&0&0&{\bf 1}\cr1&1&2&{\bf 0}\cr0&2&1&{\bf 1}\cr0&1&1&{\bf 2}\cr}\right)
+18\,\left(\matrix{2&1&1&{\bf 0}\cr1&2&1&{\bf 0}\cr0&1&2&{\bf 1}\cr1&0&0&{\bf 3}\cr}\right)
-48\,\left(\matrix{3&1&0&{\bf 0}\cr1&1&1&{\bf 1}\cr0&2&1&{\bf 1}\cr0&0&2&{\bf 2}\cr}\right)
\nonumber\\
&&-24\,\left(\matrix{3&0&0&{\bf 1}\cr1&1&1&{\bf 1}\cr0&1&1&{\bf 2}\cr0&2&2&{\bf 0}\cr}\right)
-24\,\left(\matrix{2&2&0&{\bf 0}\cr1&1&2&{\bf 0}\cr1&1&1&{\bf 1}\cr0&0&1&{\bf 3}\cr}\right)
-24\,\left(\matrix{3&1&0&{\bf 0}\cr1&1&2&{\bf 0}\cr0&1&1&{\bf 2}\cr0&1&1&{\bf 2}\cr}\right)
\nonumber\\
&&-24\,\left(\matrix{3&1&0&{\bf 0}\cr1&1&0&{\bf 2}\cr0&1&2&{\bf 1}\cr0&1&2&{\bf 1}\cr}\right)
-24\,\left(\matrix{3&0&0&{\bf 1}\cr1&2&0&{\bf 1}\cr0&1&2&{\bf 1}\cr0&1&2&{\bf 1}\cr}\right)
-24\,\left(\matrix{2&1&1&{\bf 0}\cr2&1&1&{\bf 0}\cr0&2&1&{\bf 1}\cr0&0&1&{\bf 3}\cr}\right)
\nonumber\\
&&+12\,\left(\matrix{3&0&0&{\bf 1}\cr0&2&1&{\bf 1}\cr0&1&2&{\bf 1}\cr1&1&1&{\bf 1}\cr}\right)
+24\,\left(\matrix{3&0&1&{\bf 0}\cr0&2&1&{\bf 1}\cr0&1&1&{\bf 2}\cr1&1&1&{\bf 1}\cr}\right)
+3\,\left(\matrix{2&0&0&{\bf 2}\cr2&2&0&{\bf 0}\cr0&2&2&{\bf 0}\cr0&0&2&{\bf 2}\cr}\right)
\nonumber\\
&&+12\,\left(\matrix{2&1&1&{\bf 0}\cr1&2&1&{\bf 0}\cr1&1&1&{\bf 1}\cr0&0&1&{\bf 3}\cr}\right)
-12\,\left(\matrix{2&2&0&{\bf 0}\cr2&1&1&{\bf 0}\cr0&1&1&{\bf 2}\cr0&0&2&{\bf 2}\cr}\right)
-12\,\left(\matrix{2&2&0&{\bf 0}\cr2&1&0&{\bf 1}\cr0&1&2&{\bf 1}\cr0&0&2&{\bf 2}\cr}\right)
\nonumber\\
&&+20\,\left(\matrix{2&1&0&{\bf 1}\cr0&2&2&{\bf 0}\cr1&0&2&{\bf 1}\cr1&1&0&{\bf 2}\cr}\right)
+20\,\left(\matrix{2&1&1&{\bf 0}\cr0&2&0&{\bf 2}\cr1&0&1&{\bf 2}\cr1&1&2&{\bf 0}\cr}\right)
+20\,\left(\matrix{2&0&1&{\bf 1}\cr0&2&0&{\bf 2}\cr1&2&1&{\bf 0}\cr1&0&2&{\bf 1}\cr}\right)
\nonumber\\
&&+20\,\left(\matrix{2&2&0&{\bf 0}\cr0&2&1&{\bf 1}\cr1&0&2&{\bf 1}\cr1&0&1&{\bf 2}\cr}\right)
+48\,\left(\matrix{2&0&1&{\bf 1}\cr0&2&1&{\bf 1}\cr1&1&2&{\bf 0}\cr1&1&0&{\bf 2}\cr}\right)
-84\,\left(\matrix{2&1&1&{\bf 0}\cr0&2&1&{\bf 1}\cr1&0&2&{\bf 1}\cr1&1&0&{\bf 2}\cr}\right)
\nonumber\\
&&-12\,\left(\matrix{2&1&1&{\bf 0}\cr1&2&1&{\bf 0}\cr1&0&1&{\bf 2}\cr0&1&1&{\bf 2}\cr}\right)
-12\,\left(\matrix{2&1&0&{\bf 1}\cr1&2&0&{\bf 1}\cr1&0&2&{\bf 1}\cr0&1&2&{\bf 1}\cr}\right)
-24\,\left(\matrix{2&1&0&{\bf 1}\cr1&1&0&{\bf 2}\cr1&1&2&{\bf 0}\cr0&1&2&{\bf 1}\cr}\right)
\nonumber\\
&&-24\,\left(\matrix{2&1&1&{\bf 0}\cr1&2&0&{\bf 1}\cr1&1&1&{\bf 1}\cr0&0&2&{\bf 2}\cr}\right)
-24\,\left(\matrix{2&1&0&{\bf 1}\cr1&0&1&{\bf 2}\cr1&1&1&{\bf 1}\cr0&2&2&{\bf 0}\cr}\right)
+24\,\left(\matrix{2&2&0&{\bf 0}\cr0&0&2&{\bf 2}\cr1&1&1&{\bf 1}\cr1&1&1&{\bf 1}\cr}\right)
\nonumber\\
&&+12\,\left(\matrix{2&0&1&{\bf 1}\cr2&0&1&{\bf 1}\cr0&2&1&{\bf 1}\cr0&2&1&{\bf 1}\cr}\right)
+12\,\left(\matrix{2&1&1&{\bf 0}\cr2&1&1&{\bf 0}\cr0&1&1&{\bf 2}\cr0&1&1&{\bf 2}\cr}\right)
+22\,\left(\matrix{2&1&0&{\bf 1}\cr0&2&1&{\bf 1}\cr1&0&2&{\bf 1}\cr1&1&1&{\bf 1}\cr}\right)
\nonumber\\
&&+66\,\left(\matrix{2&1&1&{\bf 0}\cr0&2&1&{\bf 1}\cr1&0&1&{\bf 2}\cr1&1&1&{\bf 1}\cr}\right)
-24\,\left(\matrix{2&0&1&{\bf 1}\cr0&2&1&{\bf 1}\cr1&1&1&{\bf 1}\cr1&1&1&{\bf 1}\cr}\right)
-24\,\left(\matrix{2&1&1&{\bf 0}\cr0&1&1&{\bf 2}\cr1&1&1&{\bf 1}\cr1&1&1&{\bf 1}\cr}\right)
\nonumber\\
&&+4\,\left(\matrix{1&1&1&{\bf 1}\cr1&1&1&{\bf 1}\cr1&1&1&{\bf 1}\cr1&1&1&{\bf 1}\cr}\right)\,.
\label{D1a}
\end{eqnarray}

\newpage
\section{Grids for $r=3$, $n=4$}

$$
\begin{array}{@{}l@{\hspace{63pt}}l@{}}
{\put(-56,-17){\grid(56,42)(14,14)}}
=\left(\matrix{3&0&0&0\cr0&3&0&0\cr0&0&3&0\cr0&0&0&3\cr}\right)\,,
&\put(-56,-17){\grid(56,42)(14,14)}
\put(-21,18){\circle{4}}\put(-7,18){\circle{4}}\put(-19,18){\line(1,0){10}}
=-12\,\left(\matrix{3&0&0&0\cr0&3&0&0\cr0&0&2&1\cr0&0&1&2\cr}\right)\,,\cr
&\cr
{\put(-56,-17){\grid(56,42)(14,14)}
\put(-21,18){\circle{4}}\put(-7,18){\circle{4}}\put(-19,18){\line(1,0){10}}
\put(-21,4){\circle{4}}\put(-7,4){\circle{4}}\put(-19,4){\line(1,0){10}}}
=6\,\left(\matrix{3&0&0&0\cr0&3&0&0\cr0&0&1&2\cr0&0&2&1\cr}\right)\,,
&{\put(-56,-17){\grid(56,42)(14,14)}
\put(-21,18){\circle{4}}\put(-7,18){\circle{4}}\put(-19,18){\line(1,0){10}}
\put(-35,4){\circle{4}}\put(-21,4){\circle{4}}\put(-33,4){\line(1,0){10}}}
=24\,\left(\matrix{3&0&0&0\cr0&2&1&0\cr0&1&1&1\cr0&0&1&2\cr}\right)\,,\cr
&\cr
{\put(-56,-17){\grid(56,42)(14,14)}
\put(-21,18){\circle{4}}\put(-7,18){\circle{4}}\put(-19,18){\line(1,0){10}}
\put(-49,4){\circle{4}}\put(-35,4){\circle{4}}\put(-47,4){\line(1,0){10}}}
=6\,\left(\matrix{2&1&0&0\cr1&2&0&0\cr0&0&2&1\cr0&0&1&2\cr}\right)\,,
&\put(-56,-17){\grid(56,42)(14,14)}
\put(-21,18){\circle{4}}\put(-7,18){\circle{4}}\put(-19,18){\line(1,0){10}}
\put(-49,18){\circle{4}}\put(-35,18){\circle{4}}\put(-47,18){\line(1,0){10}}
=6\,\left(\matrix{2&1&0&0\cr1&2&0&0\cr0&0&2&1\cr0&0&1&2\cr}\right)\,,\cr
&\cr
\put(-56,-17){\grid(56,42)(14,14)}
\put(-35,18){\circle{4}}\put(-21,18){\circle{4}}\put(-7,18){\circle{4}}
\put(-33,18){\vector(1,0){10}}\put(-19,18){\vector(1,0){10}}
=16\,\left(\matrix{3&0&0&0\cr0&2&1&0\cr0&0&2&1\cr0&1&0&2\cr}\right)\,,
&\put(-56,-17){\grid(56,42)(14,14)}
\put(-21,18){\circle{4}}\put(-7,18){\circle{4}}\put(-19,18){\line(1,0){10}}
\put(-49,18){\circle{4}}\put(-35,18){\circle{4}}\put(-47,18){\line(1,0){10}}
\put(-21,4){\circle{4}}\put(-7,4){\circle{4}}\put(-19,4){\line(1,0){10}}
=-12\,\left(\matrix{2&1&0&0\cr1&2&0&0\cr0&0&1&2\cr0&0&2&1\cr}\right)\,,\cr
&\cr
\end{array}
\label{E1a}
$$

\begin{equation}
\begin{array}{@{}l@{\hspace{63pt}}l@{}}
\put(-56,-14){\grid(56,42)(14,14)}
\put(-35,21){\circle{4}}\put(-21,21){\circle{4}}\put(-7,21){\circle{4}}
\put(-33,21){\vector(1,0){10}}\put(-19,21){\vector(1,0){10}}
\put(-49,7){\circle{4}}\put(-35,7){\circle{4}}\put(-21,7){\circle{4}}
\put(-37,7){\vector(-1,0){10}}\put(-23,7){\vector(-1,0){10}}
=24\,\left(\matrix{2&0&1&0\cr1&1&1&0\cr0&1&1&1\cr0&1&0&2\cr}\right)\,,
&\put(-56,-14){\grid(56,42)(14,14)}
\put(-49,21){\circle{4}}\put(-35,21){\circle{4}}\put(-21,21){\circle{4}}\put(-7,21){\circle{4}}
\put(-47,21){\vector(1,0){10}}\put(-33,21){\vector(1,0){10}}\put(-19,21){\vector(1,0){10}}
\put(-21,7){\circle{4}}\put(-7,7){\circle{4}}\put(-19,7){\line(1,0){10}}
=48\,\left(\matrix{2&1&0&0\cr0&2&1&0\cr0&0&1&2\cr1&0&1&1\cr}\right)\,,\cr
&\cr
\put(-56,-14){\grid(56,42)(14,14)}
\put(-49,21){\circle{4}}\put(-35,21){\circle{4}}\put(-21,21){\circle{4}}\put(-7,21){\circle{4}}
\put(-47,21){\vector(1,0){10}}\put(-33,21){\vector(1,0){10}}\put(-19,21){\vector(1,0){10}}
\put(-35,7){\circle{4}}\put(-7,7){\circle{4}}\put(-33,7){\line(1,0){24}}
=24\,\left(\matrix{2&1&0&0\cr0&1&1&1\cr0&0&2&1\cr1&1&0&1\cr}\right)\,,
&\put(-56,-14){\grid(56,42)(14,14)}
\put(-49,21){\circle{4}}\put(-35,21){\circle{4}}\put(-21,21){\circle{4}}\put(-7,21){\circle{4}}
\put(-47,21){\vector(1,0){10}}\put(-33,21){\vector(1,0){10}}\put(-19,21){\vector(1,0){10}}
\put(-49,7){\circle{4}}\put(-35,7){\circle{4}}\put(-47,7){\line(1,0){10}}
\put(-21,7){\circle{4}}\put(-7,7){\circle{4}}\put(-19,7){\line(1,0){10}}
=-24\,\left(\matrix{1&2&0&0\cr1&1&1&0\cr0&0&1&2\cr1&0&1&1\cr}\right)\,,\cr
&\cr
\end{array}
\label{E1b}
\end{equation}

\newpage
\section{Grids for $r=6$, $n=3$}

$$
\begin{array}{@{}l@{\hspace{49pt}}l@{}}
{\put(-42,-42){\grid(42,84)(14,14)}}
=\left(\matrix{6&0&0\cr0&6&0\cr0&0&6\cr}\right)\,,
&{\put(-42,-42){\grid(42,84)(14,14)}}
\put(-21,35){\circle{4}}\put(-7,35){\circle{4}}\put(-19,35){\line(1,0){10}}
=-15\,\left(\matrix{6&0&0\cr0&5&1\cr0&1&5\cr}\right)\,,\cr
&\cr
{\put(-42,-42){\grid(42,84)(14,14)}}
\put(-21,35){\circle{4}}\put(-7,35){\circle{4}}\put(-19,35){\line(1,0){10}}
\put(-21,21){\circle{4}}\put(-7,21){\circle{4}}\put(-19,21){\line(1,0){10}}
=30\,\left(\matrix{6&0&0\cr0&4&2\cr0&2&4\cr}\right)\,,
&{\put(-42,-42){\grid(42,84)(14,14)}}
\put(-21,35){\circle{4}}\put(-7,35){\circle{4}}\put(-19,35){\line(1,0){10}}
\put(-35,21){\circle{4}}\put(-21,21){\circle{4}}\put(-33,21){\line(1,0){10}}
=60\,\left(\matrix{5&1&0\cr1&4&1\cr0&1&5\cr}\right)\,,\cr
&\cr
{\put(-42,-42){\grid(42,84)(14,14)}}
\put(-35,35){\circle{4}}\put(-21,35){\circle{4}}\put(-7,35){\circle{4}}
\put(-33,35){\vector(1,0){10}}\put(-19,35){\vector(1,0){10}}
=10\,\left(\matrix{5&1&0\cr0&5&1\cr1&0&5\cr}\right)\,,
&{\put(-42,-42){\grid(42,84)(14,14)}}
\put(-21,35){\circle{4}}\put(-7,35){\circle{4}}\put(-19,35){\line(1,0){10}}
\put(-21,21){\circle{4}}\put(-7,21){\circle{4}}\put(-19,21){\line(1,0){10}}
\put(-21,7){\circle{4}}\put(-7,7){\circle{4}}\put(-19,7){\line(1,0){10}}
=-30\,\left(\matrix{6&0&0\cr0&3&3\cr0&3&3\cr}\right)\,,\cr
&\cr
{\put(-42,-42){\grid(42,84)(14,14)}}
\put(-21,35){\circle{4}}\put(-7,35){\circle{4}}\put(-19,35){\line(1,0){10}}
\put(-21,21){\circle{4}}\put(-7,21){\circle{4}}\put(-19,21){\line(1,0){10}}
\put(-35,7){\circle{4}}\put(-21,7){\circle{4}}\put(-33,7){\line(1,0){10}}
=-180\,\left(\matrix{5&1&0\cr1&3&2\cr0&2&4\cr}\right)\,,
&{\put(-42,-42){\grid(42,84)(14,14)}}
\put(-21,35){\circle{4}}\put(-7,35){\circle{4}}\put(-19,35){\line(1,0){10}}
\put(-35,21){\circle{4}}\put(-21,21){\circle{4}}\put(-33,21){\line(1,0){10}}
\put(-35,7){\circle{4}}\put(-7,7){\circle{4}}\put(-33,7){\line(1,0){24}}
=-60\,\left(\matrix{4&1&1\cr1&4&1\cr1&1&4\cr}\right)\,,\cr
&\cr
\end{array}
\label{F1a}
$$

$$
\begin{array}{@{}l@{\hspace{49pt}}l@{}}
{\put(-42,-42){\grid(42,84)(14,14)}}
\put(-21,35){\circle{4}}\put(-7,35){\circle{4}}\put(-19,35){\line(1,0){10}}
\put(-21,21){\circle{4}}\put(-7,21){\circle{4}}\put(-19,21){\line(1,0){10}}
\put(-35,7){\circle{4}}\put(-21,7){\circle{4}}\put(-33,7){\line(1,0){10}}
\put(-35,-7){\circle{4}}\put(-7,-7){\circle{4}}\put(-33,-7){\line(1,0){24}}
=180\,\left(\matrix{4&1&1\cr1&3&2\cr1&2&3\cr}\right)\,,
&{\put(-42,-42){\grid(42,84)(14,14)}}
\put(-35,35){\circle{4}}\put(-21,35){\circle{4}}\put(-7,35){\circle{4}}
\put(-33,35){\vector(1,0){10}}\put(-19,35){\vector(1,0){10}}
\put(-21,21){\circle{4}}\put(-7,21){\circle{4}}\put(-19,21){\line(1,0){10}}
\put(-21,7){\circle{4}}\put(-7,7){\circle{4}}\put(-19,7){\line(1,0){10}}
=180\,\left(\matrix{5&1&0\cr0&3&3\cr1&2&3\cr}\right)\,,\cr
&\cr
{\put(-42,-42){\grid(42,84)(14,14)}}
\put(-35,35){\circle{4}}\put(-21,35){\circle{4}}\put(-7,35){\circle{4}}
\put(-33,35){\vector(1,0){10}}\put(-19,35){\vector(1,0){10}}
\put(-21,21){\circle{4}}\put(-7,21){\circle{4}}\put(-19,21){\line(1,0){10}}
\put(-35,7){\circle{4}}\put(-21,7){\circle{4}}\put(-33,7){\line(1,0){10}}
=360\,\left(\matrix{4&2&0\cr1&3&2\cr1&1&4\cr}\right)\,,
&{\put(-42,-42){\grid(42,84)(14,14)}}
\put(-35,35){\circle{4}}\put(-21,35){\circle{4}}\put(-7,35){\circle{4}}
\put(-33,35){\vector(1,0){10}}\put(-19,35){\vector(1,0){10}}
\put(-35,21){\circle{4}}\put(-21,21){\circle{4}}\put(-7,21){\circle{4}}
\put(-33,21){\vector(1,0){10}}\put(-19,21){\vector(1,0){10}}
=20\,\left(\matrix{4&2&0\cr0&4&2\cr2&0&4\cr}\right)\,,\cr
&\cr
\end{array}
\label{F1b}
$$

$$
\begin{array}{@{}l@{\hspace{49pt}}l@{}}
{\put(-42,-42){\grid(42,84)(14,14)}}
\put(-35,35){\circle{4}}\put(-21,35){\circle{4}}\put(-7,35){\circle{4}}
\put(-33,35){\vector(1,0){10}}\put(-19,35){\vector(1,0){10}}
\put(-35,21){\circle{4}}\put(-21,21){\circle{4}}\put(-33,21){\line(1,0){10}}
\put(-35,7){\circle{4}}\put(-21,7){\circle{4}}\put(-33,7){\line(1,0){10}}
\put(-21,-7){\circle{4}}\put(-7,-7){\circle{4}}\put(-19,-7){\line(1,0){10}}
=-360\,\left(\matrix{3&3&0\cr2&2&2\cr1&1&4\cr}\right)\,,
&{\put(-42,-42){\grid(42,84)(14,14)}}
\put(-35,35){\circle{4}}\put(-21,35){\circle{4}}\put(-7,35){\circle{4}}
\put(-33,35){\vector(1,0){10}}\put(-19,35){\vector(1,0){10}}
\put(-21,21){\circle{4}}\put(-7,21){\circle{4}}\put(-19,21){\line(1,0){10}}
\put(-35,7){\circle{4}}\put(-21,7){\circle{4}}\put(-33,7){\line(1,0){10}}
\put(-35,-7){\circle{4}}\put(-7,-7){\circle{4}}\put(-33,-7){\line(1,0){24}}
=-240\,\left(\matrix{3&2&1\cr1&3&2\cr2&1&3\cr}\right)\,,\cr
&\cr
{\put(-42,-42){\grid(42,84)(14,14)}}
\put(-35,35){\circle{4}}\put(-21,35){\circle{4}}\put(-7,35){\circle{4}}
\put(-33,35){\vector(1,0){10}}\put(-19,35){\vector(1,0){10}}
\put(-35,21){\circle{4}}\put(-21,21){\circle{4}}\put(-7,21){\circle{4}}
\put(-33,21){\vector(1,0){10}}\put(-19,21){\vector(1,0){10}}
\put(-21,7){\circle{4}}\put(-7,7){\circle{4}}\put(-19,7){\line(1,0){10}}
=-180\,\left(\matrix{4&2&0\cr0&3&3\cr2&1&3\cr}\right)\,,
&{\put(-42,-42){\grid(42,84)(14,14)}}
\put(-35,35){\circle{4}}\put(-21,35){\circle{4}}\put(-7,35){\circle{4}}
\put(-33,35){\vector(1,0){10}}\put(-19,35){\vector(1,0){10}}
\put(-35,21){\circle{4}}\put(-21,21){\circle{4}}\put(-7,21){\circle{4}}
\put(-23,21){\vector(-1,0){10}}\put(-9,21){\vector(-1,0){10}}
\put(-21,7){\circle{4}}\put(-7,7){\circle{4}}\put(-19,7){\line(1,0){10}}
=-180\,\left(\matrix{4&1&1\cr1&3&2\cr1&2&3\cr}\right)\,,\cr
&\cr
\end{array}
\label{F1c}
$$

$$
\begin{array}{@{}l@{\hspace{49pt}}l@{}}
{\put(-42,-42){\grid(42,84)(14,14)}}
\put(-35,35){\circle{4}}\put(-21,35){\circle{4}}\put(-7,35){\circle{4}}
\put(-33,35){\vector(1,0){10}}\put(-19,35){\vector(1,0){10}}
\put(-35,21){\circle{4}}\put(-21,21){\circle{4}}\put(-7,21){\circle{4}}
\put(-23,21){\vector(-1,0){10}}\put(-9,21){\vector(-1,0){10}}
\put(-21,7){\circle{4}}\put(-7,7){\circle{4}}\put(-19,7){\line(1,0){10}}
\put(-35,-7){\circle{4}}\put(-21,-7){\circle{4}}\put(-33,-7){\line(1,0){10}}
=360\,\left(\matrix{3&2&1\cr2&2&2\cr1&2&3\cr}\right)\,,
&{\put(-42,-42){\grid(42,84)(14,14)}}
\put(-35,35){\circle{4}}\put(-21,35){\circle{4}}\put(-7,35){\circle{4}}
\put(-33,35){\vector(1,0){10}}\put(-19,35){\vector(1,0){10}}
\put(-35,21){\circle{4}}\put(-21,21){\circle{4}}\put(-7,21){\circle{4}}
\put(-33,21){\vector(1,0){10}}\put(-19,21){\vector(1,0){10}}
\put(-35,7){\circle{4}}\put(-21,7){\circle{4}}\put(-7,7){\circle{4}}
\put(-33,7){\vector(1,0){10}}\put(-19,7){\vector(1,0){10}}
=20\,\left(\matrix{3&3&0\cr0&3&3\cr3&0&3\cr}\right)\,,\cr
&\cr
\end{array}
\label{F1d}
$$

$$
\begin{array}{@{}l@{\hspace{49pt}}l@{}}
{\put(-42,-42){\grid(42,84)(14,14)}}
\put(-35,35){\circle{4}}\put(-21,35){\circle{4}}\put(-7,35){\circle{4}}
\put(-33,35){\vector(1,0){10}}\put(-19,35){\vector(1,0){10}}
\put(-35,21){\circle{4}}\put(-21,21){\circle{4}}\put(-7,21){\circle{4}}
\put(-33,21){\vector(1,0){10}}\put(-19,21){\vector(1,0){10}}
\put(-35,7){\circle{4}}\put(-21,7){\circle{4}}\put(-7,7){\circle{4}}
\put(-33,7){\vector(1,0){10}}\put(-19,7){\vector(1,0){10}}
\put(-21,-7){\circle{4}}\put(-7,-7){\circle{4}}\put(-19,-7){\line(1,0){10}}
\put(-21,-21){\circle{4}}\put(-7,-21){\circle{4}}\put(-19,-21){\line(1,0){10}}
=60\,\left(\matrix{3&3&0\cr0&1&5\cr3&2&1\cr}\right)\,,
&{\put(-42,-42){\grid(42,84)(14,14)}}
\put(-35,35){\circle{4}}\put(-21,35){\circle{4}}\put(-7,35){\circle{4}}
\put(-33,35){\vector(1,0){10}}\put(-19,35){\vector(1,0){10}}
\put(-35,21){\circle{4}}\put(-21,21){\circle{4}}\put(-7,21){\circle{4}}
\put(-33,21){\vector(1,0){10}}\put(-19,21){\vector(1,0){10}}
\put(-35,7){\circle{4}}\put(-21,7){\circle{4}}\put(-7,7){\circle{4}}
\put(-33,7){\vector(1,0){10}}\put(-19,7){\vector(1,0){10}}
\put(-21,-7){\circle{4}}\put(-7,-7){\circle{4}}\put(-19,-7){\line(1,0){10}}
\put(-35,-21){\circle{4}}\put(-21,-21){\circle{4}}\put(-33,-21){\line(1,0){10}}
=120\,\left(\matrix{2&4&0\cr1&1&4\cr3&1&2\cr}\right)\,,\cr
&\cr
\end{array}
\label{F1e}
$$

\begin{equation}
{\put(-42,-42){\grid(42,84)(14,14)}}
\put(-35,35){\circle{4}}\put(-21,35){\circle{4}}\put(-7,35){\circle{4}}
\put(-33,35){\vector(1,0){10}}\put(-19,35){\vector(1,0){10}}
\put(-35,21){\circle{4}}\put(-21,21){\circle{4}}\put(-7,21){\circle{4}}
\put(-33,21){\vector(1,0){10}}\put(-19,21){\vector(1,0){10}}
\put(-35,7){\circle{4}}\put(-21,7){\circle{4}}\put(-7,7){\circle{4}}
\put(-33,7){\vector(1,0){10}}\put(-19,7){\vector(1,0){10}}
\put(-35,-7){\circle{4}}\put(-21,-7){\circle{4}}\put(-7,-7){\circle{4}}
\put(-23,-7){\vector(-1,0){10}}\put(-9,-7){\vector(-1,0){10}}
\put(-35,-21){\circle{4}}\put(-21,-21){\circle{4}}\put(-7,-21){\circle{4}}
\put(-23,-21){\vector(-1,0){10}}\put(-9,-21){\vector(-1,0){10}}
=20\,\left(\matrix{1&3&2\cr2&1&3\cr3&2&1\cr}\right)\,.
\label{F1f}
\end{equation}



\begin{thebibliography}{XX}

\bibitem{ALP}A. Ac{\'\i}n, J. I. Latorre and P. Pascual, {\it Three--party entanglement from positronium}, Phys. Rev. A {\bf 63}, 042107 (2001).

\bibitem{AS1}D. G. Antzoulatos and A. A. Sawchuk, {\it Hypermatrix Algebra: Theory}, CVGIP: Image Understanding {\bf 57}, 24 (1993).

\bibitem{AS2}D. G. Antzoulatos and A. A. Sawchuk, {\it Hypermatrix Algebra: Applications in Parallel Imaging Processing}, CVGIP: Image Understanding {\bf 57}, 42 (1993).

\bibitem{As}G. S. Asanov, {\it Finsler Geometry, Relativity and Gauge Theories} (Reidel, Dordrecht, 1985).

\bibitem{Ca}A. Cayley, {\it On the Theory of Linear Transformations}, Cambridge Math. J. {\bf 4}, 193 (1845).

\bibitem{CKW}V. Coffman, J. Kundu and W. K. Wootters, {\it Distributed entanglement}, Phys. Rev. A {\bf 61}, 052306 (2000).

\bibitem{Cr}C. M. Cramlet, {\it The derivation of algebraic invariants by tensor algebra}, Bull. Am. Math. Soc. {\bf 34}, 334 (1928).

\bibitem{GKZ1}I. M. Gelfand, M. M. Kapranov and A. V. Zelevinsky, {\it Hyperdeterminants}, Adv. Math. {\bf 96}, 226 (1992).

\bibitem{GKZ2}I. M. Gelfand, M. M. Kapranov and A. V. Zelevinsky, {\it Discriminants, Resultants, and  Multidimensional Determinants} (Birkh\"auser, Boston, 1994).

\bibitem{Gu1}H. Gupta, {\it A table of partitions}, Proc. London Math. Soc. {\bf 39}, 142 (1935).

\bibitem{Gu2}H. Gupta, {\it A table of partitions (II)}, Proc. London Math. Soc. {\bf 42}, 546 (1937).

\bibitem{Hu1}C. M. Hull, {\it Strongly coupled gravity and duality}, Nucl. Phys. B {\bf 583}, 237 (2000).

\bibitem{Hu2}C. M. Hull, {\it Symetries and compactifications of $(4,0)$ conformal gravity}, J. High Energy Phys. {\bf 12}, 007 (2000).

\bibitem{Hu3}C. M. Hull, {\it Duality in gravity and higher spin gauge fields}, J. High Energy Phys. {\bf 9}, 027 (2001).

\bibitem{McM}P. A. MacMahon, {\it Combinatory Analysis} (Cambridge University Press, 1915); reprinting (Chel\-sea, New York, 1960).

\bibitem{Ru}H. Rund, {\it The Differential Geometry of Finsler Spaces} (Springer, Berlin, 1959).

\bibitem{SS}M. L. Stein and P. R. Stein, {\it Enumeration of Stochastic Matrices with Integer Elements}, Los Alamos Scientific Laboratory, report LA--4434 (1970).

\bibitem{St}R. P. Stanley, {\it Linear Homogeneous Diophantine Equations and Magic Labelings of Graphs}, Duke Math. J. {\bf 40}, 607 (1973).

\bibitem{Ta93}V. Tapia, {\it Integrable Conformal Field Theory in Four Dimensions and Fourth--Rank Geometry}, Int. J. Mod. Phys. D {\bf 3}, 413 (1993).

\bibitem{TRMC}V. Tapia, D. K. Ross, A. L. Marrakchi and M. Cataldo, {\it Renormalizable Conformally Invariant Model for the Gravitational Field}, Class. Quantum Grav. {\bf 13}, 3261 (1996).

\bibitem{TR}V. Tapia and D. K. Ross, {\it Conformal Fourth--Rank Gravity, Non--Vanishing Cosmological Constant, and Anisotropy}, Class. Quantum Grav. {\bf 15}, 245 (1998).

\bibitem{TU}V. Tapia and M. Ujevic, {\it Universal Field Equations for Metric--Affine Theories of Gravity}, Class. Quantum Grav. {\bf 15}, 3719 (1998).

\bibitem{Ta02}V. Tapia, {\it Canonical Higher--Rank Forms}, work in progress (2002).

\bibitem{WE}J. Weyman, {\it Calculating discriminants by higher direct images}, Trans. Am. Math. Soc. {\bf 343}, 367 (1994).

\bibitem{WZ1}J. Weyman and A. Zelevinsky, {\it Singularities of hyperdeterminants}, Ann. Inst. Fourier, Grenoble {\bf 46}, 591 (1996).

\bibitem{WZ2}J. Weyman and A. V. Zelevinsky, {\it Multiplicative properties of projectively dual varieties}, Manuscripta Math. {\bf 82}, 139 (1994).

\end{thebibliography}
\end{document}